\documentclass{aa}  

\usepackage{graphicx}
\usepackage{grffile}
\usepackage{txfonts}
\usepackage{xcolor}     
\usepackage{subfigure} 

\usepackage{hyperref}

\newcommand{\bd}{\begin{displaymath}}
\newcommand{\ed}{\end{displaymath}}
\newcommand{\be}{\begin{equation}}
\newcommand{\ee}{\end{equation}}
\newcommand{\beaa}{\begin{eqnarray*}}
\newcommand{\eeaa}{\end{eqnarray*}}
\newcommand{\bea}{\begin{eqnarray}}
\newcommand{\eea}{\end{eqnarray}}

\newcommand{\lens}{D_\mathrm{d}}
\newcommand{\source}{D_\mathrm{s}}
\newcommand{\ls}{D_\mathrm{ds}}

\newcommand{\lum}{{D_\mathrm{lum}}}

\newcommand{\dd}{\mathrm{d}}

\newcommand{\Rein}{R_\mathrm{Ein}}

\def\Mch{M_{\rm ch}}
\def\Msun{M_{\sun}}

\def\Ok{\Omega_{\rm k}}

\def\Om{\Omega_{\rm m}}
\def\OL{\Omega_{\Lambda}}

\def\tdist{D_{\Delta t}}

\def\Dd{D_{\rm d}}
\def\Dds{D_{\rm ds}}
\def\Ds{D_{\rm s}}
\def\tdistti{D_{\Delta t, i}^{\rm true}}
\def\Ddti{D_{{\rm d},i}^{\rm true}}
\def\tdistmi{D_{\Delta t, i}^{\rm mock}}
\def\Ddmi{D_{{\rm d},i}^{\rm mock}}
\def\tdisti{D_{\Delta t, i}}
\def\Ddi{D_{{\rm d},i}}
\def\dtdisti{\delta D_{\Delta t, i}}
\def\dDdi{\delta D_{{\rm d},i}}
\def\stdisti{\sigma_{\Delta t, i}}
\def\sDdi{\sigma_{{\rm d},i}}

\def\zd{z_{\rm d}}
\def\zs{z_{\rm s}}

\def\hst{\textit{HST}\xspace}

\def\emcee{{\sc emcee}\xspace}

\def\kmsMpc {\rm km\,s^{-1}\,Mpc^{-1}}

\def\Nsn{N_{\rm SNIa}}

\def\cosmopar{\boldsymbol{\pi}}

\begin{document}

   \title{HOLISMOKES - I. Highly Optimised Lensing Investigations of Supernovae, Microlensing Objects, and Kinematics of Ellipticals and Spirals}

   \subtitle{}

   \titlerunning{HOLISMOKES I. Programme Overview}
   \authorrunning{Suyu, Huber, Ca\~nameras et al.}

   \author{
S.~H.~Suyu\inst{\ref{mpa},\ref{tum},\ref{asiaa}} \and
S.~Huber\inst{\ref{mpa},\ref{tum}} \and
R.~Ca\~nameras\inst{\ref{mpa}} \and
M.~Kromer\inst{\ref{mpia}} \and
S.~Schuldt\inst{\ref{mpa},\ref{tum}} \and
S.~Taubenberger\inst{\ref{mpa}} \and
A.~Y{\i}ld{\i}r{\i}m\inst{\ref{mpa}} \and
V.~Bonvin\inst{\ref{epfl}} \and
J.~H.~H.~Chan\inst{\ref{epfl}} \and
F.~Courbin\inst{\ref{epfl}} \and
U.~N{\"o}bauer\inst{\ref{mpa},\ref{munre}} \and
S.~A.~Sim\inst{\ref{qub}} \and
D.~Sluse\inst{\ref{liege}}
}

   \institute{
Max-Planck-Institut f{\"u}r Astrophysik, Karl-Schwarzschild-Str.~1, 85748 Garching, Germany\\
\email{suyu@mpa-garching.mpg.de} \label{mpa} \goodbreak
\and
Physik-Department, Technische Universit\"at M\"unchen, James-Franck-Stra\ss{}e~1, 85748 Garching, Germany \label{tum} \goodbreak
\and
Academia Sinica Institute of Astronomy and Astrophysics (ASIAA), 11F
of ASMAB, No.1, Section 4, Roosevelt Road, Taipei 10617,
Taiwan \label{asiaa} \goodbreak
\and
Heidelberger Institut f\"{u}r Theoretische Studien, Schloss-Wolfsbrunnenweg 35, 69118 Heidelberg, Germany \label{mpia} \goodbreak
\and
Institute of Physics, Laboratory of Astrophysics, Ecole Polytechnique F{\'e}d{\'e}rale de Lausanne (EPFL), Observatoire de Sauverny, 1290 Versoix, Switzerland \label{epfl} \goodbreak
\and
MunichRe IT 1.6.4.1, K{\"o}niginstraße 107, 80802, Munich, Germany \label{munre} \goodbreak
\and
Astrophysics Research Centre, School of Mathematics and Physics, Queen's University Belfast, Belfast BT7 1NN, UK \label{qub} \goodbreak
\and
STAR Institute, Quartier Agora - All{\'e}e du six Ao{\^u}t, 19c B-4000
Li{\`e}ge, Belgium \label{liege} \goodbreak
             }

   \date{Received February --, 2020; accepted --}

 
   \abstract { We present the HOLISMOKES programme on strong
     gravitational lensing of supernovae as a probe of supernova (SN) physics
     and cosmology.  We investigate the effects of microlensing on
     early-phase SN Ia spectra using four different SN explosion models, and find that within 10 rest-frame
     days after SN explosion, distortions of SN Ia spectra due to
     microlensing are 
     typically negligible ($<$1\% distortion within the 1$\sigma$
     spread, and $\lesssim$$10\%$ distortion within the 2$\sigma$
     spread). 
     This shows great prospects of using lensed SNe Ia to obtain
     intrinsic early-phase SN spectra for deciphering SN Ia
     progenitors.  As a demonstration of the usefulness of lensed SNe Ia for
     cosmology, we simulate a sample of mock lensed SN Ia systems that
     are expected to have accurate and precise time-delay measurements
     in the era of the Rubin Observatory Legacy Survey of Space and Time
     (LSST). Adopting realistic yet conservative uncertainties
     on 
     their time-delay distances and lens angular diameter distances
     (of 6.6\% and 5\%, respectively), we find that a sample of 
     20 lensed SNe Ia 
     would allow a constraint on the Hubble constant ($H_0$) with
     1.3\% uncertainty in the 
     flat $\Lambda$CDM cosmology. We find a similar constraint on
     $H_0$ in an open $\Lambda$CDM cosmology, while the constraint
     degrades to $3\%$ in a flat $w$CDM cosmology.
     We anticipate lensed SNe to be an independent and powerful
     probe of SN physics and cosmology in the upcoming LSST era.}


   \maketitle
%

\section{Introduction}
\label{sec:intro}
 
In the past few years, strongly lensed supernovae (SNe)
have transformed from a theoretical fantasy to reality.  First
envisaged by \citet{refsdal1964} as a cosmological probe, a strongly
lensed SN occurs when a massive object (e.g., a galaxy) by chance lies
between the observer and the SN; the gravitational field of the
massive foreground  object acts like a lens and bends light from the
background SN, so that multiple images of the SN appear around the
foreground lensing object.  The arrival times of the light rays of the
multiple images are different, given the difference in their light
paths.  The time delays between the multiple SN images are typically
days/weeks for galaxy-scale 
foreground lenses, and years for galaxy-cluster-scale foreground
lenses. A strongly lensed SN is thus nature's orchestrated cosmic
fireworks with the same SN explosion appearing multiple times one
after another.  \citet{refsdal1964} showed that the time delays
between the multiple SN images provide a way to measure the
expansion rate of the Universe.

The first strongly lensed SN system with multiple resolved images of
the SN was discovered by \citet{kelly+2015}, half a century after the
prescient \citet{refsdal1964}.  The SN was named ``Supernova
Refsdal'', and its spectroscopy revealed that it was a core-collapse
SN \citep{kelly+2016a}.  It was first detected serendipitously when it appeared in the
galaxy cluster MACSJ1149.6+2223 in the \textit{Hubble Space Telescope}
(\hst) imaging taken as part of the Grism Lens-Amplified Survey from
Space (GLASS; PI: T.~Treu) and the Hubble Frontier Field (PI: J.~Lotz)
programmes.  While this was the first system that showed
spatially-resolved multiple SN images, \citet{quimby+2013,
  quimby+2014} had previously detected a SN in the
PanSTARRS\footnote{Panoramic Survey Telescope and Rapid Response
  System} survey \citep{kaiser+2010, chambers+2016} that was magnified
by a factor of $\sim$$30$ by a foreground intervening galaxy, although
the multiple images of the SN could not be resolved in the imaging.
Two years after the SN Refsdal event was the first discovery of a
strongly lensed Type Ia SN by \citet{goobar+2017} in the intermediate
Palomar Transient Factory \citep{law+2009}, namely the iPTF16geu
system.  This is 
particularly exciting given the standardisable nature of Type Ia
SNe for cosmological studies.

With strongly lensed SNe being discovered, we have new opportunities
of using such systems to study SN physics, particularly SN progenitors.
Strongly lensed SNe allow one to observe a SN explosion right from the
beginning, which was impossible to do in the past given both the
difficulty of finding SNe at very early phases and the time lag
to arrange follow-up observations after a SN is detected.  By
exploiting the time delay between the multiple SN images, the lens
system can be detected based on the first SN image and follow-up
(especially spectroscopic) observations can be carried out on the next
appearing SN image from its beginning.  Early-phase observations
are crucial for understanding the progenitors of SNe, especially Type
Ia SNe whose progenitors are still a puzzle after decades of debate --
are they single-degenerate (SD) systems with a white dwarf (WD)
accreting mass from a nondegenerate companion and exploding when
reaching the Chandrasekhar mass limit \citep[e.g.,][]{whelan+1973}, or
double-degenerate (DD) systems with two WDs merging
\citep[e.g.,][]{tutukov+1981,iben+1984}, a mix of the two, or other mechanisms?
A few SNe Ia now have extremely early light-curve coverage and a UV
excess is observed in some of them \citep[e.g.,][]{dimitriadis+2019},
but there are no rest-frame UV spectra at such early phases to
constrain the origin of the UV emission. A continuum-dominated UV flux
would hint at shocks and interaction of the ejecta with a companion
star or circumstellar matter, which would be prominent for
  $\sim$10\% of the viewing angles, and would favour the SD scenario \citep{Kasen:2010}. A
line-dominated early UV 
spectrum, on the other hand, would probe radioactive material close to
the surface of the SN, as predicted by some DD models \citep{maeda+2018}.

Strongly lensed SNe with time-delay measurements also provide a direct
and independent method to measure the expansion rate of the Universe,
or the Hubble constant ($H_0$), as first pointed out by
\citet{refsdal1964}.  There is currently an intriguing tension in the
measurements of $H_0$ from independent probes, particularly between
the measurement from observations of the Cosmic Microwave Background
(CMB) by the \citet{planck+2020} and the local measurement from
Cepheids distance ladder by the ``Supernovae, $H_0$, for the Equation of
State of Dark Energy'' (SH0ES) programme \citep{riess+2019}.  This
tension, if not due to any unaccounted-for measurement uncertainties,
has great implications for cosmology as it would require new physics
beyond our current standard ``flat $\Lambda$CDM'' cosmological model.
The latest $H_0$ measurement from the Megamaser Cosmology Project by
\citet{pesce+2020}, which is independent of the CMB and SH0ES, corroborates
the measurement of SH0ES, although it is within 3$\sigma$ of the
Planck measurement.
On the other hand, \citet{freedman+2019} measured $H_0$ that is right in between
the values from \citet{planck+2020} and \citet{riess+2019} through
the Carnegie-Chicago Hubble Program \citep[CCHP;][]{beaton+2016} using
a separate distance calibrator, the tip of the red giants, instead of
Cepheids.  There is ongoing debate about the method
\citep[e.g.,][]{yuan+2019, freedman+2020} and the results from CCHP and SH0ES are not
fully independent due to calibrating sources/data that are common
among the two distance ladders.  Strong-lensing time delays are
therefore highly valuable for providing a direct $H_0$ measurement,
completely independent of the CMB, the distance ladder, and megamasers
\citep{riess2019}.

Given the rarity of lensed SNe, the method of time-delay cosmography has
matured in the past two decades using lensed quasars which are more
abundant.  The H0LiCOW \citep{suyu+2017} and COSMOGRAIL
\citep{courbin+2018} collaborations have greatly refined this
technique using high-quality data and state-of-the-art analyses of
lensed quasars. The latest H0LiCOW $H_0$ measurement by
\citet{wong+2020} from the analyses of 6 lens systems
\citep{suyu+2010, suyu+2014, wong+2017, birrer+2019, 
  jee+2019, chen+2019, rusu+2020}, which include 3 systems analysed jointly
  with the SHARP collaboration
  \citep{chen+2019}, is consistent with the results from SH0ES and is
$>$$3\sigma$ higher than the value from the \citet{planck+2020},
strengthening the argument for new physics.  Analysis of new lensed
quasars is underway \citep[e.g.,][from the STRIDES collaboration]{shajib+2020}, and a detailed
account of systematic uncertainties in such measurements is presented
by \citet{millon+2020} under the new TDCOSMO organisation.  With time-delay cosmography maturing through
lensed quasar, lensed SNe are expected to be a powerful cosmological
probe.

The two known lensed SN systems, iPTF16geu and SN Refsdal, do not
have early-phase spectroscopic observations for progenitor studies, and have yet
to yield $H_0$ measurements.  The time delays between the four SN
images in iPTF16geu are short, $\lesssim$$1$ day \citep{more+2017,
  dhawan+2019}, and all four SN images were past the early phase when
the system was discovered by \citet{goobar+2017}.  The short delays
also make it difficult to obtain precise $H_0$ from this system, since
the relative uncertainties in the delays \citep[which are
$\gtrsim$$50\%$;][]{dhawan+2019} sets the lower limit on the relative
uncertainty on $H_0$.  On the other hand, SN Refsdal has one long time
delay between the SN images \citep[$\sim$$1$ year;][]{treu+2016,
  grillo+2016, kawamata+2016}, in addition to shorter delay pairs
\citep{rodney+2016}.  The reappearance of the long-delayed SN Refsdal
image was detected by \citet{kelly+2016b}, providing an approximate
time-delay measurement.  The precise measurement of the long delay using multiple
techniques is forthcoming (P.~Kelly, priv.~comm.), and this
spectacular cluster lens system with multiple sources at different
redshifts could yield the first $H_0$ measurement from a lensed SN
\citep[e.g.,][]{grillo+2018, grillo+2020}. 

Even though lensed SNe are very rare, their numbers will increase
dramatically in the coming years thanks to dedicated wide-field
cadenced imaging surveys.  In particular, \citet{goldstein+2019}
forecasted about a dozen lensed SNe from the ongoing Zwicky
Transient Facility \citep[ZTF;][]{bellm+2019, masci+2019}; most of these lensed SNe will be systems with short
time delays (days) and high magnifications, given the bright flux limit of
the ZTF survey.  The upcoming Rubin Observatory Legacy Survey of Space and
Time \citep[LSST;][]{ivezic+2019}\footnote{LSST is previously known as the Large Synoptic Survey Telescope.} that will image the
entire southern sky repeatedly 
for 10 years will yield hundreds of lensed SNe
\citep[e.g.,][]{ogurimarshall2010, goldstein+2019, wojtak+2019}.  The
efficiency of detecting these systems and measuring their time delays
depends significantly on the observing cadence strategy.
\citet{huber+2019} have carried out the first investigations of
detecting lensed SNe Ia and measuring their delays in the presence of
microlensing, with results that favour long cumulative season length
and higher cadence.

With the upcoming boom in strongly lensed SNe, we initiate the
HOLISMOKES programme: Highly Optimised Lensing Investigations of
Supernovae, Microlensing Objects, and Kinematics of Ellipticals and
Spirals.  We are developing ways to find lensed SNe
\citep[][HOLISMOKES II]{Canameras+2020} in current/future cadenced surveys and
to model the lens systems rapidly for scheduling observational
follow-up \citep[][HOLISMOKES IV]{Schuldt+2020}. We are also exploring in more detail
the microlensing of lensed SNe Ia \citep[][HOLISMOKES III]{Huber+2020} and core-collapse SNe (Bayer et al.~in prep., HOLISMOKES V) for
measuring the time delays, following the works of
\citet{goldstein+2018} and \citet{huber+2019}.

In this first paper of the HOLISMOKES series, we study and forecast
our ability to achieve two scientific goals with a sample of lenses
from the upcoming LSST: constrain SN Ia progenitors through early-phase
observations, and probe cosmology through lensing time delays.  In
Section \ref{sec:microlensSN}, we investigate microlensing effects on
SNe Ia to determine whether it is feasible to 
extract the intrinsic early-phase SN spectra that are crucial for
revealing SN Ia progenitors.  In Section \ref{sec:cosmo}, we forecast the
cosmological constraints based on an expected sample of lensed SNe
from LSST.  We summarize in Section \ref{sec:summary}.

\section{Microlensing of SNe Ia in their early phases}
\label{sec:microlensSN}

Early-phase spectra (within $\sim$$5$ rest-frame days after explosion) carry valuable information to distinguish between
different SN Ia progenitors
\citep[e.g.,][]{Kasen:2010,Rabinak:2011,Piro:2013,Piro:2014,Piro:2015,Noebauer:2017vsf}. Problems
arise when SNe are significantly influenced by microlensing
\citep{Yahalomi:2017ihe,goldstein+2018,Foxley-Marrable:2018dzu,bonvin+2019b,huber+2019},
which distorts light curves and spectra, and therefore makes them hard
to use as a probe for SN Ia progenitors. However, investigations by
\citet{goldstein+2018} and \citet{huber+2019} show that microlensing
of lensed SNe Ia is stronger in late phases than shortly after
explosion. These results raise the hope to use lensed SNe Ia for the
progenitor problem and motivates further investigation of the
influence of microlensing on early-phase spectra.

In Section \ref{sec:microlensSN:SNmodels}, we describe four explosion
models for different SN progenitor scenarios from the ARTIS
simulations \citep{Kromer:2009ce} that we use.  We then outline the
microlensing formalism in Section \ref{sec:microlensSN:micro}, before
presenting our results on the microlensed SN Ia spectra in Section
\ref{sec:microlensSN:results}.

\subsection{SN Ia models from ARTIS simulations}
\label{sec:microlensSN:SNmodels}
To probe the effect of microlensing on SNe Ia, we need the time, wavelength, and spatial
dependency of the SN radiation. 
For this, we consider four theoretical explosion models where synthetic
  observables have been calculated via \texttt{ARTIS}
  \citep{Kromer:2009ce}. We briefly describe these models below, and
  refer the readers to, e.g., \citet{Noebauer:2017vsf} for more
  details. These models allow us to explore various progenitor
  scenarios.

\renewcommand{\labelitemi}{$\bullet$}
\begin{itemize}
\setlength\itemsep{0.2\baselineskip}

\item W7 (carbon deflagration):\\ The W7 model \citep{1984:Nomoto} is
  considered one of the benchmark theoretical models for the explosion
  of a carbon-oxygen (CO) white dwarf (WD) at the Chandrasekhar mass
  limit $\Mch$ since it reproduces key observable features of SNe
  Ia. W7 is however not a self-consistent explosion model, in
  contrast to the other three models described below. The ARTIS spectral
  calculations for the W7 model is presented in \citet{Kromer:2009ce},
and we use here the calculations with 7 ionisation stages.

\item N100 (delayed detonation):\\ Following
    \citet{roepke+2012}, the $\Mch$ CO WD in this model has 100
    randomly distributed ignition spots that trigger an initial
    subsonic deflagration, which then transitions into a
    detonation. For more details on the ARTIS spectral calculations,
    see \citet{Sim+2013}.

\item subCh (sub-Chandrasekhar mass detonation):\\ This model is a
  centrally ignited detonation of a 1.06$\Msun$ CO WD \citep[model 1.06 of][]{sim+2010}.

\item merger:\\ Following \citet{pakmor+2012}, this model is a violent
  merger of a 0.9 and a 1.1 $\Msun$ CO WD, triggering a carbon
  detonation in the 1.1 $\Msun$ CO WD and disrupting the system.

\end{itemize}

We spherically average the photon packets from these simulations in
order to obtain high signal-to-noise spatial and energy distributions
of the photons from the SN Ia models.  This is valid for W7 and
subCh models that are spherically symmetric, and also a good
approximation for the N100 model which shows minimal large-scale
asymmetry \citep{roepke+2012} and low continuum polarisation of
$\sim$0.1\% \citep{bulla+2016}.  
While the merger model is inherently non-spherically symmetric, we
find the spherical averaging to be a good approximation; by following
\citet{Huber+2020} in separating portions of the photon packets
of the merger model, we find that different portions of the photon
packets yield similar results and our conclusions thus do not depend
on asymmetries in the merger model. 

\begin{figure*}
\centering
 \includegraphics[scale = 1.0]{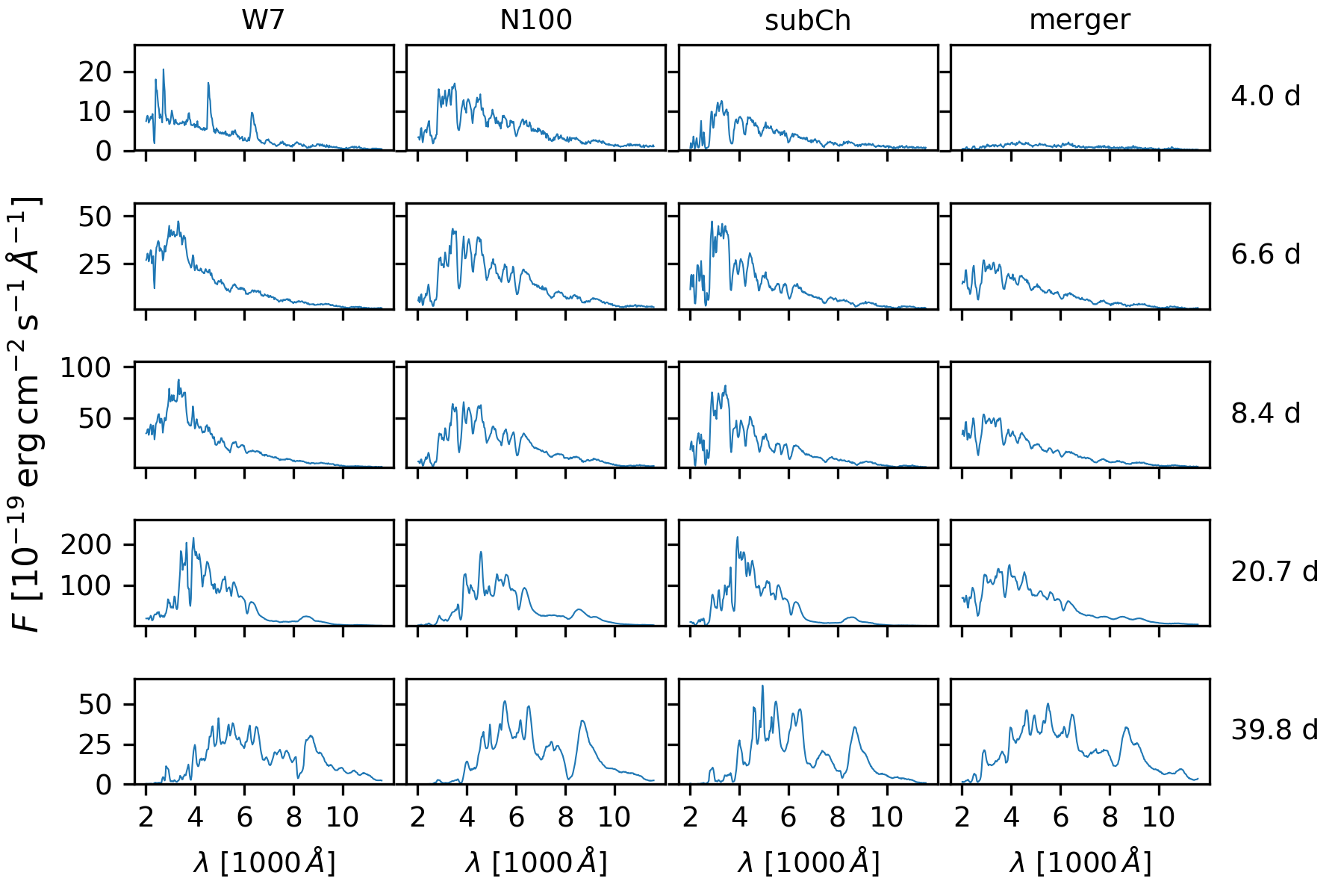}
 \caption{Spectral evolution of four different SNe Ia
     explosion models in rest-frame wavelengths. Columns from left to
     right: W7, N100, subCh and merger models.  Rows from top to
     bottom: rest-frame time after explosion in days, as indicated on
     the right of each row.  In early phases ($\lesssim$$10$ days after
     explosions), the spectra of different SNe Ia models show
     distinguishing features that depend on assumptions in the
     radiative transfer 
     calculations, whereas at later times the spectra from different models start to resemble one another.}
 \label{fig:sn_model_spectral_evol}
\end{figure*}

In Figure \ref{fig:sn_model_spectral_evol}, we show the rest-frame
spectral evolution computed from ARTIS for the four different
explosion models.  Each column corresponds to a particular model as
labelled on the top, and each row is for a rest-frame time $t$ after
explosion as indicated on the right.  The spectra at $t=$ 4.0, 6.6 and
8.4 days are significantly different amongst the models, whereas the
spectra at $t=$ 20.7 and 39.8 days are more similar to each other in
having iron-line blanketing in the UV and relatively strong absorption
lines.  In particular, at $t=$ 4.0 days, W7 has weaker absorption
lines especially in the optical and strong emission lines, whereas
N100 and subCh have strong Ca II 
absorption lines.  W7, N100 and subCh are brighter in the UV relative
to the optical, whereas merger is not.  At $t=$ 6.6 and 8.4 days, N100
has more suppression in flux at wavelengths $<$$3000\AA$ relative to
the optical due to
blended groups of Fe II, Ni II and Mg II absorption lines, whereas W7
and subCh continue to be bright in the UV and merger begins to have
more UV flux relative to the optical.

While Figure \ref{fig:sn_model_spectral_evol} displays different
relative fluxes in the UV to optical and different strengths of the
absorption features in the early SN phases amongst the explosion
models, we caution that the exact spectral shapes of these models
depend on the various approximations used in the radiative transfer
calculations \citep[e.g.,][]{Dessart+2014, Noebauer:2017vsf}.  In
particular, the number of ionisation states and metallicity of the
progenitor could affect particularly the UV spectra at levels
comparable to the differences depicted in Figure
\ref{fig:sn_model_spectral_evol} \citep[e.g.,][]{Lucy1999,
  Kromer:2009ce, Lentz+2000, Kromer+2016, Walker+2012}.
Furthermore, these spectra do not include thermal radiations from
possible interactions of the SN ejecta with a non-degenerate companion
or with circumstellar matter.  Such thermal radiation would only be in
the very early phase within $t\lesssim4$ days though
\citep[e.g.,][]{Kasen:2010}, so distinguishing spectral features after
$\sim$4 days could still be used as diagnostics.  Therefore, first
acquisitions of the early-phase UV spectra would be extremely useful
for providing clues to the explosion/progenitor scenarios, and for
guiding future directions of model developments. Lensed SNe have a further advantage in that the SN are magnified by the lensing effect, typically by a factor of $\sim$10 for lensing galaxies and even $\sim$100 for lensing galaxy clusters, which facilitates spectroscopic observations\footnote{The absolute rest-frame B-band magnitude of a SN Ia at peak is $\sim -19$.  For a SN at redshift 0.5, this corresponds to an apparent magnitude of $\sim$23 without lensing magnifications.  A lensing magnification by a factor of 10 would brighten the apparent magnitude to $\sim$20.5.}. 

\subsection{Microlensing formalism and maps}
\label{sec:microlensSN:micro}

We assume that microlensing maps and positions in the map do not vary
over typical time scales  
of a SN Ia and the 
microlensing effect is therefore just related to the spatial expansion of the SN. This  
approach is motivated by the work
of \cite{goldstein+2018} and \cite{huber+2019}. We follow closely the formalism described in \citet{huber+2019} to compute microlensing effects on a SN Ia, and briefly summarise the procedure.  The observed microlensed flux of a SN at redshift $\zs$ and
luminosity distance $\lum$ can be determined via
\begin{equation}
F_{\lambda,\mathrm{o}}(t)=\frac{1}{\lum^2(1+\zs)}\int \dd x \int \dd y \, I_{\lambda,\mathrm{e}}(t,x,y) \, \mu(x,y),
\label{eq: microlensed flux}
\end{equation}
where the emitted specific intensity $I_{\lambda,\mathrm{e}}(t,x,y)$, is multiplied with the microlensing magnification
map\footnote{Note that $\mu$ denotes the magnification factor and not
  $\cos \theta$ as usually in radiative transfer equations.}
$\mu(x,y)$ from {\tt GERLUMPH} \citep{Vernardos:2015wta, Chan+2020} and integrated over the whole size of 
the projected SN Ia. The specific intensity $I_{\lambda,\mathrm{e}}(t,x,y)$ depends
on the time since explosion $t$, the wavelength $\lambda$, and the radial 
coordinate on the source plane $p = \sqrt{x^2 + y^2}$, given our
spherical averaging of the photon packets from the models\footnote{In
equation (\ref{eq: microlensed flux}) the specific intensity is mapped onto a Cartesian grid ($x$, $y$)
to combine it with the magnification maps $\mu(x,y)$. For more details, see \cite{huber+2019}.}. The specific intensity profiles for 
different times after explosion are shown in Appendix \ref{sec: Specific intensity profiles}.
Equation (\ref{eq: microlensed flux})
is derived and explained in \cite{huber+2019}. We refer readers to \citet{huber+2019} for an example showing the effects of microlensing on spectra and light curves in detail.

For this work we focus on the spectra, particularly at early phases.  We investigate 30 different magnification maps. 
These maps depend on three main parameters: the lensing convergence
$\kappa$, the shear $\gamma$, and the smooth matter fraction
$s=1-\kappa_{\rm *}/\kappa$, where $\kappa_{\rm *}$ is the convergence of the stellar component. In our analysis we probe ($\kappa,
\gamma) = (0.29, 0.27), (0.36, 0.35), (0.43, 0.43), (0.57, 0.58), (0.70, 0.70),\\
(0.93, 0.93)$, where we test for each combination of $\kappa$ and
$\gamma$ the smooth matter fractions of $s = 0.1, 0.3, 0.5, 0.7,
0.9$. Six of these magnification maps are shown in Appendix \ref{sec: Microlensing maps}, 
where we explain also further inputs for producing the magnification maps. 
The values for the convergence and shear are calculated from the mock lens catalog of
\citet[][hereafter OM10]{ogurimarshall2010}, taking into account 416
lensed SNe Ia that adopted a singular isothermal
ellipsoid \citep{Kormann:1994} as lens mass model. The two pairs $(\kappa, \gamma) = (0.36,
0.35)$ and $(0.70, 0.70)$ correspond to the median values for type I
lensing images (time-delay minimum) and type II images (time-delay
saddle), respectively. The other $(\kappa,\gamma)$ pairs are the 16th
and 84th percentiles of the OM10 sample, taken separately for $\kappa$
and
$\gamma$. 

\subsection{Spectral distortions due to microlensing}
\label{sec:microlensSN:results}

\begin{figure*}
\centering
 \includegraphics[scale = 0.85]{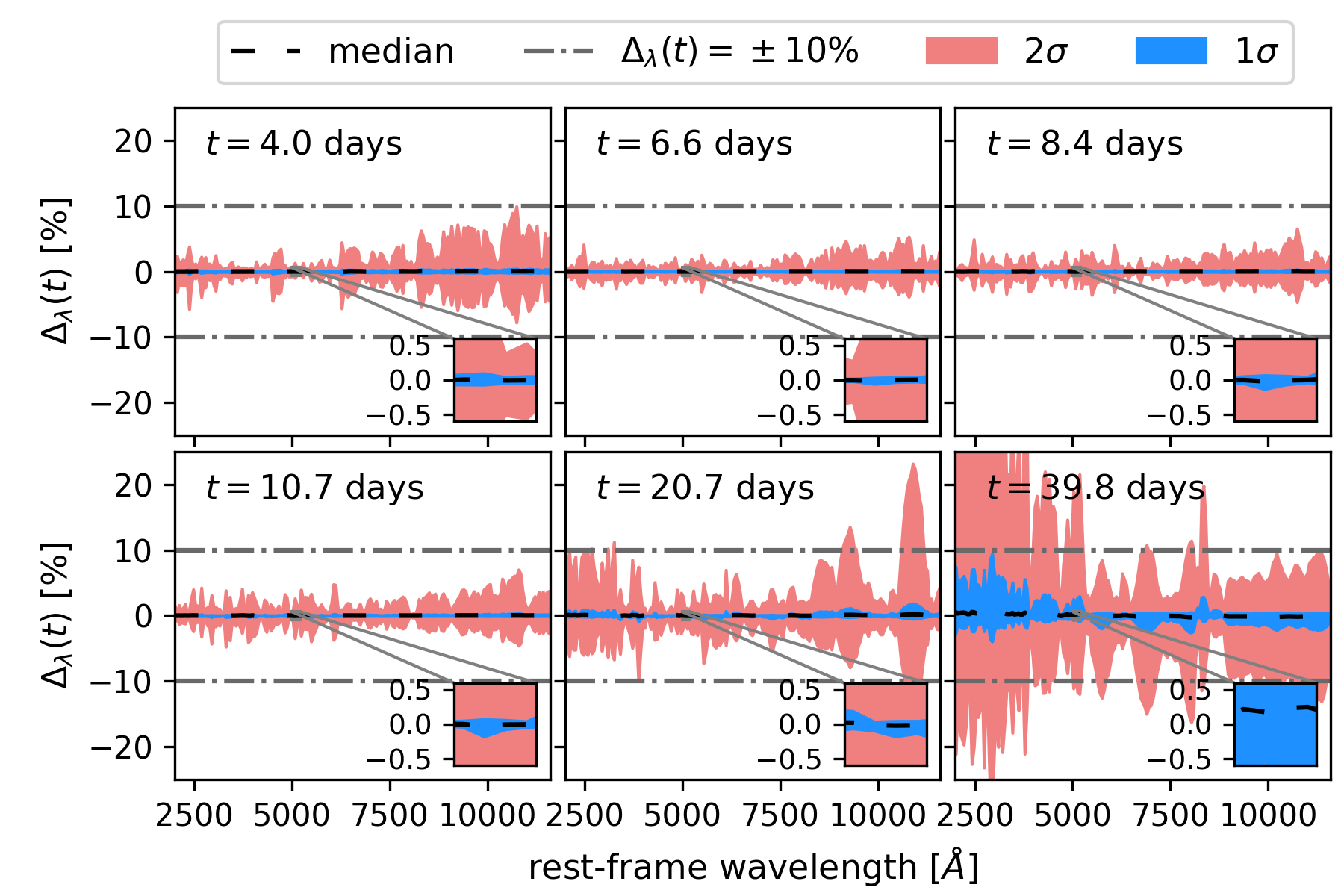}
 \caption{Deviations ($\Delta_\lambda(t)$, see equation (\ref{eq:
     deviation microlensed flux})) of the W7 SN Ia spectra due to microlensing for
   different times after explosion. The black dashed line represents
   the median, and the 1$\sigma$ and
   2$\sigma$ spreads are shown in
   blue and red shades, respectively, for a sample of 30
   different magnification maps with 10,000 random positions per
   map. The grey dot-dashed line indicates a deviation of 10\% in the spectra
   relative to the intrinsic one without microlensing
   effects. The small zoomed-in panels show a region of $150 \AA$ to illustrate the small extent
   of the 1$\sigma$ spread especially at early times. In the early
   phases within $\sim$$10$ rest-frame days after explosion, the
   2$\sigma$ ranges of the deviations are within $10\%$.}
 \label{fig:w7_deviation_of_microlensed_flux}
\end{figure*}

\begin{figure*}
\centering
 \includegraphics[scale = 0.85]{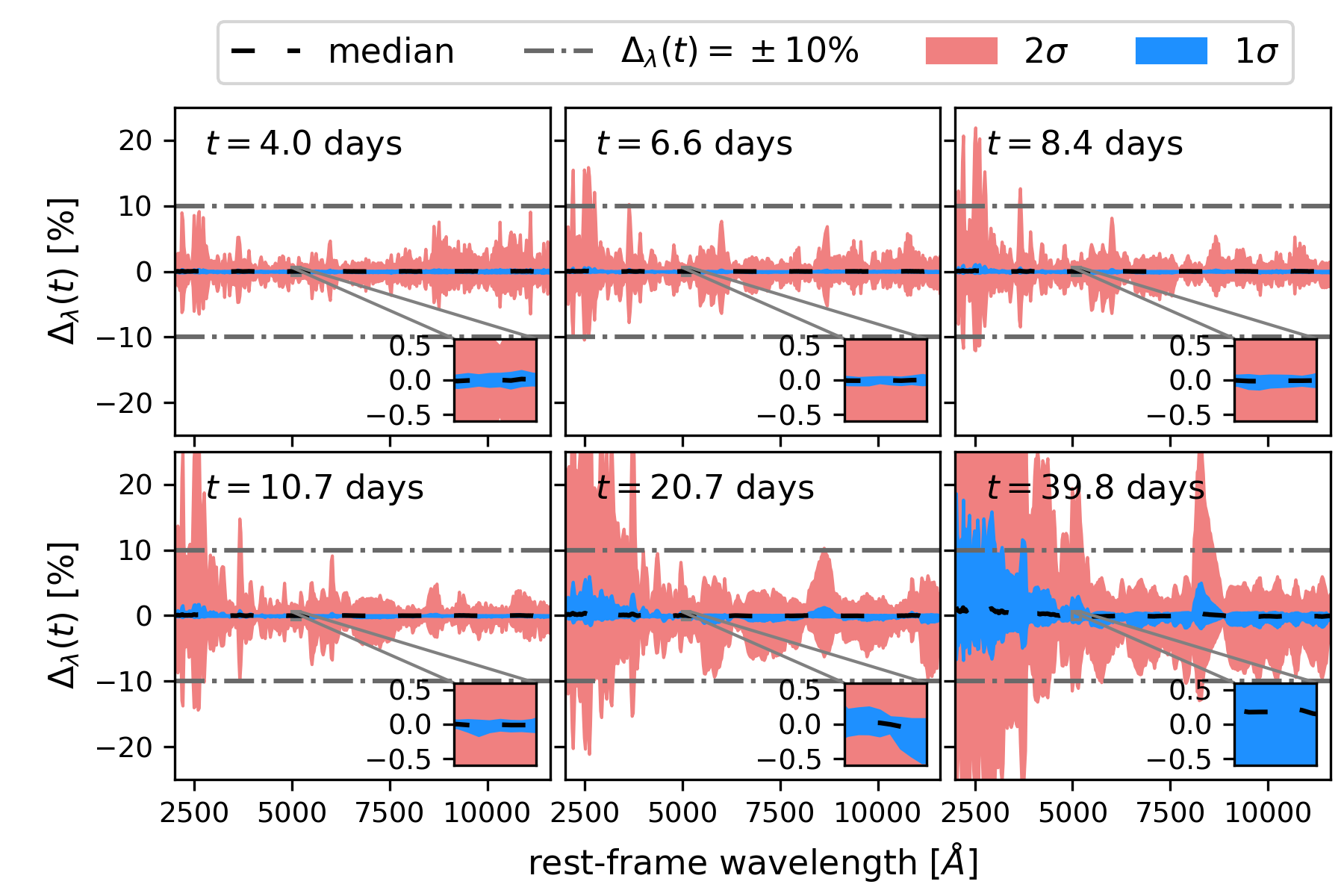}
 \caption{Deviations $\Delta_\lambda(t)$ of the N100 SN Ia spectra
   due to microlensing for different times after explosion.  The
   panels and labels are in the same format as in Figure
   \ref{fig:w7_deviation_of_microlensed_flux}. In the early phase
   ($\lesssim$$10$ rest-frame days after explosion), the 1$\sigma$ range of
   the deviation is within $1\%$ and most of the 2$\sigma$ range
   is within $10\%$ except in the UV where the 2$\sigma$ range could
   reach $\sim$$20\%$ due to the early suppression of UV flux in N100. }  
 \label{fig:n100_deviation_of_microlensed_flux}
\end{figure*}

\begin{figure*}
\centering
 \includegraphics[scale = 0.85]{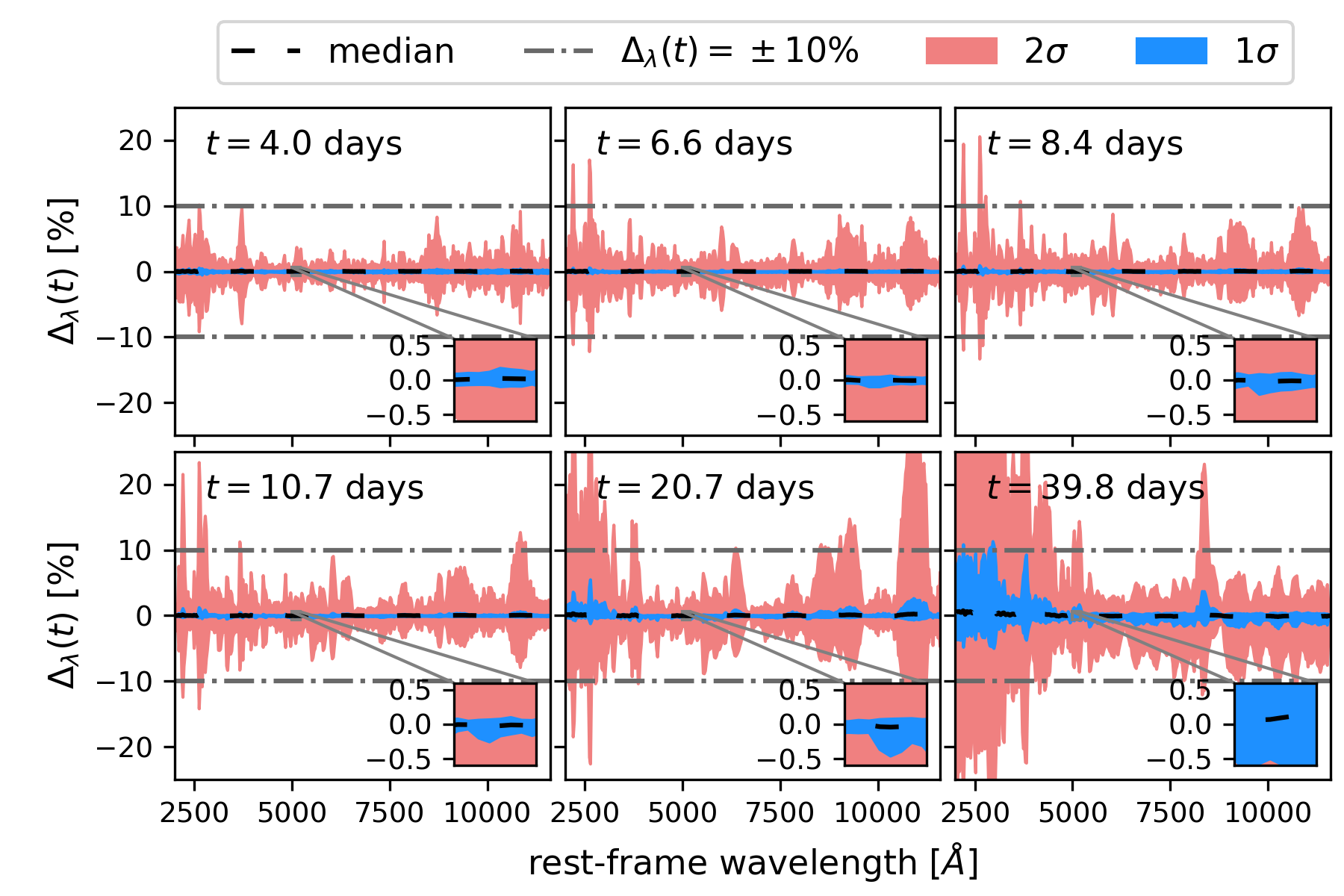}
 \caption{Deviations $\Delta_\lambda(t)$ of the subCh SN Ia spectra
   due to microlensing for different times after explosion.  The
   panels and labels are in the same format as in Figure
   \ref{fig:w7_deviation_of_microlensed_flux}.  In the early phase
   ($\lesssim$$10$ rest-frame days after explosion), the 1$\sigma$ range of
   the deviation is well within $1\%$ and most of the 2$\sigma$ range
   is within $10\%$ except at wavelengths of $\sim$2500$\AA$
   corresponding to absorption features where the 2$\sigma$ deviations could
   reach $20\%$. }
 \label{fig:subCh_deviation_of_microlensed_flux}
\end{figure*}

\begin{figure*}
\centering
 \includegraphics[scale = 0.85]{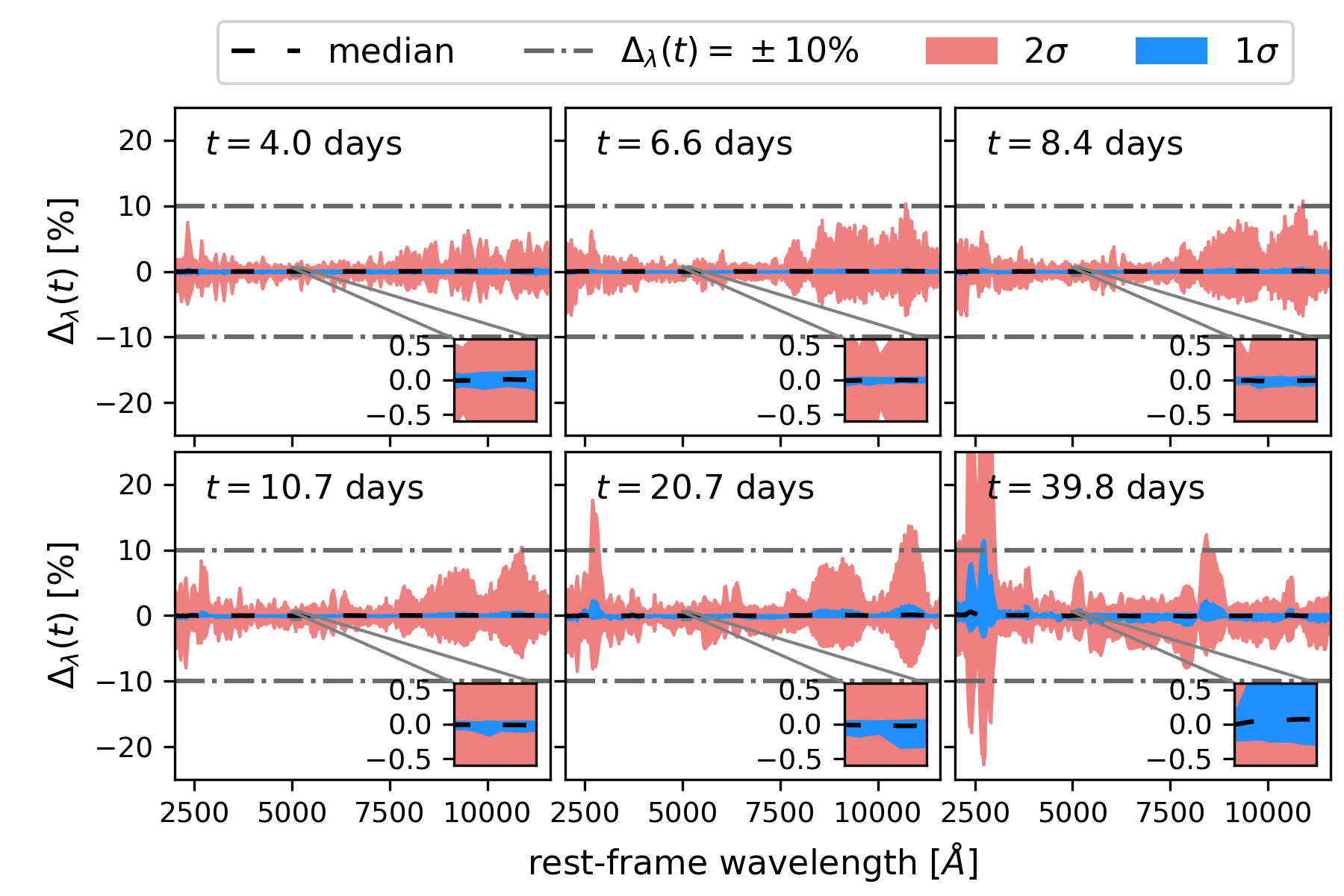}
 \caption{Deviations $\Delta_\lambda(t)$ of the merger SN Ia spectra
   due to microlensing for different times after explosion.  The
   panels and labels are in the same format as in Figure
   \ref{fig:w7_deviation_of_microlensed_flux}. In the early
   phases ($\lesssim$$10$ rest-frame days after explosion), the
   1$\sigma$ and 2$\sigma$ ranges of the deviations are within $1\%$
   and $10\%$, respectively.   }
 \label{fig:merger_deviation_of_microlensed_flux}
\end{figure*}

\begin{figure}
\centering
 \includegraphics[width=0.47\textwidth]{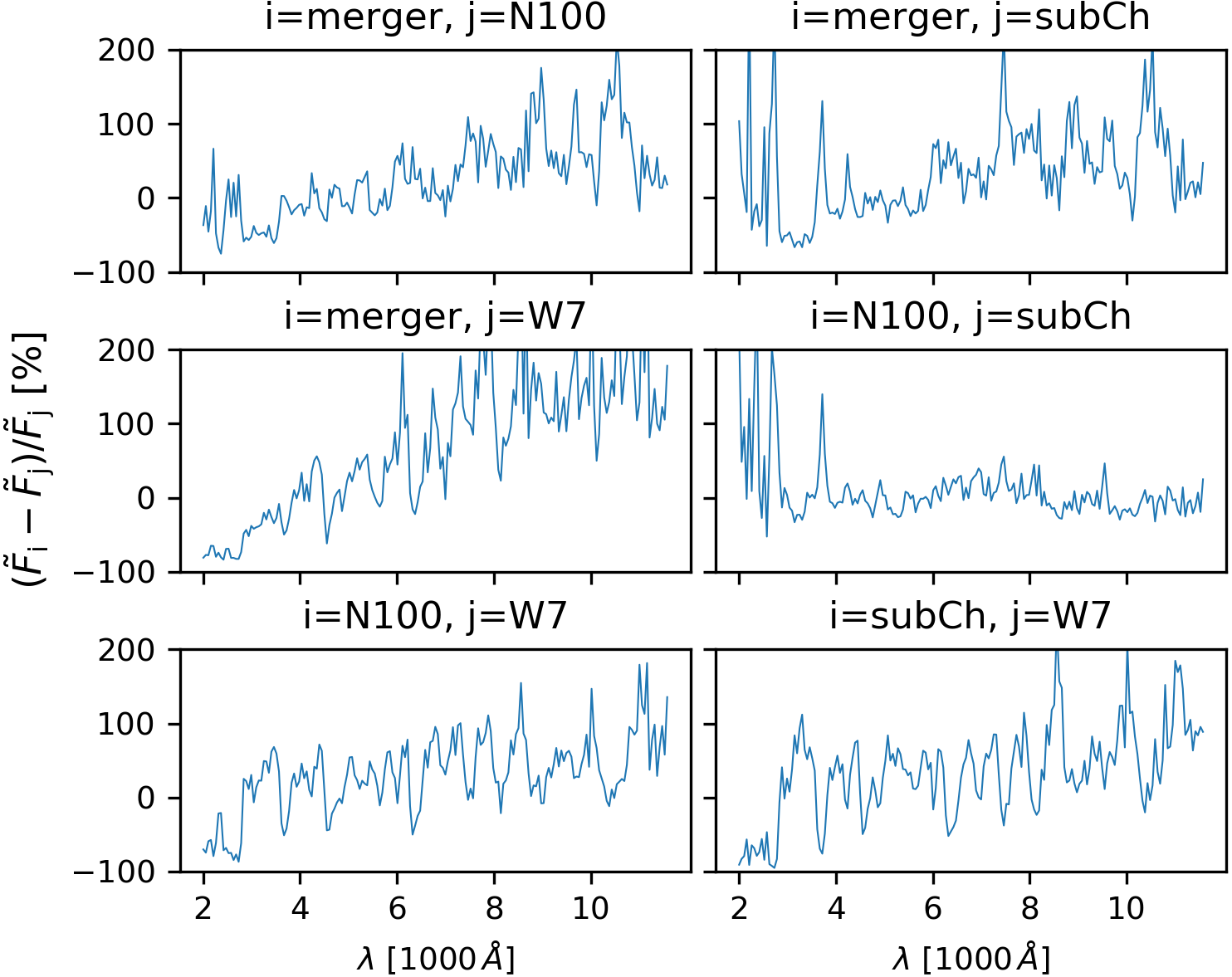}
 \caption{Deviations $\Delta_\lambda(t)$ between pairs of SN Ia
     spectra from the four SN models (W7, N100, subCh, and merger) at rest-frame
     $t=4.0$\,day after explosion.  Each panel shows a pair of SN models
     specified on the top of the panel.  The typical
     amplitude of deviations in the UV (for distinguishing progenitor
     scenarios) is $\sim$$100\%$, which is about an order of magnitude
     higher than the deviations due to microlensing.  Distortions due
     to microlensing will therefore not affect our ability to
     distinguish between the four SN explosion
     models for various progenitor scenarios.}
 \label{fig:deviations_SN_models_4}
\end{figure}

For each of the 30 magnification maps, we draw 10,000 random positions
in the map to quantify the effect of microlensing on the SN
spectra. For each position we calculate the microlensed flux
$F_\mathrm{micro}$ via equation (\ref{eq: microlensed flux}) and
compare it to the case without microlensing $F_\mathrm{no \, micro}$
($\mu = 1$) of a given SN Ia model. 
From this, we can calculate the deviation
$\Delta_\lambda(t)$ from the macro magnification as:
\begin{equation}
 \Delta_\lambda(t) = \frac{\tilde{F}_\mathrm{micro} - \tilde{F}_\mathrm{no \,
     micro}} {\tilde{F}_\mathrm{no\,micro} } =
 \frac{\tilde{F}_\mathrm{micro}}{\tilde{F}_\mathrm{no\,micro}} - 1,
 \label{eq: deviation microlensed flux}
\end{equation}
where $\tilde{F}$ is the normalised flux over a given
wavelength range such that the integrated flux over the wavelength
range yields the same 
sum for both the microlensed and the non-microlensed spectra (i.e., $\tilde{F}_\mathrm{no\,micro}= F_\mathrm{no\,micro}$, $\tilde{F}_\mathrm{micro}=K F_\mathrm{micro}$ and the normalisation constant $K$ is set such that  $\int {\rm d}\lambda\, \tilde{F}_\mathrm{micro} = \int {\rm d}\lambda\, \tilde{F}_\mathrm{no\,micro} $).  The
deviation $\Delta_\lambda$ quantifies distortions in the
spectra of a microlensed SN relative to the intrinsic SN without
microlensing, i.e., a deviation of 0 across all wavelengths implies no
chromatic microlensing distortion on the intrinsic SN spectra. We note
that a constant (macro)magnification across all wavelengths from
macrolensing without microlensing yields also a deviation of 0, since $F_\mathrm{micro}/F_\mathrm{no\,micro}$ is then the macrolensing magnification (a constant that is independent of wavelength) and $\tilde{F}_\mathrm{micro} = \tilde{F}_\mathrm{no\,micro}$ after normalisation.   We refer readers
to Figures A.1 and A.3 of \citet{huber+2019} for examples of
$\frac{F_\mathrm{micro}}{F_\mathrm{no \, micro}}$, which is
$\frac{\tilde{F}_\mathrm{micro}}{\tilde{F}_\mathrm{no \, micro}}$ up to a
constant factor. From the 30 $\times$
10,000 random configurations, we determine the median deviation  of $ \Delta_\lambda(t)$  
with the 1$\sigma$ range (68\% interval, from the 16 percentile to the
84 percentile of the microlensed spectra) and 2$\sigma$
range (95\% interval, from the 2.5 percentile to the 97.5
  percentile).  

We show the deviations at different times after explosions for the W7,
N100, subCh and merger models in Figures
\ref{fig:w7_deviation_of_microlensed_flux} to
\ref{fig:merger_deviation_of_microlensed_flux}, respectively.  The
median deviation (dashed black lines) is zero within $\ll$$1\%$ for all
models, indicating that the microlensing effect does not induce a
systematic distortion on the spectra overall, even though each
microlensed spectrum can indeed be distorted as shown by the 2$\sigma$
spreads.  We find that at early times (within $t\sim$10 rest-frame
days after explosion), the 2$\sigma$ spread of $\Delta_\lambda(t)$ for
most wavelength regions is well within the $10\%$ level, especially
for the W7 and merger models.  In the N100 and subCh models,
deviations in the UV can reach up to $\sim$$20\%$ due to absorption
features and suppression of UV flux in their spectra (see Figure
\ref{fig:sn_model_spectral_evol}). The insets in each of the panels
clearly show that the 1$\sigma$ spread of $\Delta_\lambda(t)$ for all
SN models is within $1\%$ in the early phases (rest-frame $t \lesssim
10$\,days). Therefore, microlensing would not distort the spectra of
SNe beyond $1\%$ at any wavelength in 68\% of all strongly lensed SNe
Ia at early phases. At later times ($t\sim$ 20 to 40\,days),
the influence of microlensing becomes substantially larger, as also
visible in the increased 1$\sigma$ spread, but the 1$\sigma$ spread is
still mostly below the $10\%$ level.

We find that deviations $\Delta_\lambda(t)$ due to microlensing tend
to be larger at wavelengths where the relative flux is low in the
spectra, either due to absorption features or low-level continuum (as
illustrated in Figures \ref{fig:sn_model_spectral_evol} to
\ref{fig:merger_deviation_of_microlensed_flux}).  This is because
microlensing distorts spectra across all wavelengths, so wavelengths
that have lower fluxes would have relatively larger variations in
fluxes due to microlensing, and thus higher deviations, after
normalising the spectra in equation (\ref{eq: deviation microlensed
  flux}).  We refer the reader to Appendix \ref{sec:covariance_matrix}
for an example of the covariance matrix of the deviations which
further illustrates this. 

The overall trend for all four SN models
shows that both the 1$\sigma$ and 2$\sigma$ spreads are
increasing over time. There are two reasons for this.  The first reason is that the specific intensity profiles for different
filters deviate more strongly from each other at later phases \citep[and Appendix \ref{sec: Specific intensity profiles}]{goldstein+2018,huber+2019}, 
which leads to higher deviations in the spectrum between different wavelength
regions.  The second reason is that
 SNe Ia are expanding over time and therefore
it is much more likely to cross a micro caustic at later times. 
Consequently,
  the effect of microlensing should be lower at the earliest phases
  ($t<4$\,days) compared to the first epoch from our simulations at
  $t=4$\,days; SNe Ia have smaller sizes and thus 
  smaller chances to be chromatically microlensed in these earliest phases.

To assess the impact of microlensing on our ability to distinguish
between SN explosion models from early-phase spectra, particularly in
the UV wavelengths, we compare the deviations due to microlensing to
the deviations due to model differences.  In Figure
\ref{fig:deviations_SN_models_4}, we show the deviations (equation
\ref{eq: deviation microlensed flux}) between all 6 possible pairs of
the four SN Ia models we considered at rest-frame $t=4.0$\,days after
explosion.  In the UV-wavelength range that is useful for progenitor
studies (Section \ref{sec:microlensSN:SNmodels}), the deviations are
typically $\sim$$100\%$, which is an order of magnitude larger than
the deviations due to microlensing shown in Figures
\ref{fig:w7_deviation_of_microlensed_flux} to
\ref{fig:merger_deviation_of_microlensed_flux}, respectively.  At
later times, the deviations between the spectra from different models
are shown in Appendix \ref{sec:deviations_mod_later}, and continue to
have high amplitudes.  Therefore, deviations in the spectra due to
microlensing are negligible compared to the deviations in the spectra
arising from different SN Ia progenitor scenarios and from
uncertainties in the spectra due to approximations in radiative transfer calculations.
 
To summarise, at early times ($\lesssim 10$ rest-frame days) we have
very good prospects to collect good quality spectra with negligible
distortions from microlensing, which is necessary to address the SN Ia
progenitor problem. Nevertheless, we would like to point
out that there are extreme cases where microlensing can significantly
influence even very early spectra. These extreme microlensing cases could potentially allow
one to probe the specific intensity distribution of SNe.
A comparison showing the dependency of microlensing effects on different parameters, 
such as $s$, is presented in \citet{Huber+2020}, but especially for high magnification
cases with both values of $\kappa$ and $\gamma$ close to 0.5, we find that
chromatic microlensing distortion is more likely. This can be understood by looking at the
magnification maps shown in Appendix \ref{sec: Microlensing maps}, where more caustics and higher gradients
exist for $(\kappa,\gamma) = (0.43,0.43)$ and $(\kappa,\gamma) = (0.57,0.58)$
in comparison to the other cases. 
Fortunately, in practice we can
always estimate for a given lensed SN Ia image the likelihood of it
being microlensed, to determine whether it is suitable for obtaining a
``clean'' SN spectrum that has little distortion from microlensing.

\section{Forecasted cosmological constraints from strongly lensed SNe}
\label{sec:cosmo}

Each lensed SN provides an opportunity to measure two distances: the
time-delay distance $\tdist$ and the angular diameter distance to the
deflector/lens $\Dd$ \citep[e.g.,][]{refsdal1964, suyu+2010,
  paraficzhjorth2009, birrer+2016, jee+2019}.  The time-delay distance
is defined by \citet{suyu+2010} as
\be 
\tdist = (1+\zd) \frac{\Dd \Ds}{\Dds}, 
\ee 
where $\Dds$ and $\Ds$ are angular diameter distances to the source
from the deflector and from the observer, respectively.  
Measuring $\tdist$ requires three ingredients: (1)
time delays, (2) strong lens mass model, and (3) characterisation of
the mass environment along the line of sight to the source.  All three
parts contribute to the uncertainties on $\tdist$.  The
measurement of $\Dd$ depends on (1), (2) and also the stellar velocity
dispersion of the foreground deflector, but not on (3), as shown by
\citet{jee+2015} and \citet{jee+2019}. We refer readers to reviews by,
e.g., \citet{treumarshall2016}, \citet{suyu+2018} and
\citet{oguri2019}, for more details on time-delay cosmography.

The distances $\tdist$ and $\Dd$ to lensed quasars have been
successfully measured using the time-delay method
\citep[e.g.,][]{chen+2019, jee+2019, rusu+2020, wong+2020}.  Lensed
SNe have several advantages over lensed quasars: (1) the time delays
are easier to measure with simple and sharply varying light-curve
shapes that are less prone to strong microlensing effects, (2) the lens mass distribution is easier to model without
strong contamination by quasar light that typically outshines
everything else in the lens system (SNe are bright as well, but they
fade in months, revealing their host galaxy and lens galaxy light
clearly), (3) some SNe are standardisable candles and their
intrinsic luminosities could mitigate lens model degeneracies in cases
when microlensing effects are negligible, and (4) the effect of
microlensing time delay, pointed out by \citet{tiekochanek2018} for
lensed quasars, is negligible for typical lensed SNe
\citep{bonvin+2019b}. 

We create a mock sample of lensed SNe Ia expected from the
upcoming LSST, with simulated $\tdist$ and $\Dd$ measurements in
Section \ref{sec:cosmo:mockdist}, and
forecast the resulting cosmological constraints based on the sample in
Section \ref{sec:cosmo:constraints}.

\subsection{Mock distance measurements from lensed SNe Ia}
\label{sec:cosmo:mockdist}
\begin{figure}
\centering
 \includegraphics[scale = 0.55]{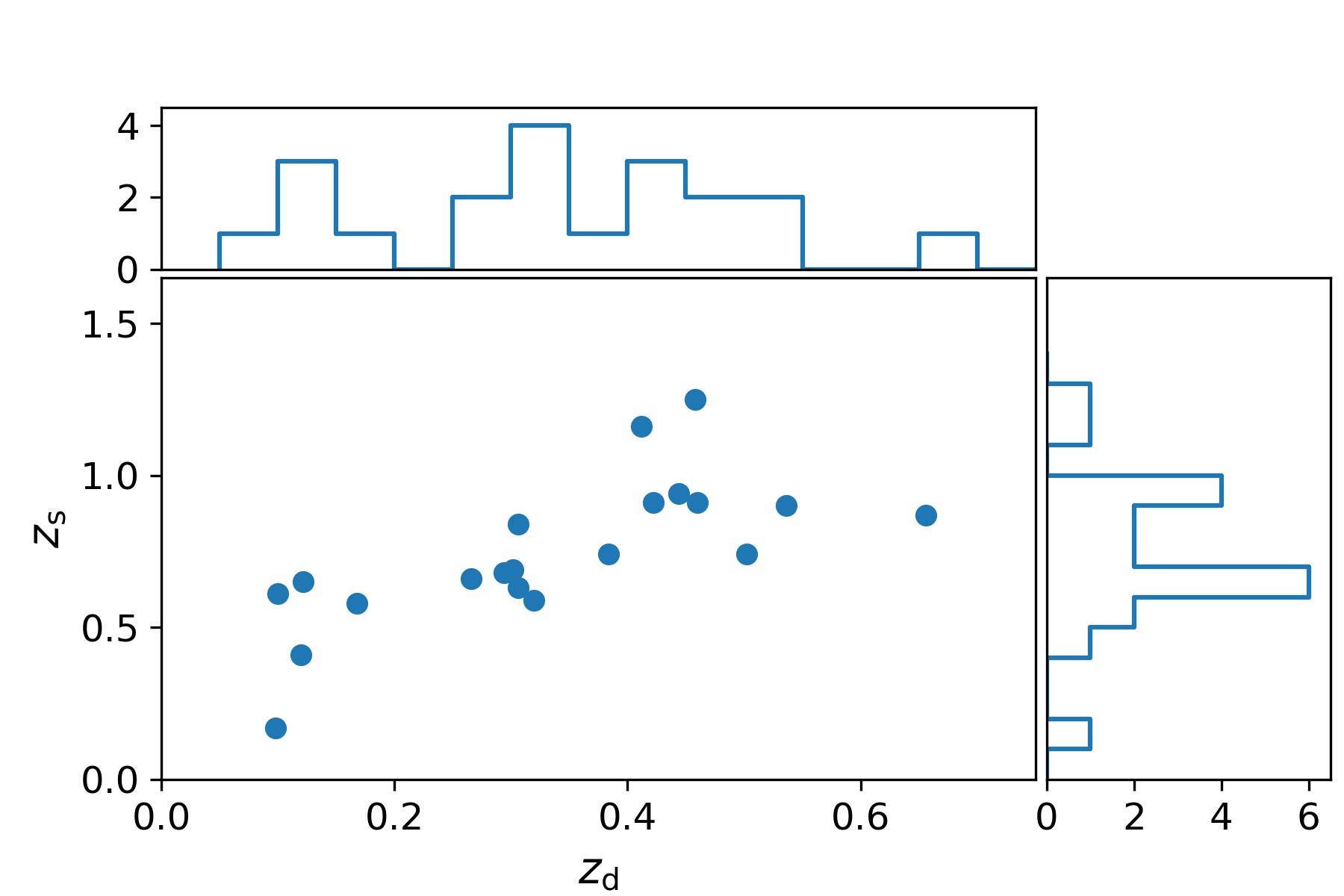}
 \caption{Source redshift $\zs$ and lens redshift $\zd$ distribution
   of the mock sample of lensed SNe Ia from LSST with good time-delay
   measurements (accuracy better than 1\% and precision better than
   5\%).  The top and right panels show the histograms of the
 number of systems in each lens-redshift and source-redshift bin, respectively. }
 \label{fig:mocklenses}
\end{figure}

We focus on a sample of lensed SNe Ia that would have
``good'' time-delay measurements even in the presence of microlensing,
i.e., those systems with accuracy 
better than 1\% 
and precision better than 5\% in their time-delay measurements.  
From the investigations of \citet{huber+2019}, the expected
number of spatially-resolved lensed SNe Ia is $\sim$$75$ for 10
years of LSST survey with 
baseline-like LSST cadence strategies.  Accounting for the effects of
microlensing, lensed SN Ia systems that
have delays longer than 20 days could yield accuracy better than 1\%,
whereas shorter delays could suffer from inaccuracy \citep[see Figure
13 of][]{huber+2019}.  SNe Ia at lower redshifts, $\zs<0.7$ are
brighter and would yield good delays (i.e., delays with accuracy and
precision within the target), whereas for SNe Ia at $\zs>0.7$,
only about half of the systems could yield good delays with deep
follow-up imaging \citep[see Figure 15 of][]{huber+2019}.  Using these
results, we start with the mock sample of lensed SNe Ia expected for
LSST from OM10 \citep{ogurimarshall2010}, and select the fraction of
lensed SN systems with at least one time delay (relative to
  the first appearing image) that is longer than 20 days, resulting in 30
lensed SNe Ia systems.\footnote{The OM10 catalog of lensed SNe is
  oversampled by a factor of 10, i.e., OM10 boosted the number of
  lensed SNe Ia by a factor of 10, in order to reduce shot noise, and
  accounted for this in their analysis.  We note
  that by using the more 
  recent LSST cadence strategies \citep{huber+2019} instead of the
  assumed detection/cadence criteria in OM10, the expected
  number of lensed SN Ia systems (75) is higher than the forecasted
  number by OM10. To account for this, 
  we determine the fraction of systems
  with at least one time delay longer than 20 days in OM10 ($41\%$), and use this
  fraction of the expected number of systems ($75$) to get 30
  systems ($=0.41\times75$) with delays longer than 20 days. These 30
  are randomly selected from the OM10 oversampled catalog of systems
  with delays longer than 20 days.}
Of these 30 systems, 10 have $\zs<0.7$ which
we keep, whereas 20 have $\zs>0.7$ and we randomly select half of
them.  This leads to a final sample of $\Nsn$ = 20 mock lensed SNe Ia that we
expect to have good delays.  Figure \ref{fig:mocklenses} shows the
redshift distributions of these mock lens systems.

To estimate the precision for $\tdist$ measurements, we conservatively
adopt 5\% for the time-delay uncertainties based on the work
  of \citet{huber+2019}, who showed that ground-based follow-up
  observations every other night in 3 filters $g$, $r$ and $i$ (with
  5$\sigma$ depth of 25.6\,mag, 25.2\,mag and 24.7\,mag, respectively)
  for our 20 mock systems would yield good time-delays (i.e.,
  precisions better than 5\%, even in the presence of
  microlensing). This is compatible with the estimated uncertainties
  by \citet{goldstein+2018} based on mock \hst\ observations, and also
  with the measured time delays of iPTF16geu by \citet{dhawan+2019}
  who obtained uncertainties of $\lesssim$$1$ day on the delays, which
  would be $\lesssim$$5\%$ for time delays longer than 20 days.
  We further adopt 3\% for the lens mass
modelling uncertainties, and 3\% for the lens environment
uncertainties, which are realistic given current lensed quasar
constraints \citep[e.g.,][]{suyu+2010, greene+2013, collett+2013,
  suyu+2014, rusu+2017, wong+2017, tihhonova+2018, bonvin+2019, chen+2019, millon+2020}.  Adding
 the values for these three sources of uncertainties in quadrature, we assign 6.6\% uncertainty to $\tdist$ from each
lensed SN Ia system.
For the precision on $\Dd$, we consider the scenario of having
spatially-resolved kinematics of the foreground lens
\citep[e.g.,][]{czoske+2008, barnabe+2009, barnabe+2011}, such that we
can measure $\Dd$ with its uncertainty essentially dominated by the
time-delay uncertainty \citep{yildirim+2020}. Spatially-resolved
kinematic observations of the lens systems would be relatively
straightforward to obtain after all the multiple SN images fade away in
$\lesssim$$1$ year.  
We thus adopt $5\%$
uncertainties on $\Dd$ for each lensed SN Ia system.

To generate mock $\tdist$ and $\Dd$ measurements for the $\Nsn$ (= 20) lensed SN
Ia systems, we adopt as input a flat $\Lambda$CDM cosmology with
$H_0=72\,\kmsMpc$ and $\Om=1-\OL=0.32$.  Given the deflector and SN
source redshifts from the OM10 catalog, we compute the $\tdistti$ and
$\Ddti$ of lensed SN Ia system $i$, where $i$=1..$\Nsn$.  Using the
estimated 1$\sigma$ uncertainty of 6.6\% for $\tdist$ and 5\% for
$\Dd$, which we denote as $\stdisti$ and $\sDdi$, respectively, we draw random Gaussian deviates, $\dtdisti$ and
$\dDdi$, to obtain the mock measurements for
lensed SN Ia system $i$ as follows,
\be
\tdistmi = \tdistti + \dtdisti
\ee
and
\be
\Ddmi = \Ddti + \dDdi.
\ee
From this, we get the following mock distance measurements for our
lensed SN Ia sample: \{$\tdistmi\pm\stdisti, \Ddmi\pm\sDdi$\} where
$i$=1..$\Nsn$. 

\subsection{Cosmological constraints from the mock lensed SN Ia sample}
\label{sec:cosmo:constraints}

To obtain the cosmological constraints, we sample the posterior
distribution of the cosmological parameters $\cosmopar$ in the same
way as we do for the analysis of lensed quasars \citep{bonvin+2017,
  wong+2020, millon+2020}.  We first describe our
likelihoods and priors for the cosmological parameters that enter the
posterior probability distribution function.  

For lensed SN Ia system $i$, we assume Gaussian likelihoods for
$\tdistmi$ and $\Ddmi$ with their corresponding
uncertainties $\stdisti$ and $\sDdi$ as the Gaussian standard
deviations.  That is, the likelihood for $(\tdistmi,\Ddmi)$ is
\bea
\label{eq:Pi}
P_i(\tdistmi, \Ddmi | \tdisti, \Ddi) & = & G(\tdistmi,\stdisti,
\tdisti) \times \\
& & G(\Ddmi,\sDdi, \Ddi),
\eea
where
\be
G(\mu_{\rm G}, \sigma_{\rm G}, x) = \frac{1}{\sqrt{2\pi\sigma_{\rm G}^2}}
\exp \left [ -\frac{(x-\mu_{\rm G})^2}{2\sigma_{\rm G}^2}\right ].
\ee
We then multiply the likelihoods of the individual mock lenses together to compute
the joint likelihood for the sample,
\be
\label{eq:Pjoint}
P_{\rm joint} = \prod_{i=1}^{\Nsn} P_i.
\ee
We adopt uniform priors on the cosmological parameters in the sampling.

We consider three background cosmological models as listed in Table
\ref{tab:cosmo_mod}, and sample the cosmological parameters in the
models.  The first cosmological model is the flat $\Lambda$CDM model
with two variable cosmological parameters $H_0$ and the matter density
$\Om$.  The second model is open $\Lambda$CDM where the variable
parameters are $H_0$, $\Om$ and the curvature density $\Ok$ (with the
dark energy density $\OL=1-\Om-\Ok > 0$). The third model is flat
$w$CDM with three variable parameters $H_0$, $\Om$ and the dark energy
equation-of-state parameter $w$ (where $w=-1$ corresponds to the
cosmological constant $\Lambda$ for dark energy).  The priors for
these parameters are summarised in Table \ref{tab:cosmo_mod}.

\renewcommand{\arraystretch}{1.5} 
\begin{table*}
\caption{Cosmological models and constraints from 20 mock lensed SNe
  Ia in the LSST era.  The input cosmological model is flat
  $\Lambda$CDM with $H_0=72\,\kmsMpc$ and $\Om=1-\OL=0.32$.}             
\label{tab:cosmo_mod}      
\centering          
\begin{tabular}{c c c c c  }     
\hline\hline       
cosmological model & parameter  & prior range & \multicolumn{2}{c}{marginalised
  constraints on cosmological parameters}
\\ 
  &  &  & from $\tdist$ only &  from $\tdist$ and $\Dd$ \\
\hline \hline 
flat $\Lambda$CDM & $H_0$ [$\kmsMpc$] & $[0,150]$ & $72.3\pm1.1$ &
$72.5^{+1.0}_{-0.9}$  \\ 
   & $\Om$ & $[0.05,0.5]$ & $0.28\pm0.15$ & $0.35^{+0.07}_{-0.06}$
    \\ 
\hline 
open $\Lambda$CDM & $H_0$ [$\kmsMpc$] & $[0,150]$ &
$72.5^{+1.2}_{-1.3}$ & $72.7\pm1.0$ \\
   & $\Om$ & $[0.05,0.5]$ & $0.29^{+0.14}_{-0.15}$ & $0.27^{+0.14}_{-0.13}$ \\
   & $\Ok$ & $[-0.5,0.5]$ & $0.14^{+0.25}_{-0.31}$ & $0.15^{+0.21}_{-0.24}$ \\
\hline 
flat $w$CDM & $H_0$ [$\kmsMpc$] & $[0,150]$  & $73.9^{+3.2}_{-2.7}$ &
$74.0^{+2.3}_{-2.1}$ \\
   & $\Om$ & $[0.05,0.5]$ & $0.33^{+0.12}_{-0.17}$ & $0.40^{+0.06}_{-0.10}$ \\
   & $w$ & $[-2.5,0.5]$ & $-1.31^{+0.89}_{-0.58}$  & $-1.38^{+0.48}_{-0.55}$ \\
\hline                  
\end{tabular}
\end{table*}

For each background cosmological model, we sample the cosmological
parameters $\cosmopar$ by computing the posterior probability which is
the joint likelihood $P_{\rm joint}$ multiplied with the prior.
Specifically, for a given set of $\cosmopar$ values, we can compute
$\tdisti$ and $\Ddi$ for system $i$ of the mock lensed SN Ia sample to
calculate $P_i$ in equation (\ref{eq:Pi}), and thus $P_{\rm joint}$
via equation (\ref{eq:Pjoint}).  Given our uniform priors on
$\cosmopar$, our posterior is, up to a constant factor, $P_{\rm
  joint}$.  We then sample the posterior probability distribution
using \emcee \citep{Foremanmackey+2013} with 32 walkers and 40,000
samples.  To compare the constraining power of the two distance
measurements on the cosmological parameters, we also consider the
constraints from only $\tdist$ and only $\Dd$ measurements.

\begin{figure}
\centering
 \includegraphics[scale = 0.7]{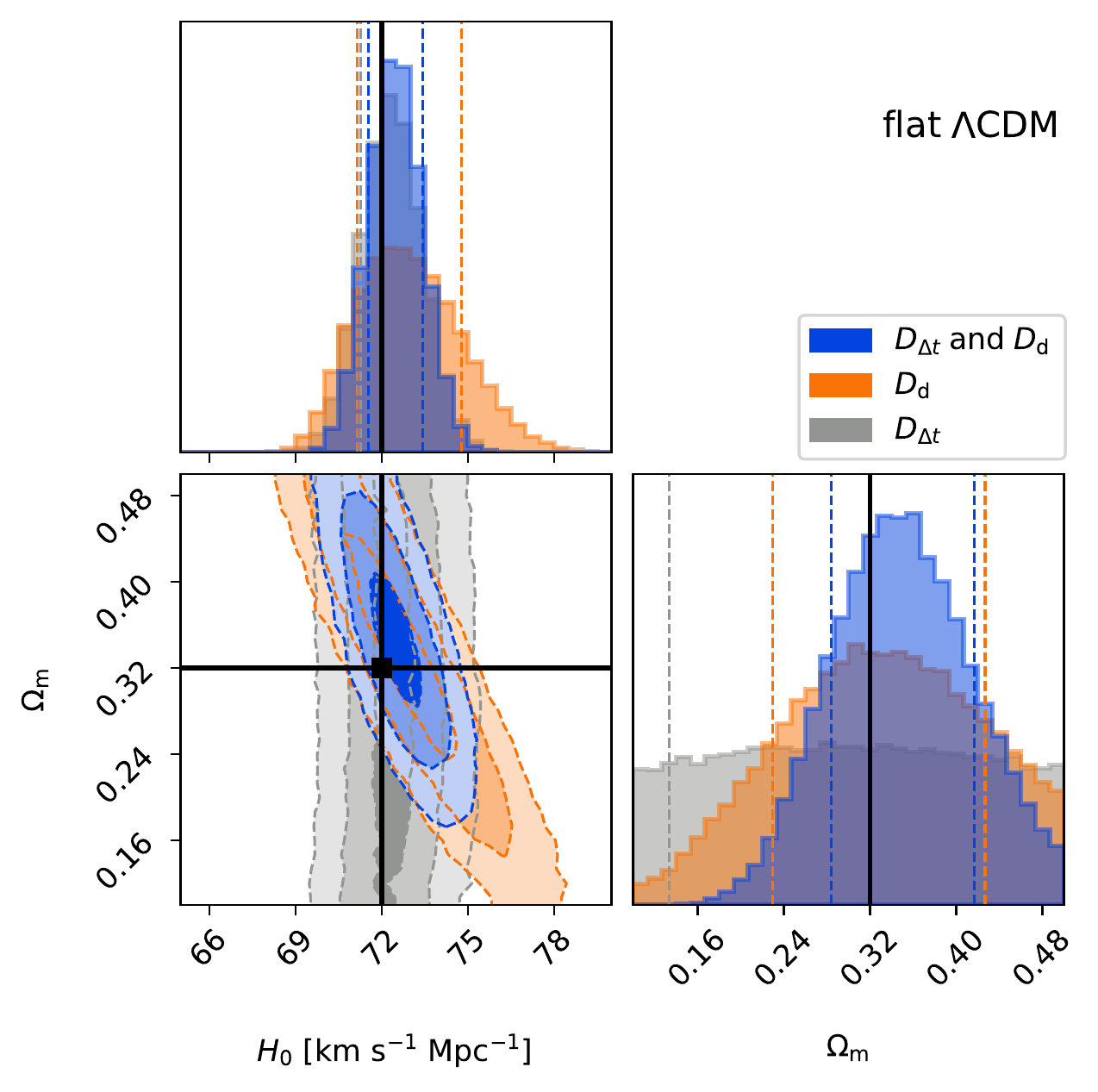}
 \caption{Cosmological constraints in flat $\Lambda$CDM from the mock
   sample of 20 lensed SN Ia systems with good time-delay
   measurements. Grey contours/histograms show the marginalised constraints from
   having only $\tdist$ measurements, orange are those from only
   $\Dd$ measurements, and blue are the results based on both
   $\tdist$ and $\Dd$ measurements.  The dashed contours mark the
   68\%, 95\% and 99.7\% credible regions, whereas the vertical dashed
   lines in the histograms mark the 68\% credible intervals.  The
   input values are marked in black, and are recovered within the
   marginalised 68\% credible intervals in 1D (or within the 95\%
   credible regions in the 2D $H_0$-$\Om$ plane). In flat
   $\Lambda$CDM, even a modest sample of 20 lensed SNe Ia could
   constrain $H_0$ and $\Om$ with precisions of 1.3\% and 19\%,
   respectively. }
 \label{fig:cosmo:flcdm}
\end{figure}

The results of the sampling in flat $\Lambda$CDM are shown in Figure
\ref{fig:cosmo:flcdm} with the marginalised cosmological constraints
listed in Table \ref{tab:cosmo_mod}. The time-delay distances $\tdist$
provide tight constraints on $H_0$ but little information on $\Om$
(grey contours). Since $\Dd$ has a different dependence on cosmological
parameters from that of $\tdist$ (the orange contours from $\Dd$ are
tilted with respect to the grey contours from $\tdist$), the
combination of the two distance constraints tightens slightly the
constraint on $H_0$ and substantially the constraint on $\Om$.  The
input cosmological parameter values (marked in black) are recovered
within the marginalised 68\% credible intervals.  With the two
distances from the forecasted sample of 20 mock lensed SN Ia systems,
we expect to measure $H_0$ with uncertainties of $1.3\%$.

\begin{figure}
\centering
 \includegraphics[scale = 0.5]{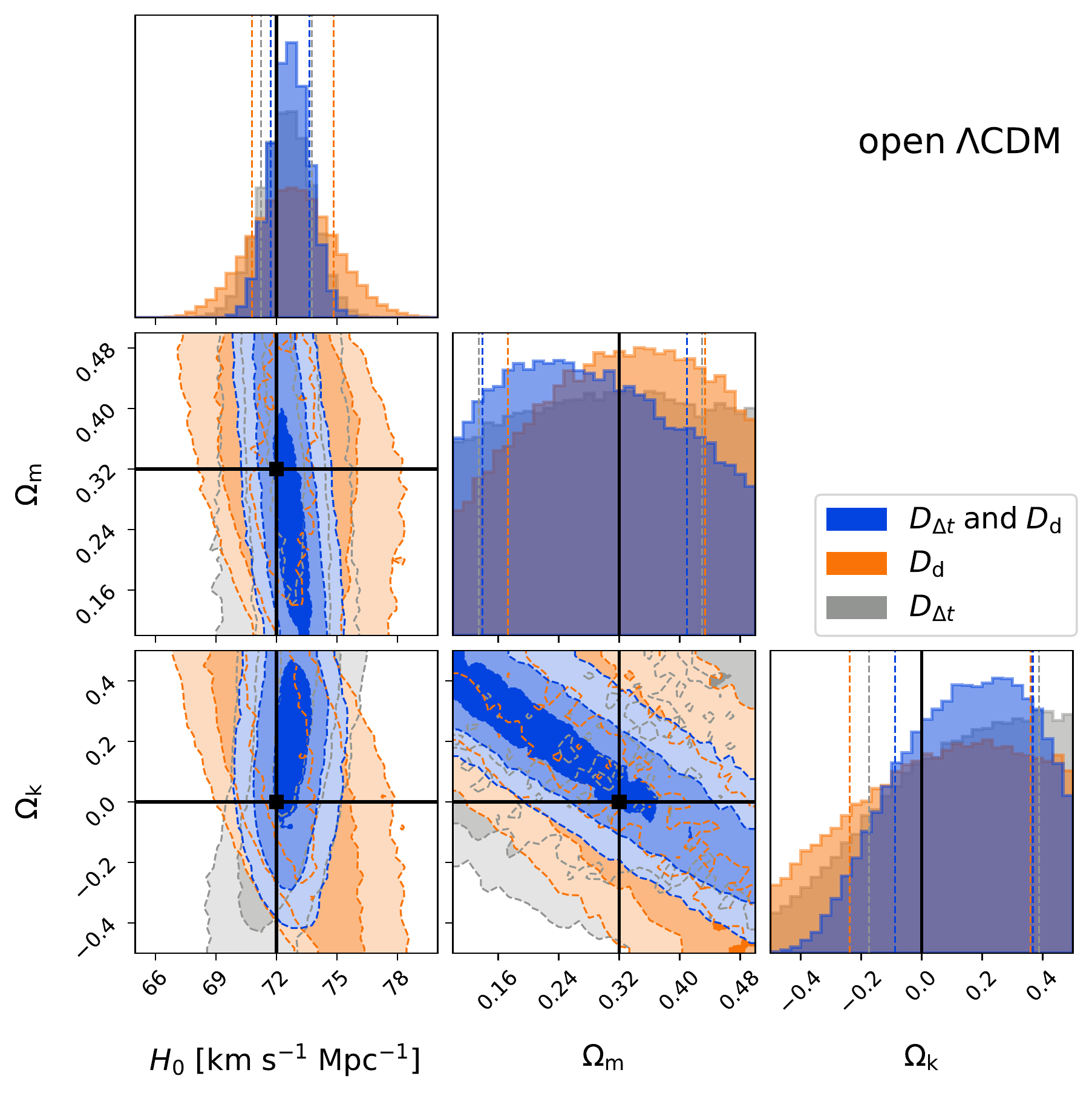}
 \caption{Cosmological constraints in open $\Lambda$CDM from the mock
   sample of 20 lensed SN Ia systems with good time-delay
   measurements. Panels and labels are in the same format as in Figure
   \ref{fig:cosmo:flcdm}. 
   The input values, marked in black, are recovered within the
   marginalised 68\% credible intervals/regions. 
   The combination of
   $\tdist$ and $\Dd$ makes the $H_0$ constraint relatively
   insensitive to other cosmological parameters, as shown in the 2D
   contours in the left panels. }
 \label{fig:cosmo:olcdm}
\end{figure}

We show in Figure \ref{fig:cosmo:olcdm} the results in open
$\Lambda$CDM, with the marginalised constraints in Table
\ref{tab:cosmo_mod}.  We see in the bottom-left panel of the figure
that the parameter degeneracies between 
$H_0$ and $\Ok$ are in different directions from the $\tdist$ and
$\Dd$ constraints, and the combination of the two helps to reduce the
degeneracies.  As a result, the inferred $H_0$ from both $\tdist$ and $\Dd$
measurements is relatively insensitive to other cosmological parameters
(the blue contours are nearly vertical in the left column of Figure
\ref{fig:cosmo:olcdm}).  In fact, the marginalised $H_0$ constraint of
$72.7\pm1.0$ is comparable in precision to that in flat $\Lambda$CDM
(see Table \ref{tab:cosmo_mod}), while the constraint on $\Om$
degrades substantially by a factor of 2 compared to that in flat
$\Lambda$CDM.

\begin{figure}
\centering
 \includegraphics[scale = 0.5]{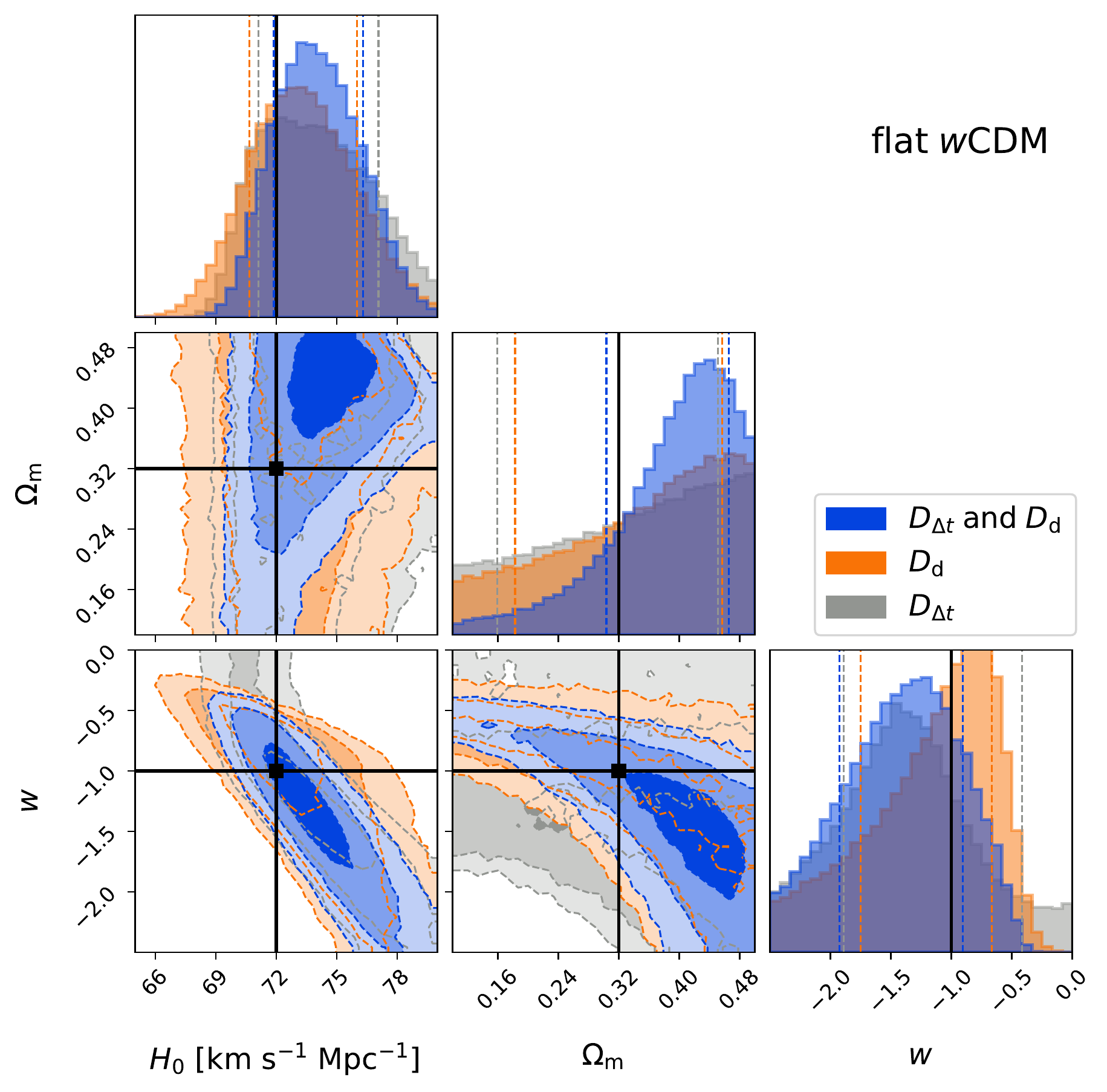}
 \caption{Cosmological constraints in flat $w$CDM from the mock sample
   of 20 lensed SN Ia systems with good time-delay measurements. 
   Panels and labels are in the same format as in Figure
   \ref{fig:cosmo:flcdm}. 
   The input values, marked in
   black, are recovered within the marginalised 68\% credible
   intervals.  When the dark energy equation-of-state
   parameter $w$ is allowed to vary, significant parameter degeneracy between
   $H_0$ and $w$ exists which weakens the constraint on $H_0$.
   }
 \label{fig:cosmo:fwcdm}
\end{figure}

For the background cosmological model of flat $w$CDM, the cosmological
constraints are shown in Figure \ref{fig:cosmo:fwcdm} and summarised
in  Table \ref{tab:cosmo_mod}.  When the dark
energy equation-of-state parameter is allowed to vary, this
substantially weakens the cosmological constraint on $H_0$ (to 3\%
uncertainty), given the strong parameter degeneracy between $H_0$ and $w$.  Having
$\Dd$ measurements is important for constraining $w$ and thus limit
the possible range of $H_0$ values, as also previously shown by e.g.~\citet{jee+2016}. We see clearly here that while
$\tdist$ is mainly sensitive to $H_0$, it does depend on the assumed
background cosmological model.  The dependence of $H_0$ inference on
the cosmological model can be reduced by making use of the Type Ia SN
relative distance scale and anchoring the distance scale with the
$\tdist$ and $\Dd$ measurements, as recently illustrated by, e.g.,
\citet{jee+2019}, \citet{arendse+2019} and \citet{taubenberger+2019}.

For our cosmological forecast, we have adopted uncertainties on the
lensing distances that are based partly on analyses of lensed quasars.
Recently, \citet{kochanek2020a} has suggested that systematic
uncertainties on $H_0$ are $\gtrsim$10\% from the lens mass modelling.
Modelling uncertainties due to lensing degeneracies, particularly the
mass-sheet degeneracy, have been previously investigated in detail
\citep[e.g.,][]{falco+1985, schneidersluse2013, suyu+2014,
  wertz+2018}. In particular, \citet{millon+2020} have carried out an
extensive study on possible systematic effects in the analysis of
lensed quasars, and do not find evidence for systematic uncertainties
larger than the quoted values from the COSMOGRAIL, H0LiCOW, SHARP and
STRIDES collaborations \citep{chen+2019, wong+2020, shajib+2020}.
\citet{kochanek2020b} has further suggested that elliptical mass
distributions with few angular degrees of freedom would lead to biased
estimates of $H_0$.  This is not applicable to the latest
state-of-the-art analyses of lensed quasars that used the so-called
composite model of baryons and dark matter (as one of the two main
families of lens models used), where the total mass distribution was
not restricted to be elliptical given that the baryonic and dark
matter components could have different ellipticities and position
angles \citep[e.g.,][]{chen+2019, rusu+2020, shajib+2020}.

Recently, \citet{birrer+2020} have allowed for the full mass-sheet
degeneracy that is  
maximally degenerate with $H_0$ in their mass models.  The mass-sheet contribution is
constrained through aperture-averaged stellar velocity dispersion measurements of the lens galaxies, and the $H_0$
precision from a sample of 7 lenses in TDCOSMO degrades to $\sim$$8\%$
\citep[with similar median value in comparison to
the][results]{wong+2020}.  Therefore, any residual systematics due to
possible 
mass-sheet transformation is no more than $\sim$$8\%$.  By combining
the 7 lensed quasars with a sample of lenses matched in property to
the lensed quasars, \citet{birrer+2020} constrain $H_0$ at the 5\%
level through a Bayesian hierarchical framework, with a shift in $H_0$
to a lower value from the inclusion of the external lens sample.
Therefore, while there is no evidence of residual systematic
uncertainties affecting the COSMOGRAIL, H0LiCOW, SHARP and STRIDES
measurements \citep{millon+2020}, \citet{birrer+2020} conclusively
shows that any potential residual systematic uncertainty due to the
mass-sheet degeneracy is
at most 5\%.  The results of \citet{birrer+2020} are based on single
aperture-averaged lens velocity dispersions, and spatially-resolved
kinematic measurements of the lens galaxy will be more powerful in
alleviating mass-model degeneracies \citep[e.g.,][in
  prep.]{barnabe+2011, barnabe+2012, yildirim+2020}.
Future
investigations in these directions are warranted. In this respect,
lensed SNe have the advantage over lensed quasars in that
spatially-resolved kinematic maps of the lens systems would be easier to
acquire after the SNe fade, without the presence of bright quasars
that contaminate the kinematic signatures of the lens galaxy.  

New methods of SN time-delay measurement
techniques \citep[e.g.,][]{pierel+2019} could yield even more
precise time-delay measurements, improving our forecasted cosmological
constraints. In addition, we used a conservative estimate
for the number of lensed SNe Ia with good time delays; more optimistic
estimates could triple the number of systems \citep[see Appendix C
of][]{huber+2019}.  While SNe Ia are standardisable candles, we have
not incorporated potential measurements of the SN Ia luminosity
distances; the multiple SN images are likely microlensed (by
stars in the foreground lens galaxy) and millilensed (by mass
substructures within the lens galaxy), as evidenced by the lensed SN
Ia system iPTF16geu \citep{more+2017, Yahalomi:2017ihe}, which make it
difficult to extract the unlensed SN fluxes and thus the SN luminosity
distance.  Nonetheless, first studies by
\citet{Foxley-Marrable:2018dzu} showed that microlensing could be
negligible for SN images that are located far from the lens galaxy,
showing promise in obtaining the unlensed SN fluxes in situations
where millilensing effects are small.  Given the uncertainties
associated with millilensing, we conservatively exclude possible
measurements of SN Ia luminosity distances when forecasting our
cosmological constraints. Furthermore, 
lensed core-collapse SNe, not considered
here,\footnote{It is beyond the scope of this work to quantify
  realistic measurement uncertainties of time delays in lensed
  core-collapse SNe.} provide additional $\tdist$ and $\Dd$
measurements,  
and studies indicate more numerous lensed core-collapse SNe than
lensed SNe Ia \citep[e.g.,][]{ogurimarshall2010, goldstein+2019,
  wojtak+2019}.  Therefore, we expect that a measurement of $H_0$ with
1\% uncertainty in flat $\Lambda$CDM from lensed SNe in the LSST era
to be achievable.


\section{Summary}
\label{sec:summary}

We initiate the HOLISMOKES programme to conduct Highly Optimised Lensing
Investigations of Supernovae, Microlensing Objects, and Kinematics of
Ellipticals and Spirals.  In this first paper of the programme, we
investigate the feasibility of achieving two scientific goals in the
LSST era: (1) early-phase SN observations for progenitor studies, and
(2) cosmology through the time-delay method. We summarise as follows.
\begin{itemize}
\item The time delays between the multiple appearances of a lensed SN
  would allow us to observe the SN in its early phases.  We find that
  microlensing distortions of early-phase SN Ia spectra (within
  $\sim$10 rest-frame days) are typically negligible, with distortions
  within $1\%$ (1$\sigma$ spread) and within $\sim$$10\%$
  (2$\sigma$ spread).  In contrast, the deviations in the spectra
between the four SN Ia progenitor models that we have considered are
typically at the $\sim$$100\%$ level. This
  provides excellent prospects for acquiring intrinsic early-phase SN
  Ia spectra, effectively free of microlensing distortions, to shed light on the
  progenitors of SNe Ia.
\item We forecast the cosmological parameters constraints from a
  sample of 20 lensed SNe Ia in the LSST era.  We
  assume that $\tdist$ and $\Dd$ to these systems could be constrained
  with uncertainties of $6.6\%$ and $5\%$, respectively.  From this
  sample, we expect to measure $H_0$ in flat
  $\Lambda$CDM with a precision of 1.3\%
  including (known) systematics, completely independent of any other cosmological
  probes.  In an open $\Lambda$CDM cosmology, we find a similar
  constraint on $H_0$, while in the flat $w$CDM cosmology, the
  constraint loosens to 3\%. 
\item Given the additional lensed core-collapse SNe, we expect
  a measurement of $H_0$ with 1\% uncertainty in flat
  $\Lambda$CDM from lensed SNe to be achievable in the LSST era.
\end{itemize}

With ongoing wide-field cadence surveys like ZTF and the upcoming
LSST, we are entering an exciting time of catching and watching SNe
being strongly lensed.  While the next systems from ZTF are likely to
have short time delays ($\lesssim$$10$ days), which could limit their
use for cosmological and supernova studies, as the surveys like LSST
image deeper, lensed SNe with longer time delays are expected to
appear \citep{wojtak+2019}.  Each one of these systems will provide an excellent
opportunity for studying SN physics and cosmology.  The cosmological analyses of lensed SNe
will be complementary to the growing sample of lensed quasars,
and the combination of the two types of lensed transients will be an
even more powerful probe of cosmology. The challenges associated with
lensed SNe 
will be to find these systems amongst the millions of daily transient
alerts from LSST, and to analyse them quickly.  Methods based on
machine learning are being developed to overcome such challenges
\citep[e.g.,][]{jacobs+2019, avestruz+2019, hezaveh+2017,
  levasseur+2017, pearson+2019, Canameras+2020}, and we are exploring these avenues in
our forthcoming publications.

\begin{acknowledgements}
We thank M.~Oguri and P.~Marshall for the useful lens catalog from
\citet{ogurimarshall2010}, E.~Komatsu, S.~Jha and S.~Rodney for
helpful discussions, and the anonymous referee for the constructive
comments. 
SHS thanks M.~Barnab{\`e} for the animated discussion in conjuring up the programme acronym.  
SHS, RC, SS and AY thank the Max Planck Society for support through the Max
Planck Research Group for SHS.  
This project has received funding from the European Research Council (ERC)
under the European Union’s Horizon 2020 research and innovation
programme (LENSNOVA: grant agreement No 771776; COSMICLENS: grant
agreement No 787886).
This research is supported in part by the Excellence Cluster ORIGINS which is funded by the Deutsche Forschungsgemeinschaft (DFG, German Research Foundation) under Germany's Excellence Strategy -- EXC-2094 -- 390783311.
This work is supported by the Swiss National Science Foundation
(SNSF). 
\end{acknowledgements}

\bibliographystyle{aa}
\bibliography{holismokesI}

\begin{thebibliography}{112}
\expandafter\ifx\csname natexlab\endcsname\relax\def\natexlab#1{#1}\fi

\bibitem[{{Arendse} {et~al.}(2019){Arendse}, {Agnello}, \&
  {Wojtak}}]{arendse+2019}
{Arendse}, N., {Agnello}, A., \& {Wojtak}, R.~J. 2019, \aap, 632, A91

\bibitem[{{Avestruz} {et~al.}(2019){Avestruz}, {Li}, {Zhu}, {Lightman},
  {Collett}, \& {Luo}}]{avestruz+2019}
{Avestruz}, C., {Li}, N., {Zhu}, H., {et~al.} 2019, \apj, 877, 58

\bibitem[{{Barnab{\`e}} {et~al.}(2011){Barnab{\`e}}, {Czoske}, {Koopmans},
  {Treu}, \& {Bolton}}]{barnabe+2011}
{Barnab{\`e}}, M., {Czoske}, O., {Koopmans}, L. V.~E., {Treu}, T., \& {Bolton},
  A.~S. 2011, \mnras, 415, 2215

\bibitem[{{Barnab{\`e}} {et~al.}(2009){Barnab{\`e}}, {Czoske}, {Koopmans},
  {Treu}, {Bolton}, \& {Gavazzi}}]{barnabe+2009}
{Barnab{\`e}}, M., {Czoske}, O., {Koopmans}, L. V.~E., {et~al.} 2009, \mnras,
  399, 21

\bibitem[{{Barnab{\`e}} {et~al.}(2012){Barnab{\`e}}, {Dutton}, {Marshall},
  {Auger}, {Brewer}, {Treu}, {Bolton}, {Koo}, \& {Koopmans}}]{barnabe+2012}
{Barnab{\`e}}, M., {Dutton}, A.~A., {Marshall}, P.~J., {et~al.} 2012, \mnras,
  423, 1073

\bibitem[{{Beaton} {et~al.}(2016){Beaton}, {Freedman}, {Madore}, {Bono},
  {Carlson}, {Clementini}, {Durbin}, {Garofalo}, {Hatt}, {Jang}, {Kollmeier},
  {Lee}, {Monson}, {Rich}, {Scowcroft}, {Seibert}, {Sturch}, \&
  {Yang}}]{beaton+2016}
{Beaton}, R.~L., {Freedman}, W.~L., {Madore}, B.~F., {et~al.} 2016, \apj, 832,
  210

\bibitem[{{Bellm} {et~al.}(2019){Bellm}, {Kulkarni}, {Graham}, {Dekany},
  {Smith}, {Riddle}, {Masci}, {Helou}, {Prince}, {Adams}, {Barbarino},
  {Barlow}, {Bauer}, {Beck}, {Belicki}, {Biswas}, {Blagorodnova}, {Bodewits},
  {Bolin}, {Brinnel}, {Brooke}, {Bue}, {Bulla}, {Burruss}, {Cenko}, {Chang},
  {Connolly}, {Coughlin}, {Cromer}, {Cunningham}, {De}, {Delacroix}, {Desai},
  {Duev}, {Eadie}, {Farnham}, {Feeney}, {Feindt}, {Flynn}, {Franckowiak},
  {Frederick}, {Fremling}, {Gal-Yam}, {Gezari}, {Giomi}, {Goldstein},
  {Golkhou}, {Goobar}, {Groom}, {Hacopians}, {Hale}, {Henning}, {Ho}, {Hover},
  {Howell}, {Hung}, {Huppenkothen}, {Imel}, {Ip}, {Ivezi{\'c}}, {Jackson},
  {Jones}, {Juric}, {Kasliwal}, {Kaspi}, {Kaye}, {Kelley}, {Kowalski},
  {Kramer}, {Kupfer}, {Landry}, {Laher}, {Lee}, {Lin}, {Lin}, {Lunnan},
  {Giomi}, {Mahabal}, {Mao}, {Miller}, {Monkewitz}, {Murphy}, {Ngeow},
  {Nordin}, {Nugent}, {Ofek}, {Patterson}, {Penprase}, {Porter}, {Rauch},
  {Rebbapragada}, {Reiley}, {Rigault}, {Rodriguez}, {van Roestel}, {Rusholme},
  {van Santen}, {Schulze}, {Shupe}, {Singer}, {Soumagnac}, {Stein}, {Surace},
  {Sollerman}, {Szkody}, {Taddia}, {Terek}, {Van Sistine}, {van Velzen},
  {Vestrand}, {Walters}, {Ward}, {Ye}, {Yu}, {Yan}, \& {Zolkower}}]{bellm+2019}
{Bellm}, E.~C., {Kulkarni}, S.~R., {Graham}, M.~J., {et~al.} 2019, \pasp, 131,
  018002

\bibitem[{{Birrer} {et~al.}(2016){Birrer}, {Amara}, \&
  {Refregier}}]{birrer+2016}
{Birrer}, S., {Amara}, A., \& {Refregier}, A. 2016, \jcap, 8, 020

\bibitem[{{Birrer} {et~al.}(2020){Birrer}, {Shajib}, {Galan}, {Millon}, {Treu},
  {Agnello}, {Auger}, {Chen}, {Christensen}, {Collett}, {Courbin}, {Fassnacht},
  {Koopmans}, {Marshall}, {Park}, {Rusu}, {Sluse}, {Spiniello}, {Suyu},
  {Wagner-Carena}, {Wong}, {Barnab{\`e}}, {Bolton}, {Czoske}, {Ding},
  {Frieman}, \& {Van de Vyvere}}]{birrer+2020}
{Birrer}, S., {Shajib}, A.~J., {Galan}, A., {et~al.} 2020, arXiv e-prints,
  arXiv:2007.02941

\bibitem[{{Birrer} {et~al.}(2019){Birrer}, {Treu}, {Rusu}, {Bonvin},
  {Fassnacht}, {Chan}, {Agnello}, {Shajib}, {Chen}, {Auger}, {Courbin},
  {Hilbert}, {Sluse}, {Suyu}, {Wong}, {Marshall}, {Lemaux}, \&
  {Meylan}}]{birrer+2019}
{Birrer}, S., {Treu}, T., {Rusu}, C.~E., {et~al.} 2019, \mnras, 484, 4726

\bibitem[{{Bonvin} {et~al.}(2017){Bonvin}, {Courbin}, {Suyu}, {Marshall},
  {Rusu}, {Sluse}, {Tewes}, {Wong}, {Collett}, {Fassnacht}, {Treu}, {Auger},
  {Hilbert}, {Koopmans}, {Meylan}, {Rumbaugh}, {Sonnenfeld}, \&
  {Spiniello}}]{bonvin+2017}
{Bonvin}, V., {Courbin}, F., {Suyu}, S.~H., {et~al.} 2017, \mnras, 465, 4914

\bibitem[{{Bonvin} {et~al.}(2019{\natexlab{a}}){Bonvin}, {Millon}, {Chan},
  {Courbin}, {Rusu}, {Sluse}, {Suyu}, {Wong}, {Fassnacht}, {Marshall}, {Treu},
  {Buckley-Geer}, {Frieman}, {Hempel}, {Kim}, {Lachaume}, {Rabus}, {Chao},
  {Chijani}, {Gilman}, {Gilmore}, {Rojas}, {Williams}, {Anguita}, {Kochanek},
  {Morgan}, {Motta}, {Tewes}, \& {Meylan}}]{bonvin+2019}
{Bonvin}, V., {Millon}, M., {Chan}, J.~H.~H., {et~al.} 2019{\natexlab{a}},
  \aap, 629, A97

\bibitem[{{Bonvin} {et~al.}(2019{\natexlab{b}}){Bonvin}, {Tihhonova}, {Millon},
  {Chan}, {Savary}, {Huber}, \& {Courbin}}]{bonvin+2019b}
{Bonvin}, V., {Tihhonova}, O., {Millon}, M., {et~al.} 2019{\natexlab{b}}, \aap,
  621, A55

\bibitem[{{Bulla} {et~al.}(2016){Bulla}, {Sim}, {Kromer}, {Seitenzahl}, {Fink},
  {Ciaraldi-Schoolmann}, {R{\"o}pke}, {Hillebrandt}, {Pakmor}, {Ruiter}, \&
  {Taubenberger}}]{bulla+2016}
{Bulla}, M., {Sim}, S.~A., {Kromer}, M., {et~al.} 2016, \mnras, 462, 1039

\bibitem[{{Ca\~nameras} {et~al.}(2020){Ca\~nameras}, {Schuldt}, {Suyu},
  {Taubenberger}, {Meinhardt}, {Leal-Taixe}, {Lemon}, {Rojas}, \&
  {Savary}}]{Canameras+2020}
{Ca\~nameras}, R., {Schuldt}, S., {Suyu}, S.~H., {et~al.} 2020, A\&A in press,
  arXiv e-prints, arXiv:2004.13048

\bibitem[{{Chambers} {et~al.}(2016){Chambers}, {Magnier}, {Metcalfe},
  {Flewelling}, {Huber}, {Waters}, {Denneau}, {Draper}, {Farrow}, {Finkbeiner},
  {Holmberg}, {Koppenhoefer}, {Price}, {Rest}, {Saglia}, {Schlafly}, {Smartt},
  {Sweeney}, {Wainscoat}, {Burgett}, {Chastel}, {Grav}, {Heasley}, {Hodapp},
  {Jedicke}, {Kaiser}, {Kudritzki}, {Luppino}, {Lupton}, {Monet}, {Morgan},
  {Onaka}, {Shiao}, {Stubbs}, {Tonry}, {White}, {Ba{\~n}ados}, {Bell},
  {Bender}, {Bernard}, {Boegner}, {Boffi}, {Botticella}, {Calamida},
  {Casertano}, {Chen}, {Chen}, {Cole}, {Deacon}, {Frenk}, {Fitzsimmons},
  {Gezari}, {Gibbs}, {Goessl}, {Goggia}, {Gourgue}, {Goldman}, {Grant},
  {Grebel}, {Hambly}, {Hasinger}, {Heavens}, {Heckman}, {Henderson}, {Henning},
  {Holman}, {Hopp}, {Ip}, {Isani}, {Jackson}, {Keyes}, {Koekemoer}, {Kotak},
  {Le}, {Liska}, {Long}, {Lucey}, {Liu}, {Martin}, {Masci}, {McLean}, {Mindel},
  {Misra}, {Morganson}, {Murphy}, {Obaika}, {Narayan}, {Nieto-Santisteban},
  {Norberg}, {Peacock}, {Pier}, {Postman}, {Primak}, {Rae}, {Rai}, {Riess},
  {Riffeser}, {Rix}, {R{\"o}ser}, {Russel}, {Rutz}, {Schilbach}, {Schultz},
  {Scolnic}, {Strolger}, {Szalay}, {Seitz}, {Small}, {Smith}, {Soderblom},
  {Taylor}, {Thomson}, {Taylor}, {Thakar}, {Thiel}, {Thilker}, {Unger},
  {Urata}, {Valenti}, {Wagner}, {Walder}, {Walter}, {Watters}, {Werner},
  {Wood-Vasey}, \& {Wyse}}]{chambers+2016}
{Chambers}, K.~C., {Magnier}, E.~A., {Metcalfe}, N., {et~al.} 2016, arXiv
  e-prints, arXiv:1612.05560

\bibitem[{{Chan} {et~al.}(2020){Chan}, {Rojas}, {Millon}, {Courbin}, {Bonvin},
  \& {Jauffret}}]{Chan+2020}
{Chan}, J.~H.~H., {Rojas}, K., {Millon}, M., {et~al.} 2020, arXiv e-prints
  (arXiv:2007.14416), arXiv:2007.14416

\bibitem[{{Chen} {et~al.}(2019){Chen}, {Fassnacht}, {Suyu}, {Rusu}, {Chan},
  {Wong}, {Auger}, {Hilbert}, {Bonvin}, {Birrer}, {Millon}, {Koopmans},
  {Lagattuta}, {McKean}, {Vegetti}, {Courbin}, {Ding}, {Halkola}, {Jee},
  {Shajib}, {Sluse}, {Sonnenfeld}, \& {Treu}}]{chen+2019}
{Chen}, G. C.~F., {Fassnacht}, C.~D., {Suyu}, S.~H., {et~al.} 2019, \mnras,
  490, 1743

\bibitem[{{Collett} {et~al.}(2013){Collett}, {Marshall}, {Auger}, {Hilbert},
  {Suyu}, {Greene}, {Treu}, {Fassnacht}, {Koopmans}, {Brada{\v c}}, \&
  {Blandford}}]{collett+2013}
{Collett}, T.~E., {Marshall}, P.~J., {Auger}, M.~W., {et~al.} 2013, \mnras,
  432, 679

\bibitem[{{Courbin} {et~al.}(2018){Courbin}, {Bonvin}, {Buckley-Geer},
  {Fassnacht}, {Frieman}, {Lin}, {Marshall}, {Suyu}, {Treu}, {Anguita},
  {Motta}, {Meylan}, {Paic}, {Tewes}, {Agnello}, {Chao}, {Chijani}, {Gilman},
  {Rojas}, {Williams}, {Hempel}, {Kim}, {Lachaume}, {Rabus}, {Abbott}, {Allam},
  {Annis}, {Banerji}, {Bechtol}, {Benoit-L{\'e}vy}, {Brooks}, {Burke}, {Carnero
  Rosell}, {Carrasco Kind}, {Carretero}, {D'Andrea}, {da Costa}, {Davis},
  {DePoy}, {Desai}, {Flaugher}, {Fosalba}, {Garc{\'{\i}}a-Bellido},
  {Gaztanaga}, {Goldstein}, {Gruen}, {Gruendl}, {Gschwend}, {Gutierrez},
  {Honscheid}, {James}, {Kuehn}, {Kuhlmann}, {Kuropatkin}, {Lahav}, {Lima},
  {Maia}, {March}, {Marshall}, {McMahon}, {Menanteau}, {Miquel}, {Nord},
  {Plazas}, {Sanchez}, {Scarpine}, {Schindler}, {Schubnell}, {Sevilla-Noarbe},
  {Smith}, {Soares-Santos}, {Sobreira}, {Suchyta}, {Tarle}, {Tucker}, {Walker},
  \& {Wester}}]{courbin+2018}
{Courbin}, F., {Bonvin}, V., {Buckley-Geer}, E., {et~al.} 2018, \aap, 609, A71

\bibitem[{{Czoske} {et~al.}(2008){Czoske}, {Barnab{\`e}}, {Koopmans}, {Treu},
  \& {Bolton}}]{czoske+2008}
{Czoske}, O., {Barnab{\`e}}, M., {Koopmans}, L. V.~E., {Treu}, T., \& {Bolton},
  A.~S. 2008, \mnras, 384, 987

\bibitem[{{Dessart} {et~al.}(2014){Dessart}, {Hillier}, {Blondin}, \&
  {Khokhlov}}]{Dessart+2014}
{Dessart}, L., {Hillier}, D.~J., {Blondin}, S., \& {Khokhlov}, A. 2014, \mnras,
  441, 3249

\bibitem[{{Dhawan} {et~al.}(2019){Dhawan}, {Johansson}, {Goobar}, {Amanullah},
  {M{\"o}rtsell}, {Cenko}, {Cooray}, {Fox}, {Goldstein}, {Kalender},
  {Kasliwal}, {Kulkarni}, {Lee}, {Nayyeri}, {Nugent}, {Ofek}, \&
  {Quimby}}]{dhawan+2019}
{Dhawan}, S., {Johansson}, J., {Goobar}, A., {et~al.} 2019, \mnras, 2578

\bibitem[{{Dimitriadis} {et~al.}(2019){Dimitriadis}, {Foley}, {Rest}, {Kasen},
  {Piro}, {Polin}, {Jones}, {Villar}, {Narayan}, {Coulter}, {Kilpatrick},
  {Pan}, {Rojas-Bravo}, {Fox}, {Jha}, {Nugent}, {Riess}, {Scolnic}, {Drout},
  {K2 Mission Team}, {Barentsen}, {Dotson}, {Gully-Santiago}, {Hedges}, {Cody},
  {Barclay}, {Howell}, {KEGS}, {Garnavich}, {Tucker}, {Shaya}, {Mushotzky},
  {Olling}, {Margheim}, {Zenteno}, {Kepler spacecraft Team}, {Coughlin}, {Van
  Cleve}, {Cardoso}, {Larson}, {McCalmont-Everton}, {Peterson}, {Ross},
  {Reedy}, {Osborne}, {McGinn}, {Kohnert}, {Migliorini}, {Wheaton}, {Spencer},
  {Labonde}, {Castillo}, {Beerman}, {Steward}, {Hanley}, {Larsen},
  {Gangopadhyay}, {Kloetzel}, {Weschler}, {Nystrom}, {Moffatt}, {Redick},
  {Griest}, {Packard}, {Muszynski}, {Kampmeier}, {Bjella}, {Flynn},
  {Elsaesser}, {Pan-STARRS}, {Chambers}, {Flewelling}, {Huber}, {Magnier},
  {Waters}, {Schultz}, {Bulger}, {Lowe}, {Willman}, {Smartt}, {Smith}, {DECam},
  {Points}, {Strampelli}, {ASAS-SN}, {Brimacombe}, {Chen}, {Mu{\~n}oz},
  {Mutel}, {Shields}, {Vallely}, {Villanueva}, {PTSS/TNTS}, {Li}, {Wang},
  {Zhang}, {Lin}, {Mo}, {Zhao}, {Sai}, {Zhang}, {Zhang}, {Zhang}, {Wang},
  {Zhang}, {Baron}, {DerKacy}, {Li}, {Chen}, {Xiang}, {Rui}, {Wang}, {Huang},
  {Li}, {Cumbres Observatory}, {Hosseinzadeh}, {Howell}, {Arcavi}, {Hiramatsu},
  {Burke}, {Valenti}, {ATLAS}, {Tonry}, {Denneau}, {Heinze}, {Weiland},
  {Stalder}, {Konkoly}, {Vink{\'o}}, {S{\'a}rneczky}, {P{\'a}l}, {B{\'o}di},
  {Bogn{\'a}r}, {Cs{\'a}k}, {Cseh}, {Cs{\"o}rnyei}, {Hanyecz}, {Ign{\'a}cz},
  {Kalup}, {K{\"o}nyves-T{\'o}th}, {Kriskovics}, {Ordasi}, {Rajmon},
  {S{\'o}dor}, {Szab{\'o}}, {Szak{\'a}ts}, {Zsidi}, {ePESSTO}, {Williams},
  {Nordin}, {Cartier}, {Frohmaier}, {Galbany}, {Guti{\'e}rrez}, {Hook},
  {Inserra}, {Smith}, {Arizona}, {Sand}, {Andrews}, {Smith}, \&
  {Bilinski}}]{dimitriadis+2019}
{Dimitriadis}, G., {Foley}, R.~J., {Rest}, A., {et~al.} 2019, \apjl, 870, L1

\bibitem[{{Falco} {et~al.}(1985){Falco}, {Gorenstein}, \&
  {Shapiro}}]{falco+1985}
{Falco}, E.~E., {Gorenstein}, M.~V., \& {Shapiro}, I.~I. 1985, \apjl, 289, L1

\bibitem[{{Foreman-Mackey} {et~al.}(2013){Foreman-Mackey}, {Hogg}, {Lang}, \&
  {Goodman}}]{Foremanmackey+2013}
{Foreman-Mackey}, D., {Hogg}, D.~W., {Lang}, D., \& {Goodman}, J. 2013, \pasp,
  125, 306

\bibitem[{Foxley-Marrable {et~al.}(2018)Foxley-Marrable, Collett, Vernardos,
  Goldstein, \& Bacon}]{Foxley-Marrable:2018dzu}
Foxley-Marrable, M., Collett, T.~E., Vernardos, G., Goldstein, D.~A., \& Bacon,
  D. 2018, \mnras, 478, 5081

\bibitem[{{Freedman} {et~al.}(2019){Freedman}, {Madore}, {Hatt}, {Hoyt},
  {Jang}, {Beaton}, {Burns}, {Lee}, {Monson}, {Neeley}, {Phillips}, {Rich}, \&
  {Seibert}}]{freedman+2019}
{Freedman}, W.~L., {Madore}, B.~F., {Hatt}, D., {et~al.} 2019, \apj, 882, 34

\bibitem[{{Freedman} {et~al.}(2020){Freedman}, {Madore}, {Hoyt}, {Jang},
  {Beaton}, {Lee}, {Monson}, {Neeley}, \& {Rich}}]{freedman+2020}
{Freedman}, W.~L., {Madore}, B.~F., {Hoyt}, T., {et~al.} 2020, \apj, 891, 57

\bibitem[{{Goldstein} {et~al.}(2019){Goldstein}, {Nugent}, \&
  {Goobar}}]{goldstein+2019}
{Goldstein}, D.~A., {Nugent}, P.~E., \& {Goobar}, A. 2019, \apjs, 243, 6

\bibitem[{{Goldstein} {et~al.}(2018){Goldstein}, {Nugent}, {Kasen}, \&
  {Collett}}]{goldstein+2018}
{Goldstein}, D.~A., {Nugent}, P.~E., {Kasen}, D.~N., \& {Collett}, T.~E. 2018,
  \apj, 855, 22

\bibitem[{{Goobar} {et~al.}(2017){Goobar}, {Amanullah}, {Kulkarni}, {Nugent},
  {Johansson}, {Steidel}, {Law}, {M{\"o}rtsell}, {Quimby}, {Blagorodnova},
  {Brandeker}, {Cao}, {Cooray}, {Ferretti}, {Fremling}, {Hangard}, {Kasliwal},
  {Kupfer}, {Lunnan}, {Masci}, {Miller}, {Nayyeri}, {Neill}, {Ofek},
  {Papadogiannakis}, {Petrushevska}, {Ravi}, {Sollerman}, {Sullivan}, {Taddia},
  {Walters}, {Wilson}, {Yan}, \& {Yaron}}]{goobar+2017}
{Goobar}, A., {Amanullah}, R., {Kulkarni}, S.~R., {et~al.} 2017, Science, 356,
  291

\bibitem[{{Greene} {et~al.}(2013){Greene}, {Suyu}, {Treu}, {Hilbert}, {Auger},
  {Collett}, {Marshall}, {Fassnacht}, {Blandford}, {Brada{\v c}}, \&
  {Koopmans}}]{greene+2013}
{Greene}, Z.~S., {Suyu}, S.~H., {Treu}, T., {et~al.} 2013, \apj, 768, 39

\bibitem[{{Grillo} {et~al.}(2016){Grillo}, {Karman}, {Suyu}, {Rosati},
  {Balestra}, {Mercurio}, {Lombardi}, {Treu}, {Caminha}, {Halkola}, {Rodney},
  {Gavazzi}, \& {Caputi}}]{grillo+2016}
{Grillo}, C., {Karman}, W., {Suyu}, S.~H., {et~al.} 2016, \apj, 822, 78

\bibitem[{{Grillo} {et~al.}(2018){Grillo}, {Rosati}, {Suyu}, {Balestra},
  {Caminha}, {Halkola}, {Kelly}, {Lombardi}, {Mercurio}, {Rodney}, \&
  {Treu}}]{grillo+2018}
{Grillo}, C., {Rosati}, P., {Suyu}, S.~H., {et~al.} 2018, \apj, 860, 94

\bibitem[{{Grillo} {et~al.}(2020){Grillo}, {Rosati}, {Suyu}, {Caminha},
  {Mercurio}, \& {Halkola}}]{grillo+2020}
{Grillo}, C., {Rosati}, P., {Suyu}, S.~H., {et~al.} 2020, \apj, 898, 87

\bibitem[{{Hezaveh} {et~al.}(2017){Hezaveh}, {Perreault Levasseur}, \&
  {Marshall}}]{hezaveh+2017}
{Hezaveh}, Y.~D., {Perreault Levasseur}, L., \& {Marshall}, P.~J. 2017, \nat,
  548, 555

\bibitem[{{Huber} {et~al.}(2019){Huber}, {Suyu}, {Noebauer}, {Bonvin},
  {Rothchild}, {Chan}, {Awan}, {Courbin}, {Kromer}, {Marshall}, {Oguri},
  {Ribeiro}, \& {LSST Dark Energy Science Collaboration}}]{huber+2019}
{Huber}, S., {Suyu}, S.~H., {Noebauer}, U.~M., {et~al.} 2019, \aap, 631, A161

\bibitem[{{Huber} {et~al.}(2020){Huber}, {Suyu}, {Noebauer}, {Chan}, {Kromer},
  {Sim}, {Sluse}, \& {Taubenberger}}]{Huber+2020}
{Huber}, S., {Suyu}, S.~H., {Noebauer}, U.~M., {et~al.} 2020, arXiv e-prints
  (arXiv:2008.10393), arXiv:2008.10393

\bibitem[{{Iben} \& {Tutukov}(1984)}]{iben+1984}
{Iben}, I., J. \& {Tutukov}, A.~V. 1984, \apjs, 54, 335

\bibitem[{{Ivezi{\'c}} {et~al.}(2019){Ivezi{\'c}}, {Kahn}, {Tyson}, {Abel},
  {Acosta}, {Allsman}, {Alonso}, {AlSayyad}, {Anderson}, {Andrew}, {Angel},
  {Angeli}, {Ansari}, {Antilogus}, {Araujo}, {Armstrong}, {Arndt}, {Astier},
  {Aubourg}, {Auza}, {Axelrod}, {Bard}, {Barr}, {Barrau}, {Bartlett}, {Bauer},
  {Bauman}, {Baumont}, {Bechtol}, {Bechtol}, {Becker}, {Becla}, {Beldica},
  {Bellavia}, {Bianco}, {Biswas}, {Blanc}, {Blazek}, {Bland ford}, {Bloom},
  {Bogart}, {Bond}, {Booth}, {Borgland}, {Borne}, {Bosch}, {Boutigny},
  {Brackett}, {Bradshaw}, {Brand t}, {Brown}, {Bullock}, {Burchat}, {Burke},
  {Cagnoli}, {Calabrese}, {Callahan}, {Callen}, {Carlin}, {Carlson}, {Chand
  rasekharan}, {Charles-Emerson}, {Chesley}, {Cheu}, {Chiang}, {Chiang},
  {Chirino}, {Chow}, {Ciardi}, {Claver}, {Cohen-Tanugi}, {Cockrum}, {Coles},
  {Connolly}, {Cook}, {Cooray}, {Covey}, {Cribbs}, {Cui}, {Cutri}, {Daly},
  {Daniel}, {Daruich}, {Daubard}, {Daues}, {Dawson}, {Delgado}, {Dellapenna},
  {de Peyster}, {de Val-Borro}, {Digel}, {Doherty}, {Dubois},
  {Dubois-Felsmann}, {Durech}, {Economou}, {Eifler}, {Eracleous}, {Emmons},
  {Fausti Neto}, {Ferguson}, {Figueroa}, {Fisher-Levine}, {Focke}, {Foss},
  {Frank}, {Freemon}, {Gangler}, {Gawiser}, {Geary}, {Gee}, {Geha}, {Gessner},
  {Gibson}, {Gilmore}, {Glanzman}, {Glick}, {Goldina}, {Goldstein}, {Goodenow},
  {Graham}, {Gressler}, {Gris}, {Guy}, {Guyonnet}, {Haller}, {Harris},
  {Hascall}, {Haupt}, {Hernand ez}, {Herrmann}, {Hileman}, {Hoblitt},
  {Hodgson}, {Hogan}, {Howard}, {Huang}, {Huffer}, {Ingraham}, {Innes},
  {Jacoby}, {Jain}, {Jammes}, {Jee}, {Jenness}, {Jernigan}, {Jevremovi{\'c}},
  {Johns}, {Johnson}, {Johnson}, {Jones}, {Juramy-Gilles}, {Juri{\'c}},
  {Kalirai}, {Kallivayalil}, {Kalmbach}, {Kantor}, {Karst}, {Kasliwal},
  {Kelly}, {Kessler}, {Kinnison}, {Kirkby}, {Knox}, {Kotov}, {Krabbendam},
  {Krughoff}, {Kub{\'a}nek}, {Kuczewski}, {Kulkarni}, {Ku}, {Kurita}, {Lage},
  {Lambert}, {Lange}, {Langton}, {Le Guillou}, {Levine}, {Liang}, {Lim},
  {Lintott}, {Long}, {Lopez}, {Lotz}, {Lupton}, {Lust}, {MacArthur}, {Mahabal},
  {Mand elbaum}, {Markiewicz}, {Marsh}, {Marshall}, {Marshall}, {May},
  {McKercher}, {McQueen}, {Meyers}, {Migliore}, {Miller}, {Mills}, {Miraval},
  {Moeyens}, {Moolekamp}, {Monet}, {Moniez}, {Monkewitz}, {Montgomery},
  {Morrison}, {Mueller}, {Muller}, {Mu{\~n}oz Arancibia}, {Neill}, {Newbry},
  {Nief}, {Nomerotski}, {Nordby}, {O'Connor}, {Oliver}, {Olivier}, {Olsen},
  {O'Mullane}, {Ortiz}, {Osier}, {Owen}, {Pain}, {Palecek}, {Parejko},
  {Parsons}, {Pease}, {Peterson}, {Peterson}, {Petravick}, {Libby Petrick},
  {Petry}, {Pierfederici}, {Pietrowicz}, {Pike}, {Pinto}, {Plante}, {Plate},
  {Plutchak}, {Price}, {Prouza}, {Radeka}, {Rajagopal}, {Rasmussen},
  {Regnault}, {Reil}, {Reiss}, {Reuter}, {Ridgway}, {Riot}, {Ritz}, {Robinson},
  {Roby}, {Roodman}, {Rosing}, {Roucelle}, {Rumore}, {Russo}, {Saha},
  {Sassolas}, {Schalk}, {Schellart}, {Schindler}, {Schmidt}, {Schneider},
  {Schneider}, {Schoening}, {Schumacher}, {Schwamb}, {Sebag}, {Selvy},
  {Sembroski}, {Seppala}, {Serio}, {Serrano}, {Shaw}, {Shipsey}, {Sick},
  {Silvestri}, {Slater}, {Smith}, {Smith}, {Sobhani}, {Soldahl},
  {Storrie-Lombardi}, {Stover}, {Strauss}, {Street}, {Stubbs}, {Sullivan},
  {Sweeney}, {Swinbank}, {Szalay}, {Takacs}, {Tether}, {Thaler}, {Thayer},
  {Thomas}, {Thornton}, {Thukral}, {Tice}, {Trilling}, {Turri}, {Van Berg},
  {Vanden Berk}, {Vetter}, {Virieux}, {Vucina}, {Wahl}, {Walkowicz}, {Walsh},
  {Walter}, {Wang}, {Wang}, {Warner}, {Wiecha}, {Willman}, {Winters},
  {Wittman}, {Wolff}, {Wood-Vasey}, {Wu}, {Xin}, {Yoachim}, \&
  {Zhan}}]{ivezic+2019}
{Ivezi{\'c}}, {\v{Z}}., {Kahn}, S.~M., {Tyson}, J.~A., {et~al.} 2019, \apj,
  873, 111

\bibitem[{{Jacobs} {et~al.}(2019){Jacobs}, {Collett}, {Glazebrook}, {McCarthy},
  {Qin}, {Abbott}, {Abdalla}, {Annis}, {Avila}, {Bechtol}, {Bertin}, {Brooks},
  {Buckley-Geer}, {Burke}, {Carnero Rosell}, {Carrasco Kind}, {Carretero}, {da
  Costa}, {Davis}, {De Vicente}, {Desai}, {Diehl}, {Doel}, {Eifler},
  {Flaugher}, {Frieman}, {Garc{\'\i}a-Bellido}, {Gaztanaga}, {Gerdes},
  {Goldstein}, {Gruen}, {Gruendl}, {Gschwend}, {Gutierrez}, {Hartley},
  {Hollowood}, {Honscheid}, {Hoyle}, {James}, {Kuehn}, {Kuropatkin}, {Lahav},
  {Li}, {Lima}, {Lin}, {Maia}, {Martini}, {Miller}, {Miquel}, {Nord}, {Plazas},
  {Sanchez}, {Scarpine}, {Schubnell}, {Serrano}, {Sevilla-Noarbe}, {Smith},
  {Soares-Santos}, {Sobreira}, {Suchyta}, {Swanson}, {Tarle}, {Vikram},
  {Walker}, {Zhang}, {Zuntz}, \& {DES Collaboration}}]{jacobs+2019}
{Jacobs}, C., {Collett}, T., {Glazebrook}, K., {et~al.} 2019, \mnras, 484, 5330

\bibitem[{{Jee} {et~al.}(2015){Jee}, {Komatsu}, \& {Suyu}}]{jee+2015}
{Jee}, I., {Komatsu}, E., \& {Suyu}, S.~H. 2015, Journal of Cosmology and
  Astro-Particle Physics, 2015, 033

\bibitem[{{Jee} {et~al.}(2016){Jee}, {Komatsu}, {Suyu}, \&
  {Huterer}}]{jee+2016}
{Jee}, I., {Komatsu}, E., {Suyu}, S.~H., \& {Huterer}, D. 2016, Journal of
  Cosmology and Astro-Particle Physics, 2016, 031

\bibitem[{Jee {et~al.}(2019)Jee, Suyu, Komatsu, Fassnacht, Hilbert, \&
  Koopmans}]{jee+2019}
Jee, I., Suyu, S.~H., Komatsu, E., {et~al.} 2019, Science, 365, 1134

\bibitem[{{Kaiser} {et~al.}(2010){Kaiser}, {Burgett}, {Chambers}, {Denneau},
  {Heasley}, {Jedicke}, {Magnier}, {Morgan}, {Onaka}, \& {Tonry}}]{kaiser+2010}
{Kaiser}, N., {Burgett}, W., {Chambers}, K., {et~al.} 2010, Society of
  Photo-Optical Instrumentation Engineers (SPIE) Conference Series, Vol. 7733,
  {The Pan-STARRS wide-field optical/NIR imaging survey}, 77330E

\bibitem[{{Kasen}(2010)}]{Kasen:2010}
{Kasen}, D. 2010, \apj, 708, 1025

\bibitem[{{Kawamata} {et~al.}(2016){Kawamata}, {Oguri}, {Ishigaki},
  {Shimasaku}, \& {Ouchi}}]{kawamata+2016}
{Kawamata}, R., {Oguri}, M., {Ishigaki}, M., {Shimasaku}, K., \& {Ouchi}, M.
  2016, \apj, 819, 114

\bibitem[{{Kelly} {et~al.}(2016{\natexlab{a}}){Kelly}, {Brammer}, {Selsing},
  {Foley}, {Hjorth}, {Rodney}, {Christensen}, {Strolger}, {Filippenko}, {Treu},
  {Steidel}, {Strom}, {Riess}, {Zitrin}, {Schmidt}, {Brada{\v{c}}}, {Jha},
  {Graham}, {McCully}, {Graur}, {Weiner}, {Silverman}, \&
  {Taddia}}]{kelly+2016a}
{Kelly}, P.~L., {Brammer}, G., {Selsing}, J., {et~al.} 2016{\natexlab{a}},
  \apj, 831, 205

\bibitem[{{Kelly} {et~al.}(2015){Kelly}, {Rodney}, {Treu}, {Foley}, {Brammer},
  {Schmidt}, {Zitrin}, {Sonnenfeld}, {Strolger}, {Graur}, {Filippenko}, {Jha},
  {Riess}, {Bradac}, {Weiner}, {Scolnic}, {Malkan}, {von der Linden}, {Trenti},
  {Hjorth}, {Gavazzi}, {Fontana}, {Merten}, {McCully}, {Jones}, {Postman},
  {Dressler}, {Patel}, {Cenko}, {Graham}, \& {Tucker}}]{kelly+2015}
{Kelly}, P.~L., {Rodney}, S.~A., {Treu}, T., {et~al.} 2015, Science, 347, 1123

\bibitem[{{Kelly} {et~al.}(2016{\natexlab{b}}){Kelly}, {Rodney}, {Treu},
  {Strolger}, {Foley}, {Jha}, {Selsing}, {Brammer}, {Brada{\v{c}}}, {Cenko},
  {Graur}, {Filippenko}, {Hjorth}, {McCully}, {Molino}, {Nonino}, {Riess},
  {Schmidt}, {Tucker}, {von der Linden}, {Weiner}, \& {Zitrin}}]{kelly+2016b}
{Kelly}, P.~L., {Rodney}, S.~A., {Treu}, T., {et~al.} 2016{\natexlab{b}},
  \apjl, 819, L8

\bibitem[{{Kochanek}(2020{\natexlab{a}})}]{kochanek2020b}
{Kochanek}, C.~S. 2020{\natexlab{a}}, arXiv e-prints, arXiv:2003.08395

\bibitem[{{Kochanek}(2020{\natexlab{b}})}]{kochanek2020a}
{Kochanek}, C.~S. 2020{\natexlab{b}}, \mnras, 493, 1725

\bibitem[{{Kormann} {et~al.}(1994){Kormann}, {Schneider}, \&
  {Bartelmann}}]{Kormann:1994}
{Kormann}, R., {Schneider}, P., \& {Bartelmann}, M. 1994, A\&A, 284, 285

\bibitem[{{Kromer} {et~al.}(2016){Kromer}, {Fremling}, {Pakmor},
  {Taubenberger}, {Amanullah}, {Cenko}, {Fransson}, {Goobar}, {Leloudas},
  {Taddia}, {R{\"o}pke}, {Seitenzahl}, {Sim}, \& {Sollerman}}]{Kromer+2016}
{Kromer}, M., {Fremling}, C., {Pakmor}, R., {et~al.} 2016, \mnras, 459, 4428

\bibitem[{Kromer \& Sim(2009)}]{Kromer:2009ce}
Kromer, M. \& Sim. 2009, MNRAS, 398, 1809

\bibitem[{{Law} {et~al.}(2009){Law}, {Kulkarni}, {Dekany}, {Ofek}, {Quimby},
  {Nugent}, {Surace}, {Grillmair}, {Bloom}, {Kasliwal}, {Bildsten}, {Brown},
  {Cenko}, {Ciardi}, {Croner}, {Djorgovski}, {van Eyken}, {Filippenko}, {Fox},
  {Gal-Yam}, {Hale}, {Hamam}, {Helou}, {Henning}, {Howell}, {Jacobsen},
  {Laher}, {Mattingly}, {McKenna}, {Pickles}, {Poznanski}, {Rahmer}, {Rau},
  {Rosing}, {Shara}, {Smith}, {Starr}, {Sullivan}, {Velur}, {Walters}, \&
  {Zolkower}}]{law+2009}
{Law}, N.~M., {Kulkarni}, S.~R., {Dekany}, R.~G., {et~al.} 2009, \pasp, 121,
  1395

\bibitem[{{Lentz} {et~al.}(2000){Lentz}, {Baron}, {Branch}, {Hauschildt}, \&
  {Nugent}}]{Lentz+2000}
{Lentz}, E.~J., {Baron}, E., {Branch}, D., {Hauschildt}, P.~H., \& {Nugent},
  P.~E. 2000, \apj, 530, 966

\bibitem[{{Lucy}(1999)}]{Lucy1999}
{Lucy}, L.~B. 1999, \aap, 345, 211

\bibitem[{{Maeda} {et~al.}(2018){Maeda}, {Jiang}, {Shigeyama}, \&
  {Doi}}]{maeda+2018}
{Maeda}, K., {Jiang}, J.-a., {Shigeyama}, T., \& {Doi}, M. 2018, \apj, 861, 78

\bibitem[{{Masci} {et~al.}(2019){Masci}, {Laher}, {Rusholme}, {Shupe}, {Groom},
  {Surace}, {Jackson}, {Monkewitz}, {Beck}, {Flynn}, {Terek}, {Landry},
  {Hacopians}, {Desai}, {Howell}, {Brooke}, {Imel}, {Wachter}, {Ye}, {Lin},
  {Cenko}, {Cunningham}, {Rebbapragada}, {Bue}, {Miller}, {Mahabal}, {Bellm},
  {Patterson}, {Juri{\'c}}, {Golkhou}, {Ofek}, {Walters}, {Graham}, {Kasliwal},
  {Dekany}, {Kupfer}, {Burdge}, {Cannella}, {Barlow}, {Van Sistine}, {Giomi},
  {Fremling}, {Blagorodnova}, {Levitan}, {Riddle}, {Smith}, {Helou}, {Prince},
  \& {Kulkarni}}]{masci+2019}
{Masci}, F.~J., {Laher}, R.~R., {Rusholme}, B., {et~al.} 2019, \pasp, 131,
  018003

\bibitem[{{Millon} {et~al.}(2020){Millon}, {Galan}, {Courbin}, {Treu}, {Suyu},
  {Ding}, {Birrer}, {Chen}, {Shajib}, {Sluse}, {Wong}, {Agnello}, {Auger},
  {Buckley-Geer}, {Chan}, {Collett}, {Fassnacht}, {Hilbert}, {Koopmans},
  {Motta}, {Mukherjee}, {Rusu}, {Sonnenfeld}, {Spiniello}, \& {Van de
  Vyvere}}]{millon+2020}
{Millon}, M., {Galan}, A., {Courbin}, F., {et~al.} 2020, \aap, 639, A101

\bibitem[{{More} {et~al.}(2017){More}, {Suyu}, {Oguri}, {More}, \&
  {Lee}}]{more+2017}
{More}, A., {Suyu}, S.~H., {Oguri}, M., {More}, S., \& {Lee}, C.-H. 2017,
  \apjl, 835, L25

\bibitem[{Noebauer {et~al.}(2017)Noebauer, Kromer, Taubenberger, Baklanov,
  Blinnikov, Sorokina, \& Hillebrandt}]{Noebauer:2017vsf}
Noebauer, U.~M., Kromer, M., Taubenberger, S., {et~al.} 2017, Mon. Not. Roy.
  Astron. Soc., 472, 2787

\bibitem[{{Nomoto} {et~al.}(1984){Nomoto}, {Thielemann}, \&
  {Yokoi}}]{1984:Nomoto}
{Nomoto}, K., {Thielemann}, F.-K., \& {Yokoi}, K. 1984, ApJ, 286, 644

\bibitem[{{Oguri}(2019)}]{oguri2019}
{Oguri}, M. 2019, Reports on Progress in Physics, 82, 126901

\bibitem[{{Oguri} \& {Marshall}(2010)}]{ogurimarshall2010}
{Oguri}, M. \& {Marshall}, P.~J. 2010, \mnras, 405, 2579

\bibitem[{{Pakmor} {et~al.}(2012){Pakmor}, {Kromer}, {Taubenberger}, {Sim},
  {R{\"o}pke}, \& {Hillebrandt}}]{pakmor+2012}
{Pakmor}, R., {Kromer}, M., {Taubenberger}, S., {et~al.} 2012, \apjl, 747, L10

\bibitem[{{Paraficz} \& {Hjorth}(2009)}]{paraficzhjorth2009}
{Paraficz}, D. \& {Hjorth}, J. 2009, \aap, 507, L49

\bibitem[{{Pearson} {et~al.}(2019){Pearson}, {Li}, \& {Dye}}]{pearson+2019}
{Pearson}, J., {Li}, N., \& {Dye}, S. 2019, \mnras, 488, 991

\bibitem[{{Perreault Levasseur} {et~al.}(2017){Perreault Levasseur}, {Hezaveh},
  \& {Wechsler}}]{levasseur+2017}
{Perreault Levasseur}, L., {Hezaveh}, Y.~D., \& {Wechsler}, R.~H. 2017, \apjl,
  850, L7

\bibitem[{{Pesce} {et~al.}(2020){Pesce}, {Braatz}, {Reid}, {Riess}, {Scolnic},
  {Condon}, {Gao}, {Henkel}, {Impellizzeri}, {Kuo}, \& {Lo}}]{pesce+2020}
{Pesce}, D.~W., {Braatz}, J.~A., {Reid}, M.~J., {et~al.} 2020, \apjl, 891, L1

\bibitem[{{Pierel} \& {Rodney}(2019)}]{pierel+2019}
{Pierel}, J.~D.~R. \& {Rodney}, S. 2019, \apj, 876, 107

\bibitem[{Piro \& Morozova(2016)}]{Piro:2015}
Piro, A.~L. \& Morozova, V.~S. 2016, Astrophys. J., 826, 96

\bibitem[{{Piro} \& {Nakar}(2013)}]{Piro:2013}
{Piro}, A.~L. \& {Nakar}, E. 2013, \apj, 769, 67

\bibitem[{{Piro} \& {Nakar}(2014)}]{Piro:2014}
{Piro}, A.~L. \& {Nakar}, E. 2014, \apj, 784, 85

\bibitem[{{Planck Collaboration}(2020)}]{planck+2020}
{Planck Collaboration}. 2020, \aap, 641, A6

\bibitem[{{Quimby} {et~al.}(2014){Quimby}, {Oguri}, {More}, {More}, {Moriya},
  {Werner}, {Tanaka}, {Folatelli}, {Bersten}, {Maeda}, \&
  {Nomoto}}]{quimby+2014}
{Quimby}, R.~M., {Oguri}, M., {More}, A., {et~al.} 2014, Science, 344, 396

\bibitem[{{Quimby} {et~al.}(2013){Quimby}, {Werner}, {Oguri}, {More}, {More},
  {Tanaka}, {Nomoto}, {Moriya}, {Folatelli}, {Maeda}, \&
  {Bersten}}]{quimby+2013}
{Quimby}, R.~M., {Werner}, M.~C., {Oguri}, M., {et~al.} 2013, \apjl, 768, L20

\bibitem[{{Rabinak} \& {Waxman}(2011)}]{Rabinak:2011}
{Rabinak}, I. \& {Waxman}, E. 2011, \apj, 728, 63

\bibitem[{{Refsdal}(1964)}]{refsdal1964}
{Refsdal}, S. 1964, \mnras, 128, 307

\bibitem[{{Riess}(2019)}]{riess2019}
{Riess}, A.~G. 2019, Nature Reviews Physics, 2, 10

\bibitem[{{Riess} {et~al.}(2019){Riess}, {Casertano}, {Yuan}, {Macri}, \&
  {Scolnic}}]{riess+2019}
{Riess}, A.~G., {Casertano}, S., {Yuan}, W., {Macri}, L.~M., \& {Scolnic}, D.
  2019, \apj, 876, 85

\bibitem[{{Rodney} {et~al.}(2016){Rodney}, {Strolger}, {Kelly}, {Brada{\v{c}}},
  {Brammer}, {Filippenko}, {Foley}, {Graur}, {Hjorth}, {Jha}, {McCully},
  {Molino}, {Riess}, {Schmidt}, {Selsing}, {Sharon}, {Treu}, {Weiner}, \&
  {Zitrin}}]{rodney+2016}
{Rodney}, S.~A., {Strolger}, L.~G., {Kelly}, P.~L., {et~al.} 2016, \apj, 820,
  50

\bibitem[{{R{\"o}pke} {et~al.}(2012){R{\"o}pke}, {Kromer}, {Seitenzahl},
  {Pakmor}, {Sim}, {Taubenberger}, {Ciaraldi-Schoolmann}, {Hillebrandt},
  {Aldering}, {Antilogus}, {Baltay}, {Benitez-Herrera}, {Bongard}, {Buton},
  {Canto}, {Cellier-Holzem}, {Childress}, {Chotard}, {Copin}, {Fakhouri},
  {Fink}, {Fouchez}, {Gangler}, {Guy}, {Hachinger}, {Hsiao}, {Chen},
  {Kerschhaggl}, {Kowalski}, {Nugent}, {Paech}, {Pain}, {Pecontal}, {Pereira},
  {Perlmutter}, {Rabinowitz}, {Rigault}, {Runge}, {Saunders}, {Smadja},
  {Suzuki}, {Tao}, {Thomas}, {Tilquin}, \& {Wu}}]{roepke+2012}
{R{\"o}pke}, F.~K., {Kromer}, M., {Seitenzahl}, I.~R., {et~al.} 2012, \apjl,
  750, L19

\bibitem[{{Rusu} {et~al.}(2017){Rusu}, {Fassnacht}, {Sluse}, {Hilbert}, {Wong},
  {Huang}, {Suyu}, {Collett}, {Marshall}, {Treu}, \& {Koopmans}}]{rusu+2017}
{Rusu}, C.~E., {Fassnacht}, C.~D., {Sluse}, D., {et~al.} 2017, \mnras, 467,
  4220

\bibitem[{{Rusu} {et~al.}(2020){Rusu}, {Wong}, {Bonvin}, {Sluse}, {Suyu},
  {Fassnacht}, {Chan}, {Hilbert}, {Auger}, {Sonnenfeld}, {Birrer}, {Courbin},
  {Treu}, {Chen}, {Halkola}, {Koopmans}, {Marshall}, \& {Shajib}}]{rusu+2020}
{Rusu}, C.~E., {Wong}, K.~C., {Bonvin}, V., {et~al.} 2020, \mnras, 498, 1440

\bibitem[{{Schneider} \& {Sluse}(2013)}]{schneidersluse2013}
{Schneider}, P. \& {Sluse}, D. 2013, \aap, 559, A37

\bibitem[{{Schuldt} {et~al.}(2020){Schuldt}, {Suyu}, {Meinhardt},
  {Leal-Taix{\'e}}, {Ca{\~n}ameras}, {Taubenberger}, \&
  {Halkola}}]{Schuldt+2020}
{Schuldt}, S., {Suyu}, S.~H., {Meinhardt}, T., {et~al.} 2020, arXiv e-prints
  (arXiv:2010.00602), arXiv:2010.00602

\bibitem[{{Shajib} {et~al.}(2020){Shajib}, {Birrer}, {Treu}, {Agnello},
  {Buckley-Geer}, {Chan}, {Christensen}, {Lemon}, {Lin}, {Millon}, {Poh},
  {Rusu}, {Sluse}, {Spiniello}, {Chen}, {Collett}, {Courbin}, {Fassnacht},
  {Frieman}, {Galan}, {Gilman}, {More}, {Anguita}, {Auger}, {Bonvin},
  {McMahon}, {Meylan}, {Wong}, {Abbott}, {Annis}, {Avila}, {Bechtol}, {Brooks},
  {Brout}, {Burke}, {Carnero Rosell}, {Carrasco Kind}, {Carretero},
  {Castander}, {Costanzi}, {da Costa}, {De Vicente}, {Desai}, {Dietrich},
  {Doel}, {Drlica-Wagner}, {Evrard}, {Finley}, {Flaugher}, {Fosalba},
  {Garc{\'\i}a-Bellido}, {Gerdes}, {Gruen}, {Gruendl}, {Gschwend}, {Gutierrez},
  {Hollowood}, {Honscheid}, {Huterer}, {James}, {Jeltema}, {Krause},
  {Kuropatkin}, {Li}, {Lima}, {MacCrann}, {Maia}, {Marshall}, {Melchior},
  {Miquel}, {Ogando}, {Palmese}, {Paz-Chinch{\'o}n}, {Plazas}, {Romer},
  {Roodman}, {Sako}, {Sanchez}, {Santiago}, {Scarpine}, {Schubnell}, {Scolnic},
  {Serrano}, {Sevilla-Noarbe}, {Smith}, {Soares-Santos}, {Suchyta}, {Tarle},
  {Thomas}, {Walker}, \& {Zhang}}]{shajib+2020}
{Shajib}, A.~J., {Birrer}, S., {Treu}, T., {et~al.} 2020, \mnras, 494, 6072

\bibitem[{{Sim} {et~al.}(2010){Sim}, {R{\"o}pke}, {Hillebrandt}, {Kromer},
  {Pakmor}, {Fink}, {Ruiter}, \& {Seitenzahl}}]{sim+2010}
{Sim}, S.~A., {R{\"o}pke}, F.~K., {Hillebrandt}, W., {et~al.} 2010, \apjl, 714,
  L52

\bibitem[{{Sim} {et~al.}(2013){Sim}, {Seitenzahl}, {Kromer},
  {Ciaraldi-Schoolmann}, {R{\"o}pke}, {Fink}, {Hillebrandt}, {Pakmor},
  {Ruiter}, \& {Taubenberger}}]{Sim+2013}
{Sim}, S.~A., {Seitenzahl}, I.~R., {Kromer}, M., {et~al.} 2013, \mnras, 436,
  333

\bibitem[{{Suyu} {et~al.}(2017){Suyu}, {Bonvin}, {Courbin}, {Fassnacht},
  {Rusu}, {Sluse}, {Treu}, {Wong}, {Auger}, {Ding}, {Hilbert}, {Marshall},
  {Rumbaugh}, {Sonnenfeld}, {Tewes}, {Tihhonova}, {Agnello}, {Blandford},
  {Chen}, {Collett}, {Koopmans}, {Liao}, {Meylan}, \& {Spiniello}}]{suyu+2017}
{Suyu}, S.~H., {Bonvin}, V., {Courbin}, F., {et~al.} 2017, \mnras, 468, 2590

\bibitem[{{Suyu} {et~al.}(2018){Suyu}, {Chang}, {Courbin}, \&
  {Okumura}}]{suyu+2018}
{Suyu}, S.~H., {Chang}, T.-C., {Courbin}, F., \& {Okumura}, T. 2018, \ssr, 214,
  91

\bibitem[{{Suyu} {et~al.}(2010){Suyu}, {Marshall}, {Auger}, {Hilbert},
  {Blandford}, {Koopmans}, {Fassnacht}, \& {Treu}}]{suyu+2010}
{Suyu}, S.~H., {Marshall}, P.~J., {Auger}, M.~W., {et~al.} 2010, \apj, 711, 201

\bibitem[{{Suyu} {et~al.}(2014){Suyu}, {Treu}, {Hilbert}, {Sonnenfeld},
  {Auger}, {Blandford}, {Collett}, {Courbin}, {Fassnacht}, {Koopmans},
  {Marshall}, {Meylan}, {Spiniello}, \& {Tewes}}]{suyu+2014}
{Suyu}, S.~H., {Treu}, T., {Hilbert}, S., {et~al.} 2014, \apjl, 788, L35

\bibitem[{{Taubenberger} {et~al.}(2019){Taubenberger}, {Suyu}, {Komatsu},
  {Jee}, {Birrer}, {Bonvin}, {Courbin}, {Rusu}, {Shajib}, \&
  {Wong}}]{taubenberger+2019}
{Taubenberger}, S., {Suyu}, S.~H., {Komatsu}, E., {et~al.} 2019, \aap, 628, L7

\bibitem[{{Tie} \& {Kochanek}(2018)}]{tiekochanek2018}
{Tie}, S.~S. \& {Kochanek}, C.~S. 2018, \mnras, 473, 80

\bibitem[{{Tihhonova} {et~al.}(2018){Tihhonova}, {Courbin}, {Harvey},
  {Hilbert}, {Rusu}, {Fassnacht}, {Bonvin}, {Marshall}, {Meylan}, {Sluse},
  {Suyu}, {Treu}, \& {Wong}}]{tihhonova+2018}
{Tihhonova}, O., {Courbin}, F., {Harvey}, D., {et~al.} 2018, \mnras, 477, 5657

\bibitem[{{Treu} {et~al.}(2016){Treu}, {Brammer}, {Diego}, {Grillo}, {Kelly},
  {Oguri}, {Rodney}, {Rosati}, {Sharon}, {Zitrin}, {Balestra}, {Brada{\v{c}}},
  {Broadhurst}, {Caminha}, {Halkola}, {Hoag}, {Ishigaki}, {Johnson}, {Karman},
  {Kawamata}, {Mercurio}, {Schmidt}, {Strolger}, {Suyu}, {Filippenko}, {Foley},
  {Jha}, \& {Patel}}]{treu+2016}
{Treu}, T., {Brammer}, G., {Diego}, J.~M., {et~al.} 2016, \apj, 817, 60

\bibitem[{{Treu} \& {Marshall}(2016)}]{treumarshall2016}
{Treu}, T. \& {Marshall}, P.~J. 2016, \aapr, 24, 11

\bibitem[{{Tutukov} \& {Yungelson}(1981)}]{tutukov+1981}
{Tutukov}, A.~V. \& {Yungelson}, L.~R. 1981, Nauchnye Informatsii, 49, 3

\bibitem[{Vernardos {et~al.}(2015)Vernardos, Fluke, Bate, Croton, \&
  Vohl}]{Vernardos:2015wta}
Vernardos, G., Fluke, C.~J., Bate, N.~F., Croton, D., \& Vohl, D. 2015, ApJ
  Suppl., 217, 23

\bibitem[{{Walker} {et~al.}(2012){Walker}, {Hachinger}, {Mazzali}, {Ellis},
  {Sullivan}, {Gal Yam}, \& {Howell}}]{Walker+2012}
{Walker}, E.~S., {Hachinger}, S., {Mazzali}, P.~A., {et~al.} 2012, \mnras, 427,
  103

\bibitem[{{Wertz} {et~al.}(2018){Wertz}, {Orthen}, \& {Schneider}}]{wertz+2018}
{Wertz}, O., {Orthen}, B., \& {Schneider}, P. 2018, \aap, 617, A140

\bibitem[{{Whelan} \& {Iben}(1973)}]{whelan+1973}
{Whelan}, J. \& {Iben}, Icko, J. 1973, \apj, 186, 1007

\bibitem[{{Wojtak} {et~al.}(2019){Wojtak}, {Hjorth}, \& {Gall}}]{wojtak+2019}
{Wojtak}, R., {Hjorth}, J., \& {Gall}, C. 2019, \mnras, 487, 3342

\bibitem[{{Wong} {et~al.}(2017){Wong}, {Suyu}, {Auger}, {Bonvin}, {Courbin},
  {Fassnacht}, {Halkola}, {Rusu}, {Sluse}, {Sonnenfeld}, {Treu}, {Collett},
  {Hilbert}, {Koopmans}, {Marshall}, \& {Rumbaugh}}]{wong+2017}
{Wong}, K.~C., {Suyu}, S.~H., {Auger}, M.~W., {et~al.} 2017, \mnras, 465, 4895

\bibitem[{{Wong} {et~al.}(2020){Wong}, {Suyu}, {Chen}, {Rusu}, {Millon},
  {Sluse}, {Bonvin}, {Fassnacht}, {Taubenberger}, {Auger}, {Birrer}, {Chan},
  {Courbin}, {Hilbert}, {Tihhonova}, {Treu}, {Agnello}, {Ding}, {Jee},
  {Komatsu}, {Shajib}, {Sonnenfeld}, {Blandford}, {Koopmans}, {Marshall}, \&
  {Meylan}}]{wong+2020}
{Wong}, K.~C., {Suyu}, S.~H., {Chen}, G.~C.~F., {et~al.} 2020, \mnras, 498,
  1420

\bibitem[{Yahalomi {et~al.}(2017)Yahalomi, Schechter, \&
  Wambsganss}]{Yahalomi:2017ihe}
Yahalomi, D.~A., Schechter, P.~L., \& Wambsganss, J. 2017
  [\eprint[arXiv]{1711.07919}]

\bibitem[{{Y{\i}ld{\i}r{\i}m} {et~al.}(2020){Y{\i}ld{\i}r{\i}m}, {Suyu}, \&
  {Halkola}}]{yildirim+2020}
{Y{\i}ld{\i}r{\i}m}, A., {Suyu}, S.~H., \& {Halkola}, A. 2020, \mnras, 493,
  4783

\bibitem[{{Yuan} {et~al.}(2019){Yuan}, {Riess}, {Macri}, {Casertano}, \&
  {Scolnic}}]{yuan+2019}
{Yuan}, W., {Riess}, A.~G., {Macri}, L.~M., {Casertano}, S., \& {Scolnic},
  D.~M. 2019, \apj, 886, 61

\end{thebibliography}

\begin{appendix}

\section{Specific intensity profiles}
\label{sec: Specific intensity profiles}
\begin{figure}[ht]
\centering
\subfigure{\includegraphics[trim=5 17 15 14,clip,width=.24\textwidth]{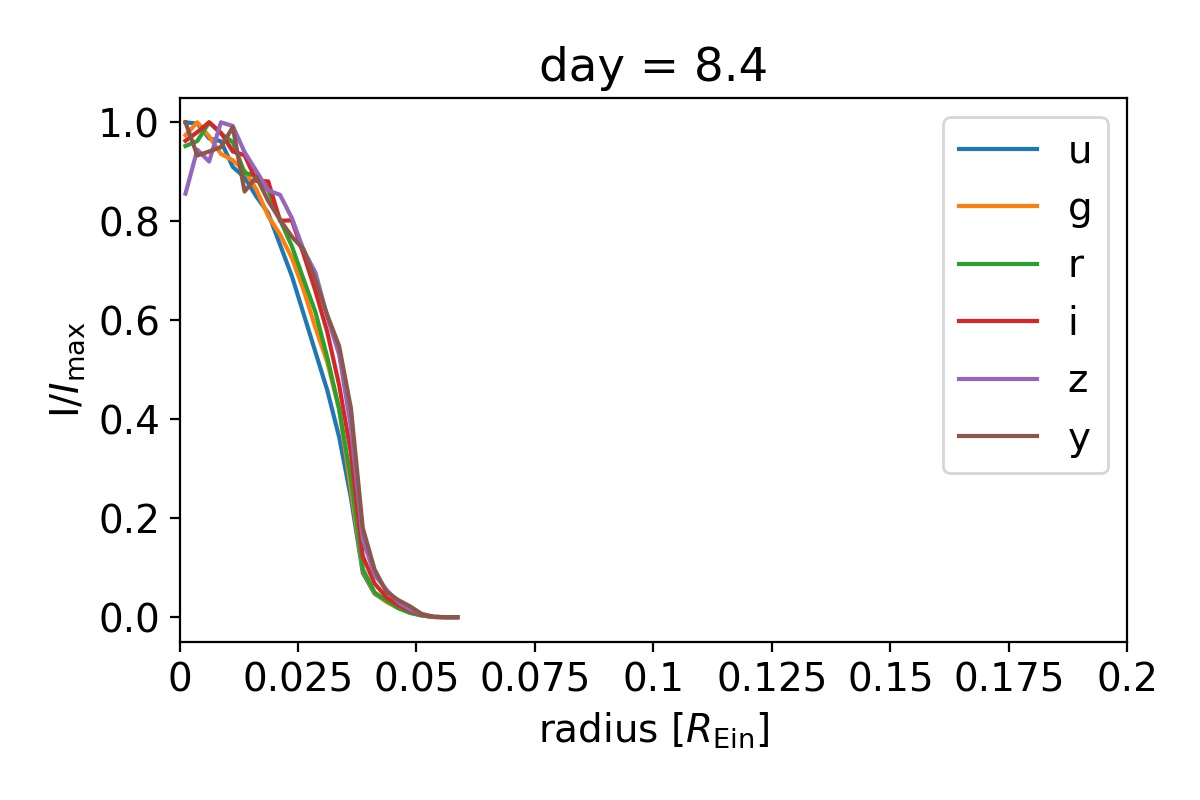}}
\subfigure{\includegraphics[trim=5 17 15 14,clip,width=.24\textwidth]{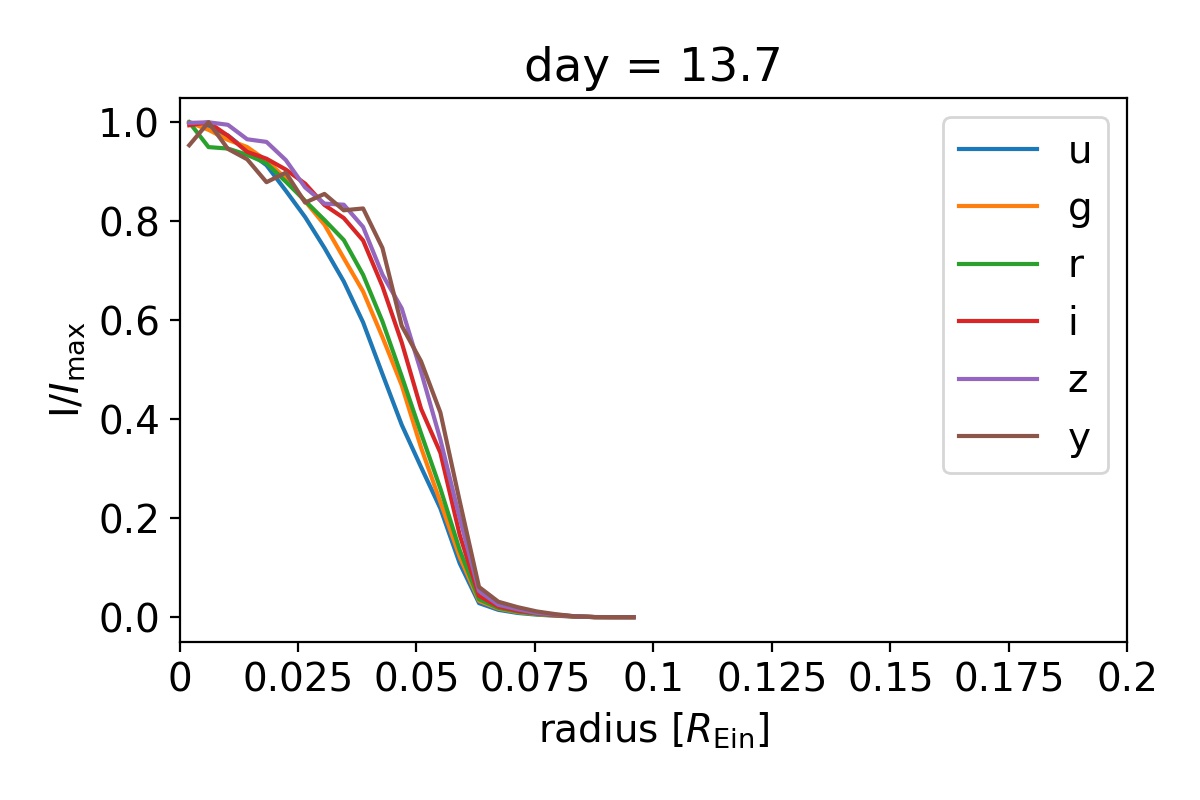}}
\subfigure{\includegraphics[trim=5 17 15 14,clip,width=.24\textwidth]{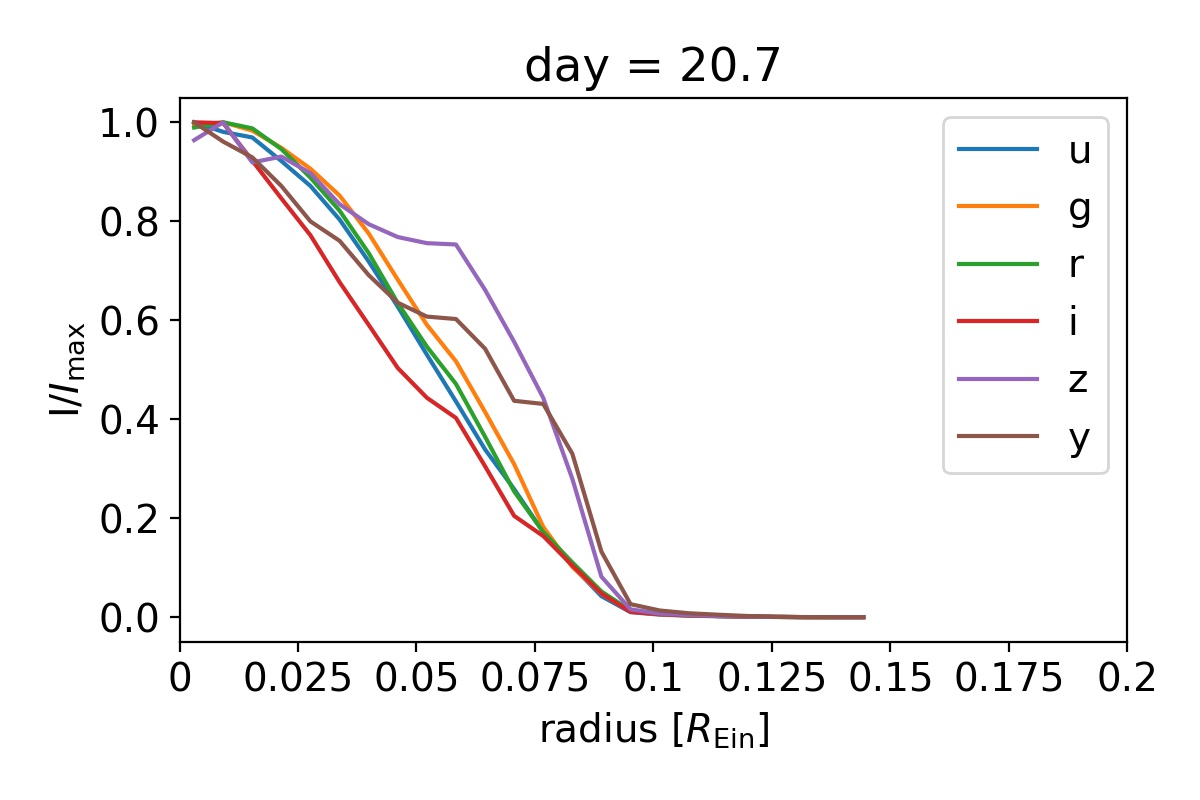}}
\subfigure{\includegraphics[trim=5 17 15 14,clip,width=.24\textwidth]{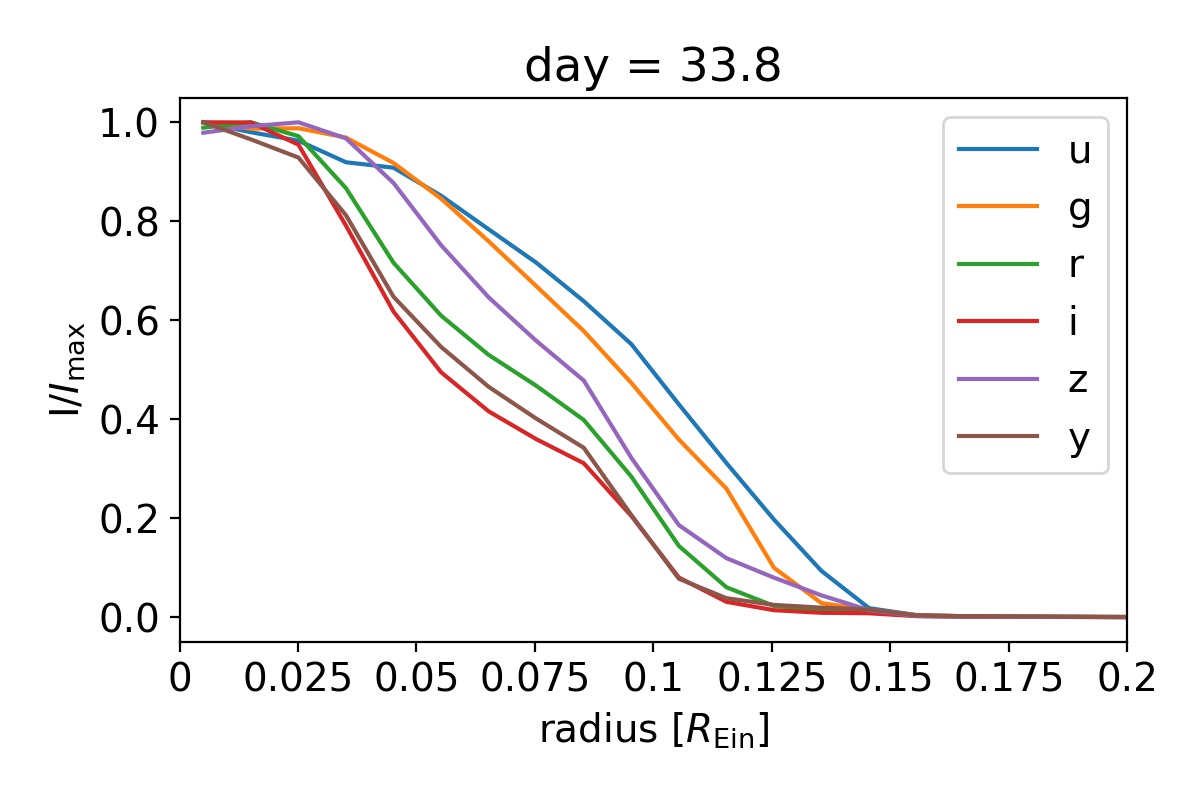}}
\caption{Normalised specific intensity profiles of the W7 model for 6 LSST filters and four different rest-frame times after explosion.}
\label{fig: normalized specific intensity profiles}
\end{figure}

In Figure \ref{fig: normalized specific intensity profiles} we show
the normalised specific intensity profiles of the W7 model for 4 different rest-frame
times after explosion for the 6 LSST filters \textit{u, g, r, i, z,}
and \textit{y}. The specific intensity profiles at early times are
more similar to each other than at later stages, which leads to the so-called achromatic phase described in \cite{goldstein+2018}.  The specific intensity profiles for the other SN explosion models show similar qualitative trend.

\section{Microlensing maps}
\label{sec: Microlensing maps}
In Figure \ref{fig: microlensing maps different kappa, gammas}, we show
examples of the microlensing magnification maps that we have used in
Section \ref{sec:microlensSN}. 
The panels on the left correspond to type I macrolensing images (i.e.,
time-delay minimum images), where as the panels on the right correspond
to type II macrolensing images (i.e., time-delay saddle images).\footnote{Lensing
    images appear at stationary points of the time-delay surface (Fermat's
    Principle).  For a typical lens system with either 4 or 2
    macrolens images, each of
    the images is
    either a minimum (type I image) or a saddle (type II image) in the
    time-delay surface.\label{footnote}}
These maps show the magnification factor $\mu(x,y)$ 
as a function of Cartesian coordinates $x$ and $y$ on the source plane in units of the Einstein radius
\begin{equation}
\Rein=\sqrt{\frac{4 G \langle M \rangle}{c^2} \frac{\source \ls}{\lens}}.
\label{eq: Einstein Radius physical coordinate in cm}
\end{equation}

We assume a Salpeter initial mass function with a mean mass of the point mass microlenses (stars 
in the foreground lens galaxy) of $\langle M \rangle = 0.35 M_\odot$. 
As defined in Section \ref{sec:cosmo}, the angular diameter distances $\source$, $\lens$ and $\ls$ are
distances from us to the source, from us to the lens (deflector), and between the
lens and the source, respectively. To calculate these distances, we assume a 
flat $\Lambda$CDM cosmology with $H_0 = 72 \, \mathrm{km} \, \mathrm{s}^{-1} \, \mathrm{Mpc}^{-1}$ and $\Om = 0.26$,
following \cite{ogurimarshall2010} whose lensed SN Ia catalog is used in this work, 
and further $z_\mathrm{d} = 0.32$ and 
$z_\mathrm{s} = 0.77$, which correspond to the median values of the OM10 sample.  For these redshifts, the Einstein radius is equal to $2.9\times 10^{16}$\,cm.
Our maps have a resolution of 20,000 $\times$ 20,000 pixels with a total size of $10 \Rein$ $\times$ $10 \Rein$.  
Therefore the size of a pixel is $0.0005\Rein = 1.5 \times 10^{13}$\,cm. Defining the radius of the SN as the radius of the projected disc that encloses 99.9\% of the specific intensity, the W7 SN radius covers at day 4.0 about 50 pixels, at day 8.4 about 100 pixels, and at day 39.8 about 400 pixels.

\begin{figure}[ht]
\centering
\subfigure{\includegraphics[trim=30 17 22 14,clip,width=.24\textwidth]{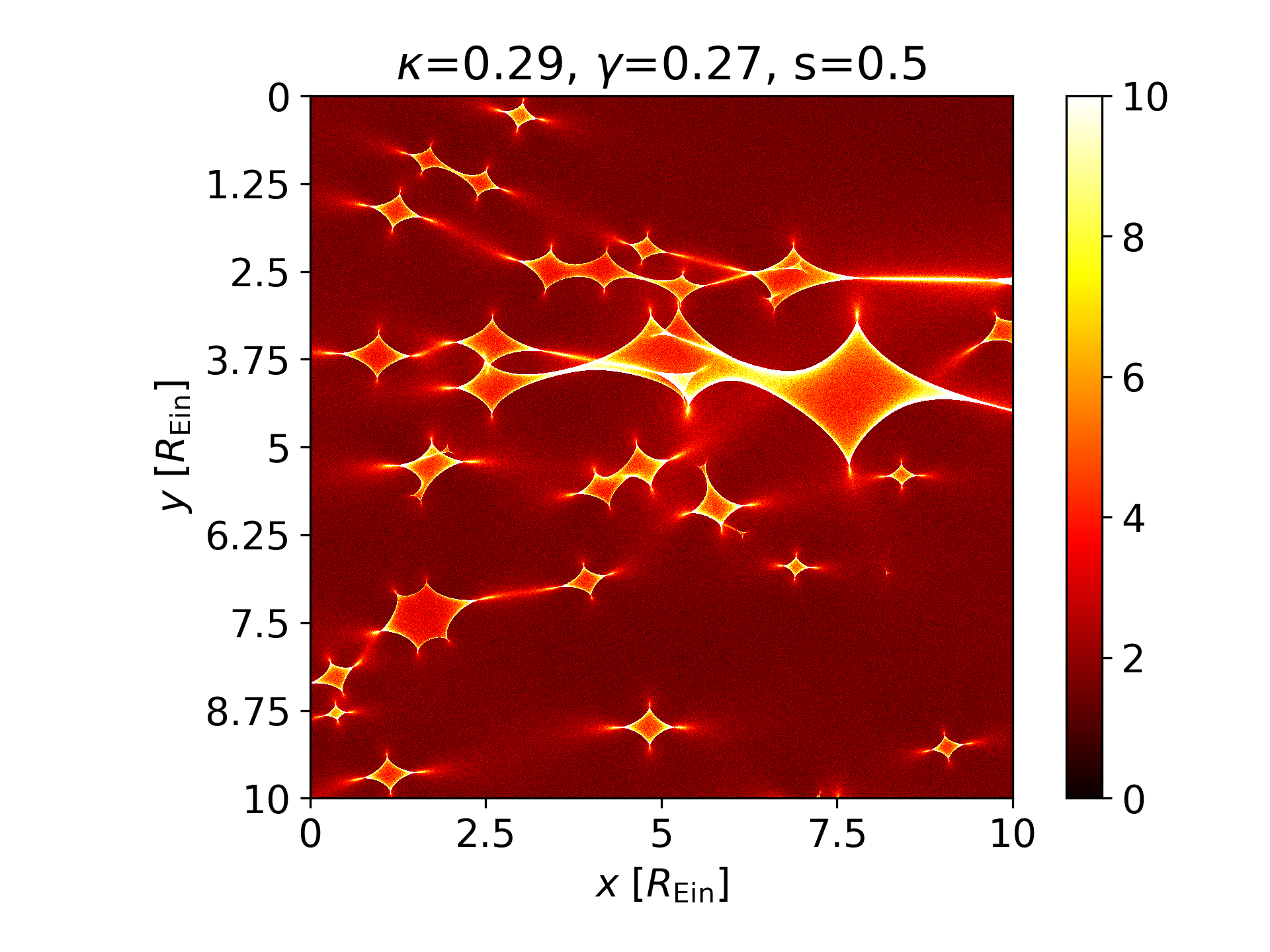}}
\subfigure{\includegraphics[trim=30 17 22 14,clip,width=.24\textwidth]{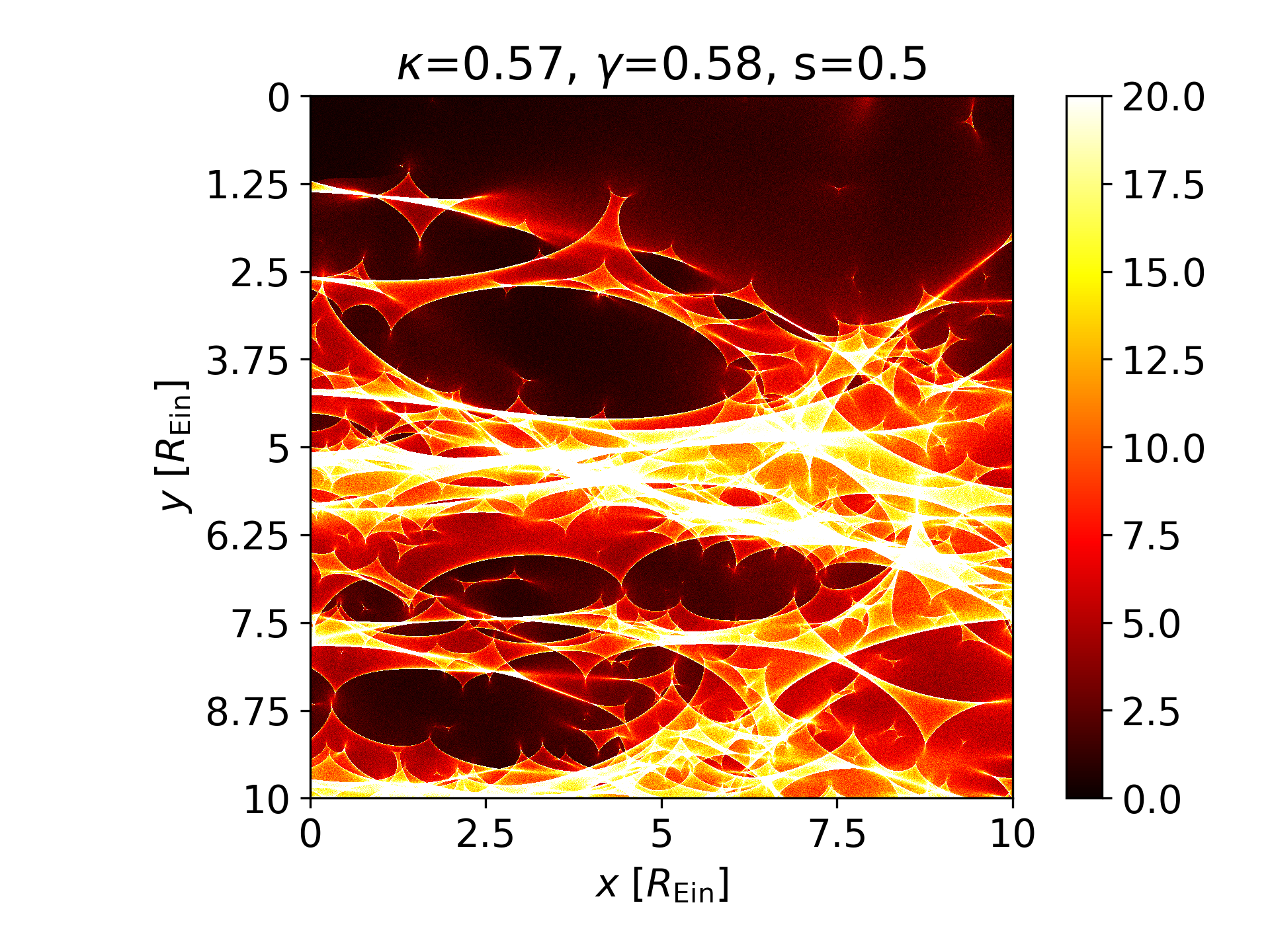}}
\subfigure{\includegraphics[trim=30 17 22 14,clip,width=.24\textwidth]{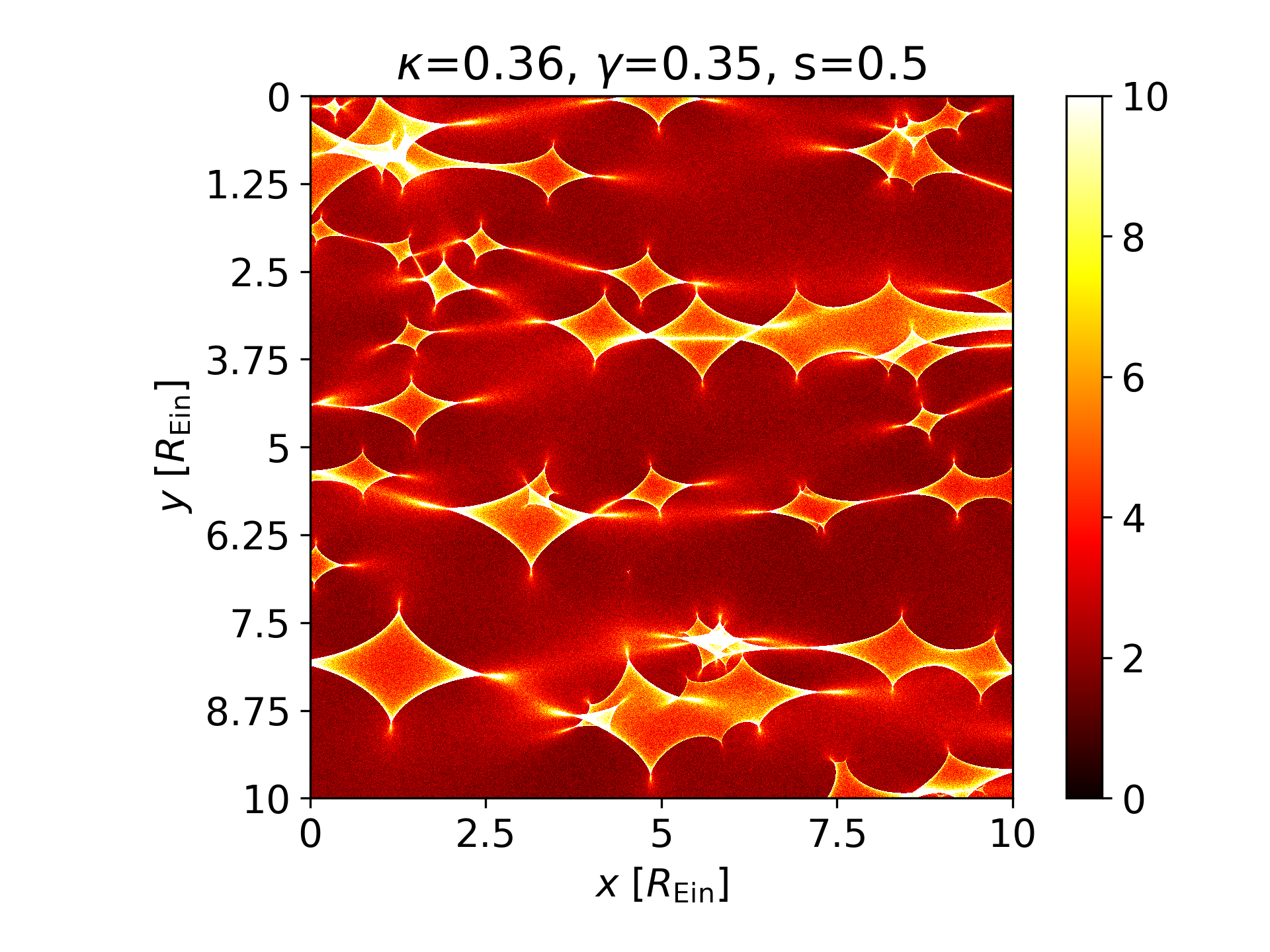}}
\subfigure{\includegraphics[trim=30 17 22 14,clip,width=.24\textwidth]{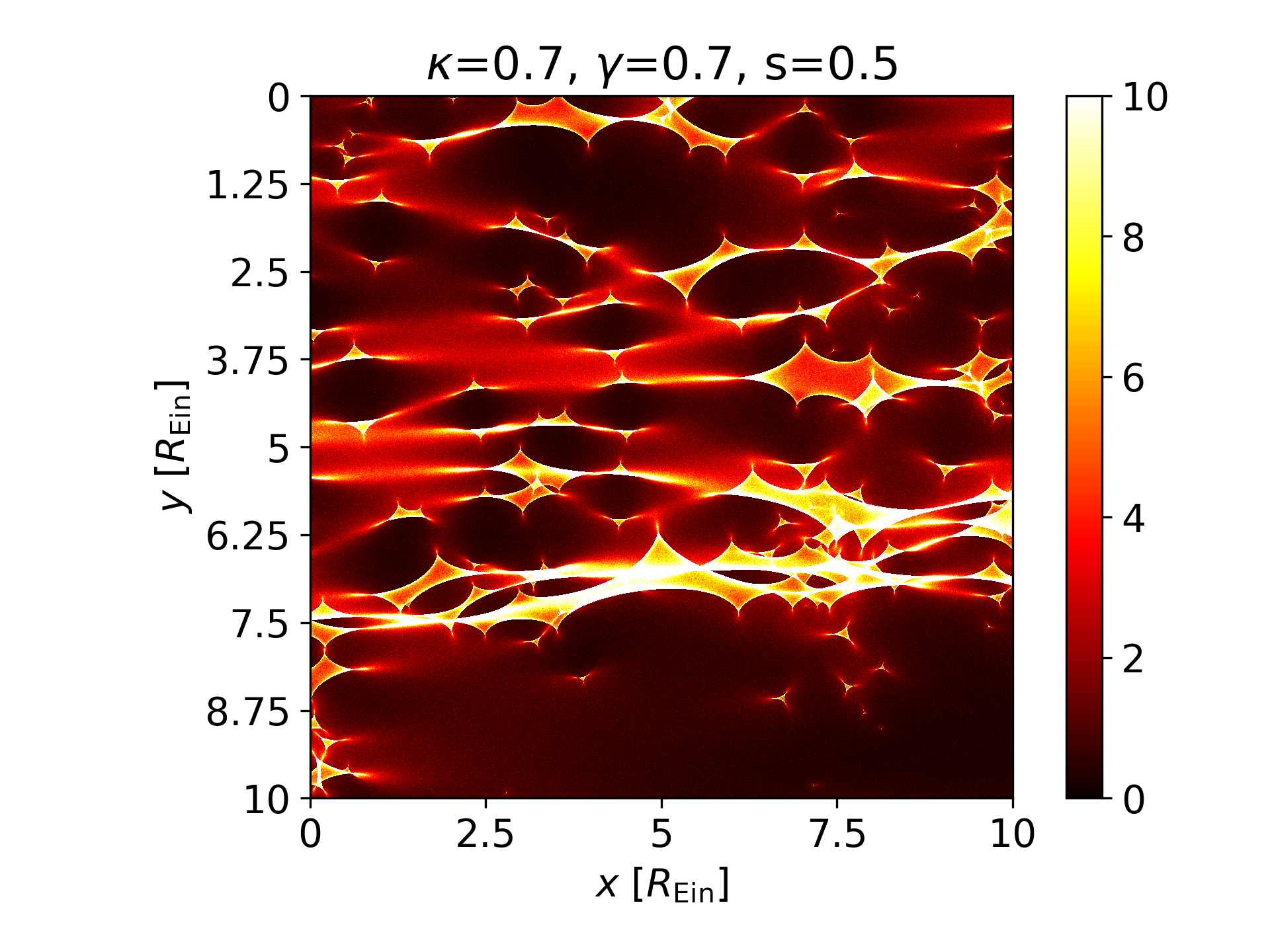}}
\subfigure{\includegraphics[trim=30 17 22 14,clip,width=.24\textwidth]{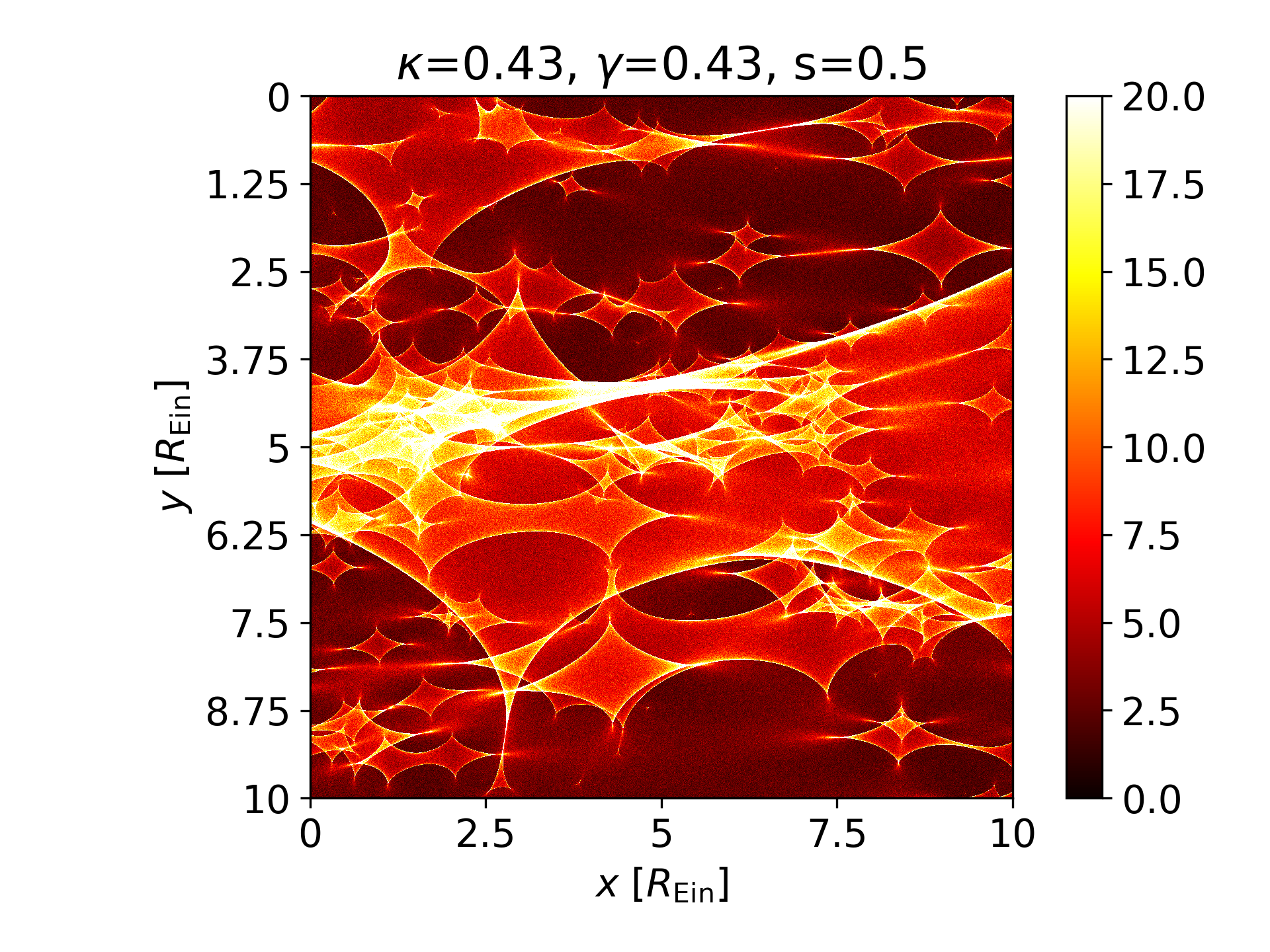}}
\subfigure{\includegraphics[trim=30 17 22 14,clip,width=.24\textwidth]{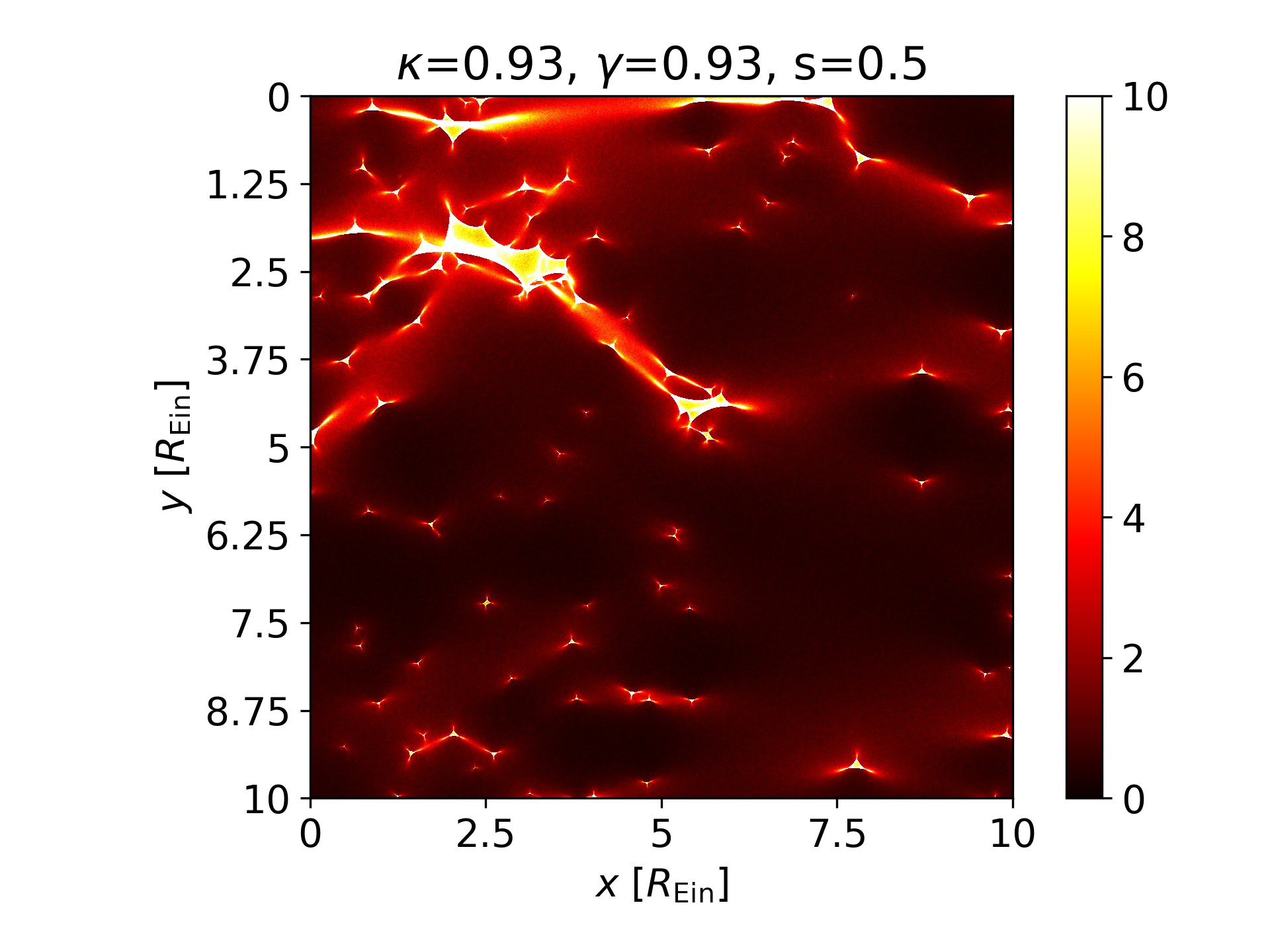}}
\caption{Magnification maps for six different ($\kappa,\gamma$) pairs
  for $s = 0.5$. The colour scale indicates for each panel the
  magnification factor $\mu(x,y)$. The panels on the left-hand side correspond to type
  I lensing images (time-delay minimum images) and the panels on the right-hand side to
  type II lensing images (time-delay saddle images).$^{\ref{footnote}}$ In all panels, 
  many micro caustics are present, separating low and high
  magnification areas.}
\label{fig: microlensing maps different kappa, gammas}
\end{figure}

\section{Covariance matrix of the spectral deviation}
\label{sec:covariance_matrix}

\begin{figure}
\centering
 \includegraphics[width=0.5\textwidth]{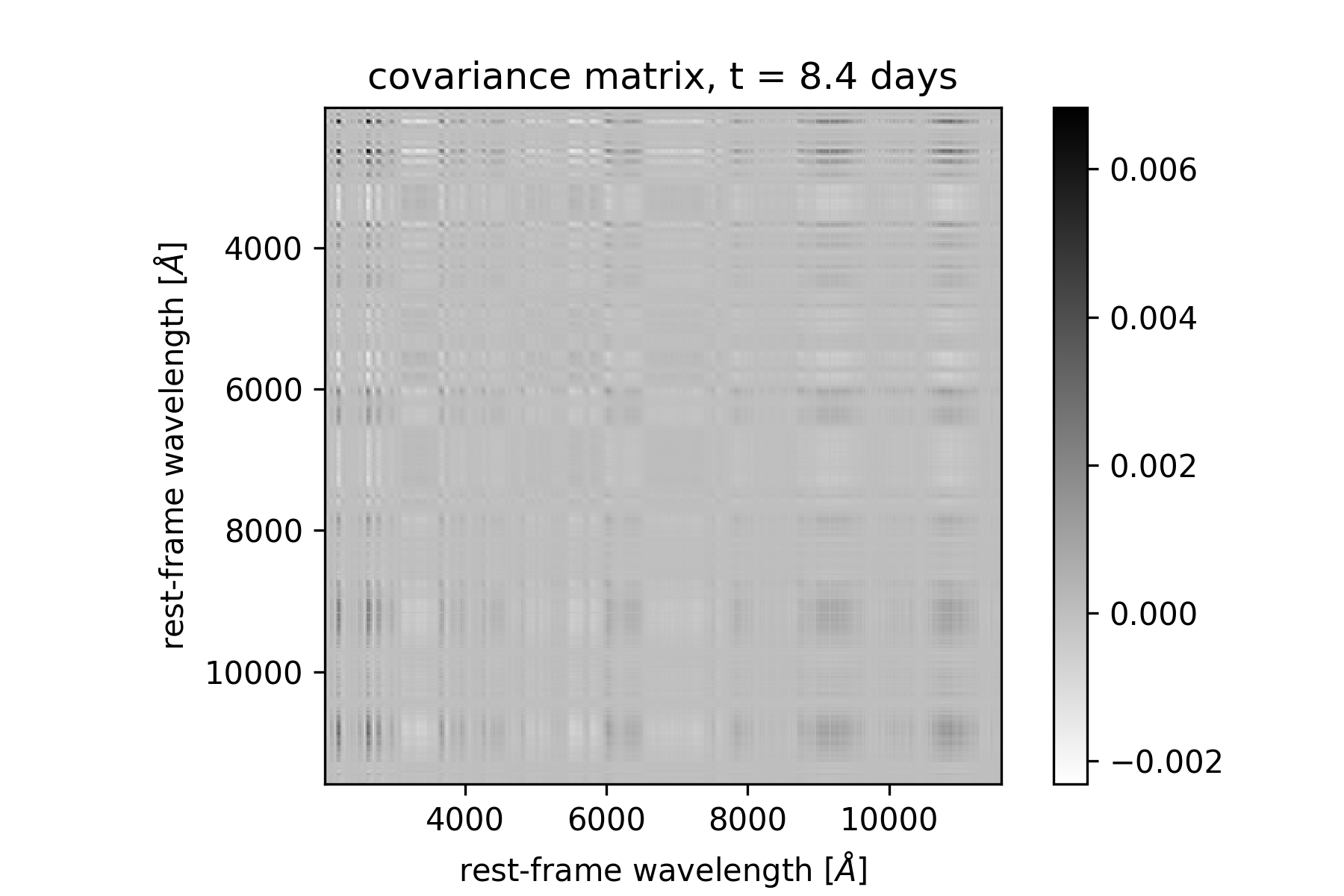}
 \caption{Covariance matrix of $\Delta_\lambda(t=8.4\,{\rm
       days})$ between different wavelength bins for the subCh model.
      Wavelengths that have high deviations (as shown in Figure
      \ref{fig:subCh_deviation_of_microlensed_flux}) tend to be
      positively correlated with each other.  }
 \label{fig:deviation_cov_matrix}
\end{figure}

As an example, we show in Figure \ref{fig:deviation_cov_matrix} the
covariance matrix of the deviations $\Delta_\lambda$ across
wavelengths for the subCh model at 8.4 days after explosion.  The covariance between wavelength bin $i$ and $j$ is defined as 
\be
{\rm cov}(\Delta_i, \Delta_j) = \frac{1}{N_{\rm spec}} \sum_{k=1}^{N_{\rm spec}} (\Delta_{i,k}-\overline{\Delta_i})(\Delta_{j,k}-\overline{\Delta_j}),
\ee
where $N_{\rm spec}=3\times10^5$ is the number of microlensed spectra (for the 30 microlensing maps and 10,000 positions per map), and $\overline{\Delta_i}$ ($\overline{\Delta_j}$) is the mean deviation of wavelength bin $i$ ($j$), averaged over $N_{\rm spec}$.
Comparing Figure \ref{fig:subCh_deviation_of_microlensed_flux} with
Figure \ref{fig:deviation_cov_matrix}, the covariance matrix shows
positive correlations for wavelengths that have higher
$\Delta_\lambda$.  Since the $\Delta_\lambda$ are typically higher at
wavelengths that have relatively lower fluxes in the spectrum due to
absorption features or lower continuum, the features in the covariance
matrix reflect the spectral evolution of the SN Ia.  As time progresses after SN
explosion and the SN fluxes are suppressed at specific wavelengths
(e.g., due to absorption features), deviations at these wavelengths
generally become stronger and are thus positively correlated.  The
covariance matrices at other epochs and of other SN Ia models show
similar behaviour.

\section{Deviations in spectra between different SN Ia progenitor models}
\label{sec:deviations_mod_later}

\begin{figure*}
\centering
 \includegraphics[width=0.97\textwidth]{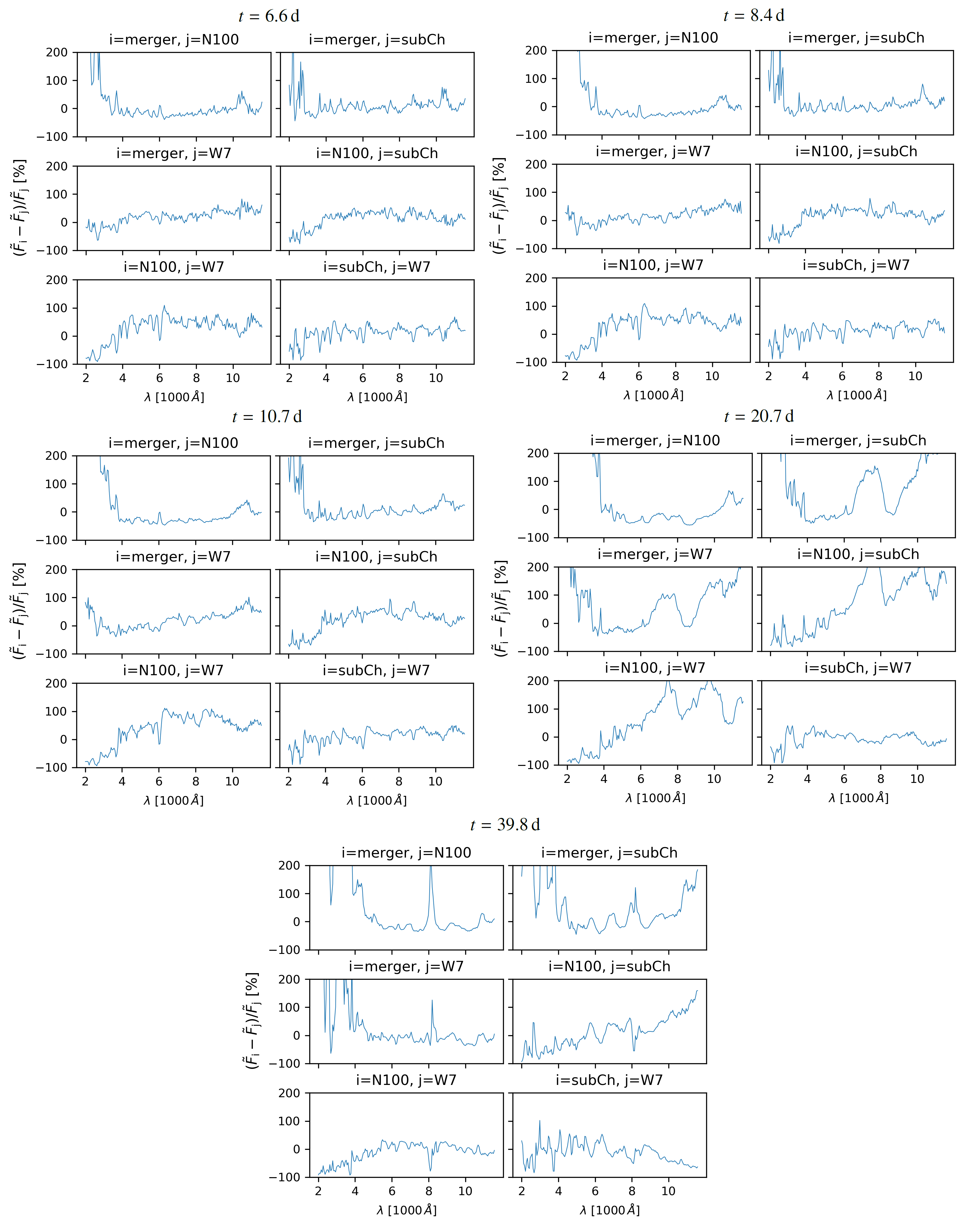}
 \caption{Deviations $\Delta_\lambda(t)$ between pairs of SN Ia
     spectra from the four SN models (W7, N100, subCh, and merger) at
     five different rest-frame times $t$ after explosion, as indicated
     on top of each panel. The panels and labels are in the
     same format as in Figure \ref{fig:deviations_SN_models_4}.}
 \label{fig:deviations_SN_models_summarized}
\end{figure*}

We show in Figure \ref{fig:deviations_SN_models_summarized} the deviations between spectra from
different pairs of SN Ia models
for rest-frame $t=6.6$, 8.4, 10.7, 20.7 and 39.8\,days after
explosion, respectively.  Deviations typically have amplitudes
$\gtrsim$100\%, much larger than the deviations due to microlensing.

\end{appendix}

\end{document}